\documentclass[10pt]{article}

\usepackage{natbib} \bibpunct[]{(}{)}{,}{a}{}{,}
\usepackage{graphicx}
\usepackage[breaklinks,colorlinks,citecolor=black,linkcolor=black,urlcolor=black]{hyperref}
\usepackage{multicol}

\usepackage[footnotesize,bf,FIGBOTCAP]{subfigure}

\usepackage[subfigure]{tocloft}
\newlistof{affiliation}{loa}{}
\newcommand*{\newaffiliation}[2]{%
\stepcounter{affiliation}%
\newcounter{#1}%
\setcounter{#1}{\value{affiliation}}%
\addtocontents{loa}{\protect$^\arabic{#1}$\hspace{0.1ex}#2\vspace{-1ex}\newline}%
}
\newcommand*{\affiliate}[1]{%
$^\arabic{#1}$%
}
\makeatletter
\def\@maketitle{%
   \newpage
   \null
   \begin{center}%
   \let \footnote \thanks
     {\LARGE \@title \par}%
     \vskip 1.5em%
     {\large
       \lineskip .5em%
       \begin{tabular}[t]{c}%
         \@author
       \end{tabular}\par}%
    \begin{flushleft}%
     \vskip -2.75em%
     {\scriptsize\listofaffiliation}%
     \end{flushleft}%
     \vskip -0em%
    {\large \@date}%
   \end{center}%
   \par
   \vskip 1.5em}
\makeatother

\newcommand*{\mysub}[2]{\ensuremath{#1_{\mathrm{#2}}}}
\newcommand*{\unit}[1]{\ensuremath{\mathrm{\, #1}}}

\newcommand*{\Omegam}{\mysub{\Omega}{m}}
\newcommand*{\fgas}{\mysub{f}{gas}}
\newcommand*{\LCDM}{\ensuremath{\Lambda}CDM}
\newcommand*{\rhocr}{\mysub{\rho}{cr}}
\newcommand*{\NH}{\mysub{N}{H}}
\newcommand*{\dA}{\mysub{d}{A}}

\newcommand*{\second}{\unit{s}}

\newcommand*{\Ms}{\unit{Ms}}
\newcommand*{\erg}{\unit{erg}}
\newcommand*{\cm}{\unit{cm}}
\newcommand*{\km}{\unit{km}}
\newcommand*{\kpc}{\unit{kpc}}
\newcommand*{\Mpc}{\unit{Mpc}}

\newcommand*{\keV}{\unit{keV}}
\newcommand*{\Msun}{\ensuremath{\, M_{\odot}}}

\newcommand*{\E}[1]{\ensuremath{\times 10^{#1}}}

\newcommand*{\ltsim}{\ {\raise-.75ex\hbox{$\buildrel<\over\sim$}}\ }
\newcommand*{\gtsim}{\ {\raise-.75ex\hbox{$\buildrel>\over\sim$}}\ }
\newcommand*{\proptosim}{\ {\raise-.75ex\hbox{$\buildrel\propto\over\sim$}}\ }

\newcommand*{\secref}{Section}

\newcommand*{\eqnref}{Equation}
\newcommand*{\figref}{Figure}
\newcommand*{\tabref}{Table}

\newcommand*{\Chandra}{{\it Chandra}}
\newcommand*{\Planck}{{\it Planck}}
\newcommand*{\Suzaku}{{\it Suzaku}}
\newcommand*{\Nel}{\mysub{N}{el}}
\newcommand*{\prog}[1]{\textsc{#1}}

\defcitealias{Allen0706.0033}{A08}
\newcommand*{\arsemf}{\citetalias{Allen0706.0033}}

\newcommand*{\cosmopaper}{Paper~II}
\newcommand*{\profilespaper}{Paper~III}
\newcommand*{\calpaper}{Paper~IV}

\newcommand*{\fel}{\mysub{f}{el}} 
\newcommand*{\bel}{\mysub{\Gamma}{el}} 
\newcommand*{\SPAc}{SPA$_\mathrm{c}$}
\newcommand*{\csb}{\mysub{c}{SB}}

\usepackage{rotating}

\begin{document}

\newaffiliation{chicago}{Department of Astronomy and Astrophysics, University of Chicago, 5640 South Ellis Avenue, Chicago, IL 60637, USA}
\newaffiliation{kicp}{Kavli Institute for Cosmological Physics, University of Chicago, 5640 South Ellis Avenue, Chicago, IL 60637, USA}
\newaffiliation{kipac}{Kavli Institute for Particle Astrophysics and Cosmology, Stanford University, 452 Lomita Mall, Stanford, CA 94305, USA}
\newaffiliation{stanford}{Department of Physics, Stanford University, 382 Via Pueblo Mall, Stanford, CA 94305, USA}
\newaffiliation{slac}{SLAC National Accelerator Laboratory, 2575 Sand Hill Road, Menlo Park, CA  94025, USA}
\newaffiliation{heidelberg}{Astronomisches Rechen-Institut, Zentrum f\"ur Astronomie der Universit\"at Heidelberg, M\"onchhofstrasse 12-14, D-69120 Heidelberg, Germany}
\newaffiliation{dark}{Dark Cosmology Centre, Niels Bohr Institute, University of Copenhagen, Juliane Maries Vej 30, 2100 Copenhagen, Denmark}

\title{Cosmology and Astrophysics from Relaxed Galaxy Clusters I: \\Sample Selection}

\author{Adam B.\ Mantz,\affiliate{chicago}$^,$\affiliate{kicp}\thanks{Corresponding author e-mail: \href{mailto:amantz@kicp.uchicago.edu}{\tt amantz@kicp.uchicago.edu}} {}
  Steven W.\ Allen,\affiliate{kipac}$^,$\affiliate{stanford}$^,$\affiliate{slac} 
  R.\ Glenn Morris,\affiliate{kipac}$^,$\affiliate{slac} 
  Robert W.\ Schmidt,\affiliate{heidelberg} \\
  Anja von der Linden,\affiliate{kipac}$^,$\affiliate{stanford}$^,$\affiliate{dark}
  Ondrej Urban\affiliate{kipac}$^,$\affiliate{stanford}
}
\date{\small Accepted 2015 January 29. Received 2015 January 27; in original form 2014 November 13}

\maketitle

\begin{abstract}
  This is the first in a series of papers studying the astrophysics and cosmology of massive, dynamically relaxed galaxy clusters. Here we present a new, automated method for identifying relaxed clusters based on their morphologies in X-ray imaging data. While broadly similar to others in the literature, the morphological quantities that we measure are specifically designed to provide a fair basis for comparison across a range of data quality and cluster redshifts, to be robust against missing data due to point-source masks and gaps between detectors, and to avoid strong assumptions about the cosmological background and cluster masses. Based on three morphological indicators -- Symmetry, Peakiness and Alignment -- we develop the SPA criterion for relaxation. This analysis was applied to a large sample of cluster observations from the \Chandra{} and ROSAT archives. Of the 361 clusters which received the SPA treatment, 57 (16 per cent) were subsequently found to be relaxed according to our criterion. We compare our measurements to similar estimators in the literature, as well as projected ellipticity and other image measures, and comment on trends in the relaxed cluster fraction with redshift, temperature, and survey selection method. Code implementing our morphological analysis will be made available on the web.\footnote{\url{http://www.slac.stanford.edu/~amantz/work/morph14/}}
\end{abstract}

\section{Introduction}

Dynamically relaxed clusters of galaxies play a special role in investigations of cluster astrophysics and cosmology. While a variety of non-equilibrium processes taking place in the intracluster medium (ICM) are of astrophysical interest, it is only in the most regular systems that the large-scale, three-dimensional properties of the ICM can be studied in detail with minimal systematic uncertainties due to projection. In addition, the masses of relaxed clusters can be estimated with high precision and minimal bias. As a result, relaxed clusters have featured in a number of prominent studies of cluster astrophysics, scaling relations and cosmology \citep{Allen0110610, Allen0205007, Allen0405340, Allen0706.0033, Schmidt0405374, Rapetti0409574, Rapetti0710.0440, Vikhlinin0412306, Vikhlinin0507092, Vikhlinin0805.2207, Vikhlinin0812.2720, Arnaud0709.1561, Schmidt0610038, Mantz0909.3098, Mantz0909.3099}.

High-resolution X-ray imaging data provide a powerful tool to assess the dynamical state of the ICM. The X-rays produced by hot clusters are primarily a combination of bremsstrahlung and line emission. Because the ICM is optically thin, X-ray data carry information about the gas at all radii, albeit in projection. Furthermore, the two-body nature of bremsstrahlung emission results in local density fluctuations producing an exaggerated contrast in surface brightness. This property has enabled studies of a variety of astrophysical features in the regions of clusters where the gas density is relatively high, including shocks and cold fronts (e.g.\ \citealt{Markevitch0001269, Markevitch0110468, Markevitch0412451, Vikhlinin0008496}; see \citealt{Markevitch0701821} for a review), gas sloshing (e.g.\ \citealt{Ascasibar0603246, Roediger1007.4209, ZuHone1108.4427, Johnson1106.3489, Simionescu1208.2990, Paterno-Mahler1306.3520}), cavities (e.g.\ \citealt{Fabian0007456, Fabian0306036, Fabian0510476, McNamara0001402, Forman0312576, Forman0604583, Hlavacek-Larrondo1110.0489}; see also reviews by \citealt{McNamara0001402} and \citealt{Fabian1204.4114}), and the cool, dense cores found in some clusters (e.g.\ \citealt{Fabian1994MNRAS.267..779F, White9707269, Peres9805122, Peterson0512549}). In cluster outskirts, gas clumping (unresolved inhomogeneities) is implicated by excess X-ray brightness observed by ROSAT and \Suzaku{}, although the very low density and emissivity of the gas at large radii makes these observations comparably difficult (e.g.\ \citealt{Simionescu1102.2429, Urban1102.2430, Urban1307.3592, Walker1205.2276, Walker1203.0486}).

The increase in surface brightness provided by cool cores significantly biases X-ray searches in favor of finding relaxed clusters. While this is an advantage in some sense, the redshift-dependent selection bias imposed by an X-ray flux limit complicates efforts to estimate the degree of relaxation of the cluster population as a whole, and particularly its evolution with time (e.g.\ \citealt{Vikhlinin0611438, Santos1008.0754}). At redshifts $z\gtsim0.5$, the bulk of high-resolution X-ray observations of clusters currently target systems discovered through the Sunyaev-Zel'dovich (SZ) effect or other means (e.g.\ association with a quasar). Within these data sets, some clusters with cool cores have been identified \citep{Allen0101162, Siemiginowska1008.1739,  McDonald1208.2962, Semler1208.3368}, but constructing a complete picture of relaxed systems within the evolving cluster population remains challenging.
 
While a number of studies have identified relaxed clusters ``by eye,'' others have proposed quantitative measurements of image features to assess dynamical state. These generally fall into two categories: those which attempt to measure bulk asymmetry on intermediate scales (e.g.\ \citealt{Mohr1993ApJ...413..492, Buote9502002, Jeltema0501360, Nurgaliev1309.7044, Rasia1211.7040}), and those which attempt to assess the presence or development of a cool core (e.g.\ \citealt{Vikhlinin0611438, Santos0802.1445, Mantz2009PhDT........18M, Bohringer0912.4667}).\footnote{Cool cores are generally thought to correlate strongly with relaxation, although there exist notable examples of merging clusters containing remnant cool cores of gas, for example  Abell\,115 \citep{Forman1981ApJ...243L.133F}, Cygnus\,A \citep{Arnaud1984MNRAS.211..981A}, and the Bullet Cluster, 1E\,0657$-$56 \citep{Markevitch0110468}.} Automated algorithms based on such simple measurements are inevitably limited compared to visual classification, but their reproducibility, objectivity and particularly their straightforward applicability to data sets from large follow-up programs make them appealing.

This series of papers explores what can be learned by exploiting the most massive, relaxed galaxy clusters. Subsequent papers focus on cosmological constraints from measurements of the gas mass fraction in relaxed clusters (\cosmopaper{}, \citealt{Mantz1402.6212}), thermodynamic profiles and scaling relations of the ICM (\profilespaper{}, Mantz et~al., in prep), and the calibration of X-ray hydrostatic mass estimates using weak gravitational lensing (\calpaper{}, Applegate et~al., in prep). Here we present a new, automatic method for identifying relaxed clusters based on X-ray imaging data, and apply it in a comprehensive search of the \Chandra{} archive, in order to produce a suitable sample for this work. Our approach broadly follows others in the literature, but with particular emphasis on wide applicability (across a range in redshift and image depth), robustness against missing data (point source masks and unexposed parts of the focal plane), and independence from cosmological assumptions. For example, these considerations lead us to forgo measurements in the literature which explicitly assume the angular diameter distance to a cluster (i.e.\ the conversion of angle to metric distance) or the cluster mass (or a radius linked to the mass), or which involve centroids (highly dependent on the treatment of missing data).

In \secref~\ref{sec:data}, we describe in detail the reduction of the \Chandra{} and ROSAT X-ray data, which are also used in our subsequent papers. \secref~\ref{sec:approach} provides a broad overview of our approach to measuring the X-ray morphologies of clusters, and \secref~\ref{sec:procedure} presents the procedure in detail. We discuss the resulting measurements, compare them to other work in the literature, and devise a criterion for relaxation in \secref~\ref{sec:results}. \secref~\ref{sec:conclusion} summarizes our findings. Where cosmological calculations are necessary, we adopt a flat \LCDM{} model with Hubble parameter $H_0=70\km\second^{-1}\Mpc^{-1}$ and matter density with respect to critical $\Omegam=0.3$.

\section{Data} \label{sec:data}

For this work, we analyzed data for a large sample of galaxy clusters which have archival \Chandra{} observations (as of 1 February, 2013). Clusters were selected from the following sources:
\begin{enumerate}
\item The ROSAT Brightest Cluster Sample (BCS; \citealt{Ebeling1998MNRAS.301..881E}), with a minimum 0.1--2.4\keV{} luminosity of $2.5\E{44}\erg\second^{-1}$.
\item The ROSAT-ESO Flux Limited X-ray (REFLEX) cluster sample \citep{Bohringer0405546}, with the same luminosity threshold.
\item The Clusters In the Zone of Avoidance (CIZA) sample \citep{Ebeling2002ApJ...580..774, Kocevski0512321}, with the same luminosity threshold.
\item The MAssive Cluster Survey \citep{Ebeling0009101, Ebeling0703394, Ebeling1004.4683}.
\item The 400 Square Degree ROSAT survey (400d; \citealt{Burenin0610739}).
\item The South Pole Telescope (SPT) SZ cluster survey (\citealt{Bleem1409.0850}).
\item The cluster sample of \citet[][, hereafter \arsemf{}]{Allen0706.0033}.
\end{enumerate}

Our reduction of the \Chandra{} data is described below in \secref~\ref{sec:chandra}. The imaged field of view prohibits the use of \Chandra{} data alone for morphological studies of very nearby clusters (redshifts $z\ltsim0.05$, in practice). For a small number of clusters at low redshifts, we have therefore analyzed ROSAT Positional Sensitive Proportional Counter (PSPC) data, as described in \secref~\ref{sec:rosat}. Due to the low resolution of PSPC, additional caveats apply to these results, as discussed in \secref~\ref{sec:rosatres}. In total, we reduced and analyzed data for 361 clusters. \tabref{}s~\ref{tab:data} and \ref{tab:rosatdata} list the clusters and observations employed here.

\subsection{Reduction of \Chandra{} Data} \label{sec:chandra}

We used version 4.4 of the \Chandra{} software analysis suite, \prog{ciao},\footnote{\url{http://cxc.harvard.edu/ciao/}}  and version 4.4.10 of the \Chandra{} calibration database, \prog{caldb},\footnote{\url{http://cxc.harvard.edu/caldb/}} throughout this work. Subsequent changes to the calibration are not expected to significantly influence the imaging analysis presented here.

In order to ensure a uniform data reduction, and to obtain the benefits of calibration updates, all data were re-reduced to create new events files. Starting from the data products in the \Chandra{} archive, the data were processed using the method outlined in the ``ACIS [Advanced CCD Imaging Spectrometer] Data Preparation'' \Chandra{} analysis guide.\footnote{\url{http://cxc.harvard.edu/ciao/guides/acis_data.html}}

The regenerated level-2 events files were screened for periods of high background by filtering their light curves using the \prog{lc\_clean} tool. In detail, we begin by selecting a CCD to use for cleaning. Normally, this is the S1 chip for ACIS-S exposures, and the I0 or I2 chip for ACIS-I exposures. There are, however, many exceptions to this, dependent on the specific configuration of each observation. In some ACIS-S exposures the S1 chip is not active, and in these cases we use a relatively source-free area of the S3 chip where possible. For some ACIS-I exposures of low-redshift clusters, where the cluster fills some or all of the detector, we use the S2 chip.

We visually inspect the chip and mask out any sources of astrophysical emission (point sources, cluster emission, etc.), and any bad pixels, cosmic rays etc.\ that were not removed during the reduction phase. We then produce a light curve, using the same parameters as were used to make the \Chandra{} blank-sky background data sets,\footnote{\url{http://cxc.cfa.harvard.edu/ciao/threads/acisbackground/}} i.e.\ for front-illuminated (FI) CCDs the energy range 0.3--12\keV{}, and a time bin of 259.28\,s; and for back-illuminated (BI) CCDs the energy range 2.5--6\keV{} (S1 chip) or 2.5--7\keV{} (S3 chip), and a time bin of 1037.12\,s. (The different sets of parameters are motivated by the different sensitivities of the FI and BI chips to background flaring.)

We then apply the \prog{lc\_clean} tool with default settings: initial mean calculated using $3 \sigma$ clipping, followed by removal of intervals where the count rate is more than a factor of 1.2 different from the mean. In all cases, we visually inspected the resulting light-curves and checked that they were reasonable. The automatic clipping algorithm is sometimes misled by periods of exceptionally high background flaring. In cases like these, we manually exclude the time period corresponding to the flare, and/or manually set the initial mean to the correct quiescent level.

\begin{figure*}
  \centering
  \includegraphics[scale=1]{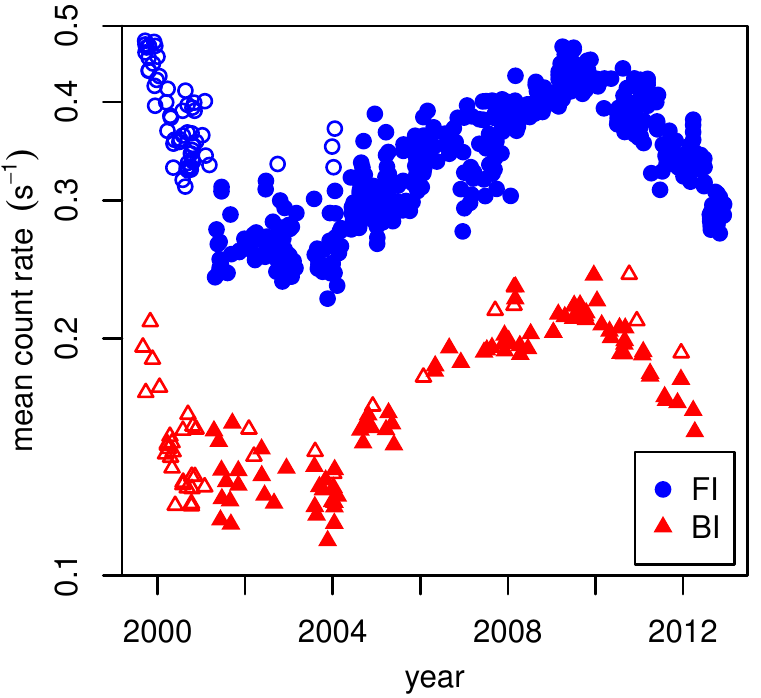}
  \hspace{1cm}
  \includegraphics[scale=1]{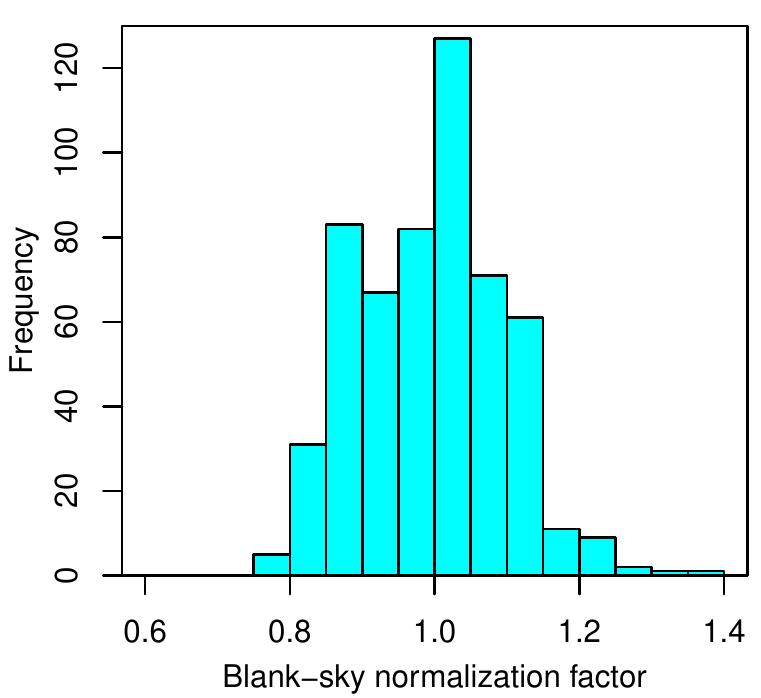}
  \caption[]{
    Left: Mean background count rates after light-curve filtering for the front-illuminated (FI) and back-illuminated (BI) ACIS CCDs, in the energy bands given in the text. Filled and open symbols respectively denote observations taken in VFAINT and FAINT modes. Observations where significant cluster emission is present on all active CCDs are excluded from the plot. Assuming that source-free regions of the detector exist, strong deviations from these trends (which are dominated by the solar cycle) would indicate extended periods of high background, allowing such observations to be excluded from the analysis.
    Right: Histogram of blank-sky scaling factors. These factors are the ratio of the 9.5--12\keV{} count rates for the science events file relative to that in the blank-sky events file. There is some chip-to-chip variation in the factor; those plotted here are from S3 for ACIS-S exposures, and I0--3 for ACIS-I exposures.
  }
  \label{fig:meanctr}
  \label{fig:bgscale}
\end{figure*}

For every exposure, we carry out this process for at least two CCDs, and check that they give consistent results. If both FI and BI chips are active, we always examine at least one of each type. Since the BI chips have a higher sensitivity to flares, the BI good-time interval (GTI) is generally applicable to the FI chips as well, but in a few cases we use separate GTIs for the FI and BI chips. As a final safety precaution, we check that the mean level of the light curve after filtering is reasonable,  since there are sometimes extended periods of high background which are difficult to detect in short exposures. These values are shown in the left panel of \figref~\ref{fig:meanctr}, as a function of the date of the observation. Values are per-CCD, corrected for any fraction of the chip area that was excluded.

The overall trend as a function of time (high at the start of the mission, before $\sim$2001, then fairly flat from 2001--2003, then rising until $\sim$2010, then declining again) is representative of the evolution of the \Chandra{} background, which is influenced by the solar cycle. In addition to this overall shape, FAINT-mode exposures tend to have a higher rate than VFAINT-mode exposures from the same epoch.

For some very extended, low-redshift clusters where there is essentially no region of the detector free from cluster emission, the rates are somewhat elevated due to cluster contamination (these are excluded from the figure). In these cases, all we can do is check that the light curve looks reasonable, and that excluding larger fractions of the chip in the direction of the cluster center reduces the normalized rate.

On the basis of these checks, we exclude a minority of obsids from further analysis, generally because they are either extensively flared or suspected to be affected by flares, and only represent a small fraction of the data that exist for the target in question (these are noted in \tabref~\ref{tab:data}). Finally, any non-cluster sources in the analyzed fields were masked out by visual inspection of the cleaned events files.

To account for possible variations between the blank-sky background exposures and the science exposures, we normalized the blank-sky files using the high-energy count rates, which should measure the overall level of the particle background. Specifically, we apply a multiplicative factor derived from the ratio of the 9.5--12\keV{} count rates in the science and blank-sky files. (Note that in background period A, and in a small number of science exposures, only events up to 10\keV{} were telemetered.) These scaling factors typically lie in the range 0.8--1.2, as shown in the right panel of \figref~\ref{fig:bgscale}. We find some evidence for chip-to-chip, and indeed node-to-node, variations in the scaling factors, but there are no clear trends. For detailed spectral analysis in subsequent papers, we use per-CCD scaling factors; here, for our basic imaging analysis, we take the more straightforward approach of adopting a single mean scaling per observation for all FI or BI chips.

Note that in background epochs A--C, the blank-sky events files are in FAINT mode. In order to use these blank-sky files with science exposures, the science events files must also be processed in FAINT mode, i.e.\ the VFAINT correction cannot be applied even if available (resulting in somewhat noisier data than would otherwise be the case; see \figref~\ref{fig:bgscale}). Such exposures are indicated by ``V*'' in \tabref~\ref{tab:data}.

\subsection{Reduction of ROSAT Data} \label{sec:rosat}

The ROSAT PSPC observations were reduced using the Extended Source Analysis Software package of \citet{Snowden1994ApJ...424..714}. In short, we identify good time intervals using a master veto threshold of 170\,counts$\second^{-1}$, to exclude times of anomalously high particle background rates, and a time delay of 15\,s, to remove the events at the beginning of each observation before the detector high voltage achieved its nominal level. We create light curves for each the seven standard ROSAT bands, and compute a list of nominal scattered solar X-ray (SSX) background count rates, under the assumption that the residual atmosphere along the line of sight is optically thin. The solar X-ray spectrum is modelled as a two temperature thermal plasma, with individual temperatures of $10^{5.7}$\,K and $10^{6.2}$\,K. By inspecting the light curves of the SSX background count rates, we identify and exclude periods of intense SSX contamination. In the remaining time intervals, we model the X-ray background in the nominal energy bands of 0.7--0.9, 0.9--1.3 and 1.3--2.0\,keV (standard ROSAT bands R5--R7), using the standard assumption that the background consists of a cosmic component, the calibrated particle background, a SSX component and a possible long-term enhancement (where required). These models are used to generate background count rate maps. Note that these background maps are not equivalent to the blank-sky maps available for \Chandra{}, since they do not account for the astrophysical background; this leads to small differences in our analysis of the ROSAT images in \secref~\ref{sec:boot}.

\section{General Approach} \label{sec:approach}

\subsection{Preliminaries}

Our procedure for characterizing the morphology of galaxy clusters, detailed in the next section, is guided by a few broad principles. (1) It should provide a fair basis to compare clusters spanning a wide range of redshift and mass, and using data of variable quality. Thus, very nearby clusters should not be penalized because we can discern detailed structure within them that would not be resolved at higher redshift. The most crucial step to achieving this is identifying comparable regions of different clusters, which is described in \secref~\ref{sec:sbradii}. Additionally, because the gaps between \Chandra{} CCDs generally mask part of the cluster emission at redshifts $\ltsim0.25$, we avoid the use of centroids and other quantities which assume complete images. (2) As much as possible, the algorithm should be insensitive to the prevalence of Poisson noise, to avoid unfairly penalizing clusters  imaged with shallow exposures or located at high redshifts. Integral to meeting this requirement is the robust estimation of measurement uncertainties, which we address by bootstrapping the input photon images, as detailed in \secref~\ref{sec:boot}. (3) Since the main purpose of this work is to identify a relaxed cluster sample to use for cosmological studies, it is also advantageous to avoid strong assumptions about either the mass (or virial radii) of the clusters, or the background cosmology.

The particular quantities that we calculate from the cluster images are designed to measure the features on which subjective determinations of relaxation are generally based. In general terms, these are:
\begin{enumerate}
\item the sharpness of the peak in surface brightness.
\item the shifting of isophotes with respect to one another (i.e.\ the appearance of sloshing).
\item the distance between the center of symmetry on large scales (a low brightness isophote) and small scales (e.g.\ the cool core, if any).
\end{enumerate}
\secref{}s~\ref{sec:peakiness} and \ref{sec:ellipses} provide more complete details of the measurements, which are carefully designed to respect the ``fair comparison'' requirement above. In practice, this suite of three relatively simple calculations performs well, and the close connection between the measurements and visible features aids their interpretation.

The particular thresholds for the measured values that we adopt to identify relaxed clusters are roughly placed with reference to prior, subjective decisions. Once in place, however, the thresholds are applied without regard to any subjective determinations. We assess the performance of the algorithm both by whether its decisions are subjectively reasonable, and, more pertinently, by comparing the measured intrinsic scatter of the gas mass fraction for the new relaxed sample with the subjectively identified sample of \arsemf{}; this comparison was made only \emph{after} the new sample was finalized. As described in \secref~\ref{sec:fgascompare}, the algorithmically identified sample has a somewhat smaller intrinsic scatter than the \arsemf{} sample. Although it is beyond the scope of this work, testing our algorithm against mock X-ray images of simulated galaxy clusters can potentially provide further refinements.

\subsection{Standardizing Cluster Surface Brightness}\label{sec:sbradii}

Outside of their central regions, the surface brightness profiles of galaxy clusters are approximately self-similar (e.g.\ \citealt{Vikhlinin0507092, Croston0801.3430}). This raises the possibility of identifying characteristic radii that are comparable across clusters via the surface brightness. To that end, we motivate a redshift- and temperature-dependent scaling of surface brightness based on the self-similar model of \citet[][, see also \citealt{Santos0802.1445}]{Kaiser1986MNRAS.222..323K}.

The average surface brightness within a circular aperture of angular radius $\theta$, corresponding to physical radius $r=\theta\,\dA(z)$, for a cluster with redshift $z$ and angular diameter distance $\dA(z)$, is
\begin{equation}
  S = \frac{F}{\pi\theta^2} \propto \frac{K(z,T,\NH) L}{(1+z)^4r^2},
\end{equation}
where $F$ and $L$ are, respectively, the observer-frame flux and rest-frame bolometric luminosity of the cluster. Here the coefficient $K$ accounts for the redshift- and temperature-dependent K-correction from bolometric flux to flux in the observed energy band, as well as any Galactic absorption (equivalent absorbing hydrogen column density, \NH{}). For self-similar profiles, this proportionality also holds for surface brightness at a given characteristic radius, $r_\Delta$, defined in terms of the cluster mass and the critical density of the Universe; $M(<r_\Delta)=(4/3)\pi \Delta \rhocr(z) r_\Delta^3$. Using scalings from the Kaiser model,
\begin{eqnarray} \label{eq:simscaling}
  L & \propto & T^2 E(z), \\
  r_\Delta & \propto & \frac{T^{1/2}}{E(z)}, \nonumber
\end{eqnarray}
where $E(z) = H(z)/H_0$ is the normalized Hubble parameter, we can eliminate $L$ and $r_\Delta$ in favor of the ICM temperature, $T$. This yields the relation $S(r_\Delta) \propto f_S$, with
\begin{equation} \label{eq:sbscal}
  f_S \equiv K(z,T,\NH) \frac{E(z)^3}{(1+z)^4} \left(\frac{kT}{\mathrm{keV}}\right) ~\mathrm{photons}\Ms^{-1}\cm^{-2}\,(0.984\,\mathrm{arcsec})^{-2},
\end{equation}
where we have assigned units which are convenient for the analysis of \Chandra{} data (see \secref~\ref{sec:boot}).
Following the argument above, surface brightness levels corresponding to constant multiples of $f_S$ should correspond to approximately the same values of $\Delta$ across all clusters, provided they fall in the self-similar part of the profile.

With this rescaling, it becomes possible to identify approximately corresponding regions of clusters with different masses and redshifts, without explicitly assuming the angular diameter distance to each or a prescription for estimating some scale radius $r_\Delta$ (equivalently $M_\Delta$). There is an implicit assumption of cosmological parameters necessary to evaluate $E(z)$, but this sensitivity is relatively mild. As input, we need only the redshifts of clusters, column densities for their positions on the sky, and rough temperature estimates for them.\footnote{Approximate temperatures are sufficient, since the product $K(z,T,\NH)\,kT \proptosim T^{1/2}$ (for temperatures characteristic of the clusters in our data set, for which there is negligible line emission at soft energies) varies slowly with $kT$.}

\begin{figure}
  \begin{minipage}[c]{0.5\textwidth}
    {\centering
      \includegraphics[scale=1]{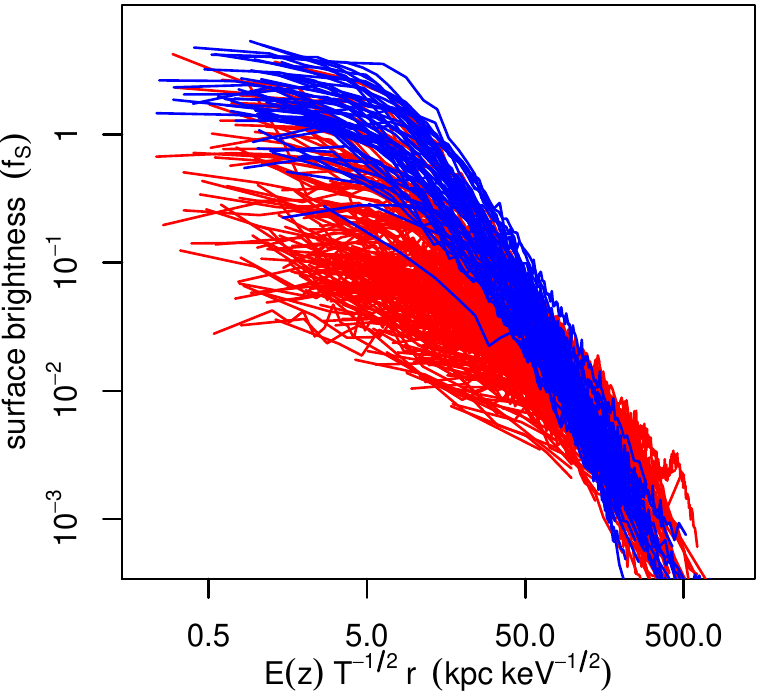}
    }
  \end{minipage}
  \begin{minipage}[c]{0.5\textwidth}
    \caption[]{
      Scaled, background subtracted surface brightness profiles from our \Chandra{} analysis. The scaling factors follow from the self-similar model and are given in \eqnref{}s~\ref{eq:simscaling}--\ref{eq:sbscal}. Clusters that are ultimately categorized as relaxed in this work are shown in blue, and others in red. Measurement errors on the individual profiles are not shown (but see \figref~\ref{fig:sbprof}).
    }
  \end{minipage}
  \label{fig:allprofiles}
\end{figure}

As an a posteriori check of how reasonable this scaling is, \figref~\ref{fig:allprofiles} shows surface brightness profiles from our analysis of \Chandra{} data (\secref~\ref{sec:sbprofile}). The surface brightness values are background-subtracted and shown in units of $f_S$, and the radial coordinate is scaled by $E(z)/\sqrt{T}$ according to \eqnref~\ref{eq:simscaling}.\footnote{Note that the conversion of angular to metric distance introduces an additional cosmology-dependent factor of $\dA(z)$ in the radial coordinates.} The intrinsic scatter among profiles is significant at small radii, tightening to a self-similar profile at large radii. The clusters that are ultimately identified as relaxed in this work form a particularly tight locus.

\section{Procedure} \label{sec:procedure}

This section describes in detail our procedure for measuring morphological indicators and their uncertainties.

\subsection{Data Preparation and Bootstrapping} \label{sec:boot}

For the \Chandra{} observations, images in the 0.6--2.0\keV{} band are extracted from both the cleaned science and blank-sky event files, and are binned by a factor of two (obtaining $\approx1$ arcsec resolution). An appropriate exposure map is generated for the same energy range. Off-chip pixels and pixels contaminated by point sources are flagged in the science images. These files, along with the blank-sky normalization factor and its statistical error, serve as input to our morphological algorithm.

All the steps described below are performed on 1000 bootstrap realizations of each observation. We bootstrap the science and blank-sky images at the level of individual counts; that is, the pixel locations of each detected photon in the original image are listed (with repetition, as appropriate), and photons are added to pixels of the bootstrap image by sampling from this list with replacement. For each bootstrap iteration, we also sample a new value of the blank-sky normalization factor, based on its statistical uncertainty.

To estimate statistical signal-to-noise throughout the analysis, we keep track of the variance in various quantities, beginning with the counts in the images. We assign the statistical variance $N+1$ to each pixel of the science and blank-sky images, where $N$ is the number of counts in the corresponding pixel. This choice is motivated by the fact that the Bayesian posterior for the average number of counts in an equal-length exposure, based on the observed counts $N$, has variance $N+1$;\footnote{More specifically, $N+1$ is the variance when the prior is chosen to be a flat Gamma distribution such that the posterior is maximized at $N$ (shape parameter $k=1$ and rate parameter $\beta=0$).} furthermore, it neatly avoids the pathological assignment of zero uncertainty to pixels with zero counts. Note that our final uncertainties are entirely characterized by the bootstrap procedure; the error maps described here only provide approximate signal-to-noise estimates for, e.g., the surface brightness profile fitting and adaptive smoothing steps below.

The blank-sky image is rescaled according to the normalization factor and subtracted from the science image (recall that each of these ingredients is bootstrapped), propagating the variance of the background-subtracted image straightforwardly. The result is divided by the exposure map, assuming no uncertainty in the latter, to create a flat-fielded image. At this stage, it is possible to straightforwardly combine images from multiple observations of a cluster by the same telescope. Finally, we convert the brightness images to intensity in units of photons\,$\Ms^{-1}\cm^{-2}\,(0.984\,\mathrm{arcsec})^{-2}$.\footnote{The particular choice of units here is purely for convenience, as it makes the intensity of a typical cluster center of order unity, and simplifies the case of $2\times2$ binned \Chandra{} images.} 

For ROSAT observations, our procedure differs in a few details. The ROSAT images cover the 0.7--2.0\keV{} energy band and have the native PSPC resolution of 14.9 arcsec. Since there are no blank-sky fields, we subtract the ROSAT particle background rate maps from the images after converting the latter to count rates but before flat fielding (since the particle background is not vignetted). A spatially constant residual background level, accounting for unresolved astrophysical sources, is fit and subtracted at a later stage (see \secref~\ref{sec:sbprofile}).

\subsection{Center Finding and Surface Brightness Profiling} \label{sec:sbprofile}

A global center for each cluster is defined by computing the median photon location in an iteratively shrinking aperture. Beginning with the entire image, the center is defined as $(\tilde{x},\tilde{y})$, where $\tilde{x}$ ($\tilde{y}$) is calculated by summing the image over columns (rows), shifting the resulting one-dimensional array to be non-negative, and computing the median of the resulting discrete function of $x$ ($y$). A new image is extracted, centered on $(\tilde{x},\tilde{y})$ but with dimensions smaller by a factor of $2/3$ (or more if the edge of the image is encountered), and the procedure is repeated until a minimum aperture size of 40 pixels square has been reached and the center is static.

In practice, this median center compromises between two widely used alternatives, the brightest pixel and the centroid. In clusters having a cool core that is offset from the center of emission on larger scales, the median center tends to be located within the cool core, although not necessarily at its center or brightest point. Like the centroid, it does respond to a degree to the weight of emission in the fainter regions of the cluster. However, the median center is much less biased by the presence of masked regions than the centroid, to the extent that ``filling in'' masked regions and gaps between detectors is generally unnecessary. Compared to simply choosing the brightest pixel, the median procedure has the clear advantage that it is less susceptible to Poisson noise or mistakenly unmasked point-source emission.

An azimuthally averaged surface brightness profile about the median center is calculated in annuli which are adaptively chosen to provide a signal-to-noise ratio $>2$ (with a single, signal-to-noise $<2$ annulus covering the largest imaged radii). A $\beta$ model \citep{Cavaliere76} plus constant background level are then fitted to the radially outermost half of the profile,\footnote{Blindly fitting the outermost half of the profiles works well in general for \Chandra{} data, where the blank-sky background subtraction is typically approximately correct. Given the adaptive binning of the profile, the outer half tends to span the power-law tail of the cluster and any residual foreground/background, e.g.\ Galactic contamination which is not included in the blank-sky maps. In rare cases where the residual constant term has very high signal-to-noise (so that the outer half of the adaptively binned profile excludes too much of the cluster), we manually choose the radial range of the profile to use in this step.} and the best-fitting constant is subtracted from the image and surface brightness profile. When brightness levels are compared to the surface brightness profile in the following sections, we compare to the measured profile at radii where it was constrained with the target signal-to-noise, and to the $\beta$ model at larger radii. Similarly, when random values are drawn to be consistent with the profile at a given radius, we scatter them according to the measurement uncertainty for the appropriate annulus, or the outermost annulus in the case of extrapolation.

\begin{figure*}
  \centering
  \includegraphics[scale=0.79]{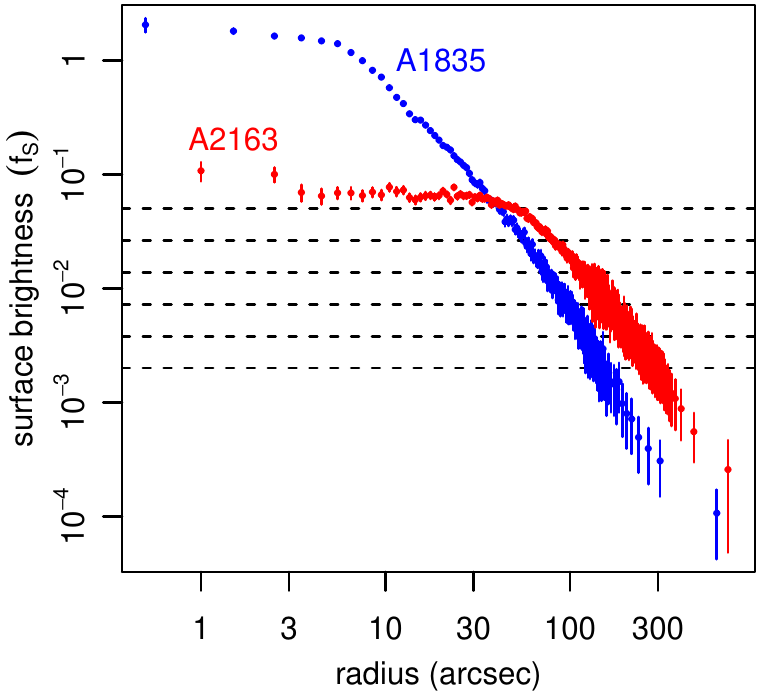}
  \includegraphics[scale=0.73]{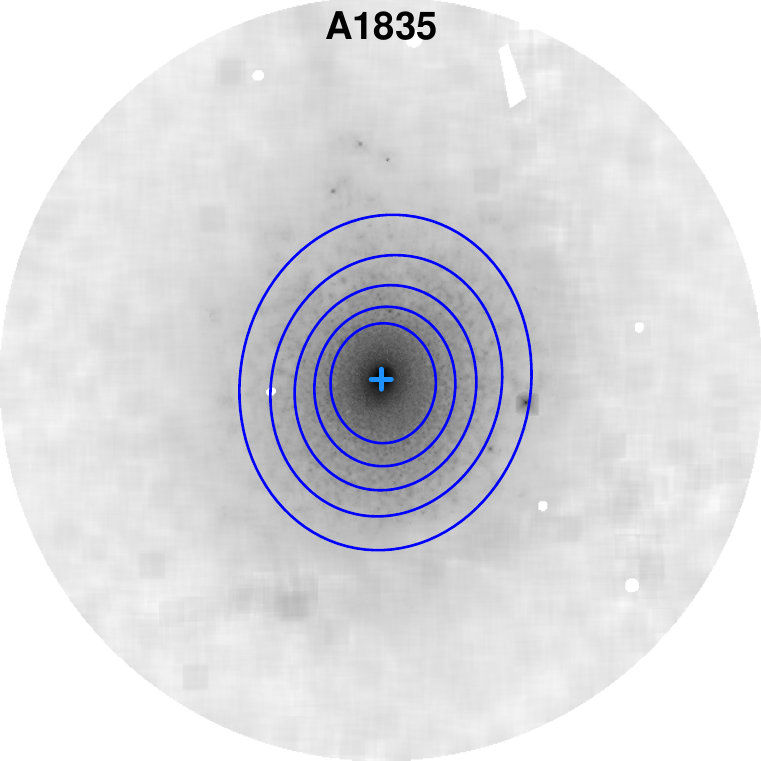}
  \includegraphics[scale=0.73]{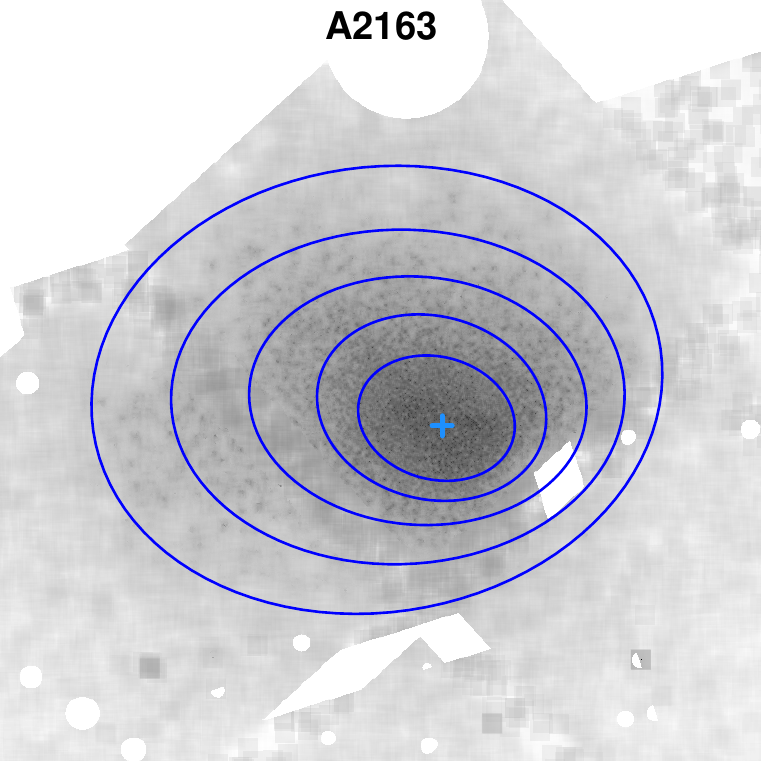}
 \caption[]{
   Left: Surface brightness profiles for Abell\,1835 and Abell\,2163, scaled according to \eqnref~\ref{eq:sbscal}. Dashed lines correspond to the brightness levels defined in \eqnref~\ref{eq:sblevels} (the lowest level corresponds to $j=0$). Our peakiness metric depends on the average surface brightness in a circular aperture whose radius is given by the intersection of the observed profile with the highest level ($j=5$).
   Center and right: Smoothed images of the same clusters, logarithmically scaled. Crosses show the global centers determined in \secref~\ref{sec:sbprofile}. Also shown are the ellipses fit to each of five isophotes, defined as the ranges between the surface brightness levels given in \eqnref~\ref{eq:sblevels}. The alignment metric reflects how close the centers of these ellipses lie to one another, while the symmetry metric reflects how well they agree with the global center.
   These clusters exemplify very relaxed (Abell\,1835) and very unrelaxed (Abell\,2163) morphologies, according to all three quantities.
  }
  \label{fig:sbprof}
  \label{fig:ellipses}
\end{figure*}

Following the argument in \secref~\ref{sec:sbradii}, we define a set of characteristic surface brightness levels in our adopted units,
\begin{equation} \label{eq:sblevels}
  S_j = 0.002 \times 10^{0.28j} f_S,
\end{equation}
where $j=0,1,\ldots,5$. The number of levels and the range in surface brightness that they span were chosen empirically to provide good performance for the measurement of our morphological estimates (described in the following sub-sections) over a wide range of data quality and cluster redshifts. These scaled surface brightness levels are shown along with example profiles for Abell\,1835 (which has a cool, bright core) and Abell\,2163 (which has a flat core) in the left panel of \figref~\ref{fig:sbprof}.

\subsection{Surface Brightness Peakiness} \label{sec:peakiness}

The presence of a core of bright, relatively cool, X-ray emitting gas in the center of a cluster is a common signature of dynamically relaxed systems \citep{Fabian1994MNRAS.267..779F, Peterson0512549}. The formation of these features is expected, and to some extent observed, to be disrupted by major mergers \citep{Burns0708.1954, Henning0903.4184, Million0910.0025, Rossetti1106.4563, Skory1211.3117, Ichinohe1410.1955}. Thus, while cool cores are not necessarily completely destroyed by major mergers once formed, requiring the presence of a core should provide an efficient way to reject unrelaxed clusters.

Although measuring a temperature decrement in the center of a cluster is relatively involved, detecting the presence of a central brightness enhancement is straightforward. Consequently, simple measurements of the sharpness of the peak in surface brightness at cluster centers have been widely employed as a proxy for the presence of cool cores. Various measurements of peak strength have been introduced. \citet{Vikhlinin0611438} used the logarithmic slope of the gas density profile at a radius of $0.04\,r_{500}$. \citet{Santos0802.1445} advocate using the ratio of fluxes contained in two metric apertures; flux ratios in apertures linked to $r_{500}$ have also been employed (e.g.\ \citealt{Mantz2009PhDT........18M, Bohringer0912.4667}).

For the present work, the explicit reliance of each of these approaches on metric distances (i.e.\ on an assumed angular diameter distance) or scale radii ($r_{500}$) is a disadvantage. Instead, we introduce a measurement which relies only on the scaled surface brightness profile in the region where it is typically very well constrained, as follows. First, we determine the angular radius, $\theta_5$, where the measured surface brightness profile is equal to $S_5$, as defined in the previous section; if the profile never exceeds this value, then the radius bounding the innermost bin of the surface brightness profile is used. We then calculate the average surface brightness at distances $\leq \theta_5$ from the global center of \secref~\ref{sec:sbprofile} in units of $f_S$, assigning to each masked pixel in this region a random value based on the surface brightness profile and its uncertainty at the appropriate radius. (This calculation is statistically equivalent to taking the area-weighted average of the surface brightness profile at radii $\leq\theta_5$.)

\begin{figure}
  \begin{minipage}[c]{0.4\textwidth}
    {\centering
      \includegraphics[scale=0.85]{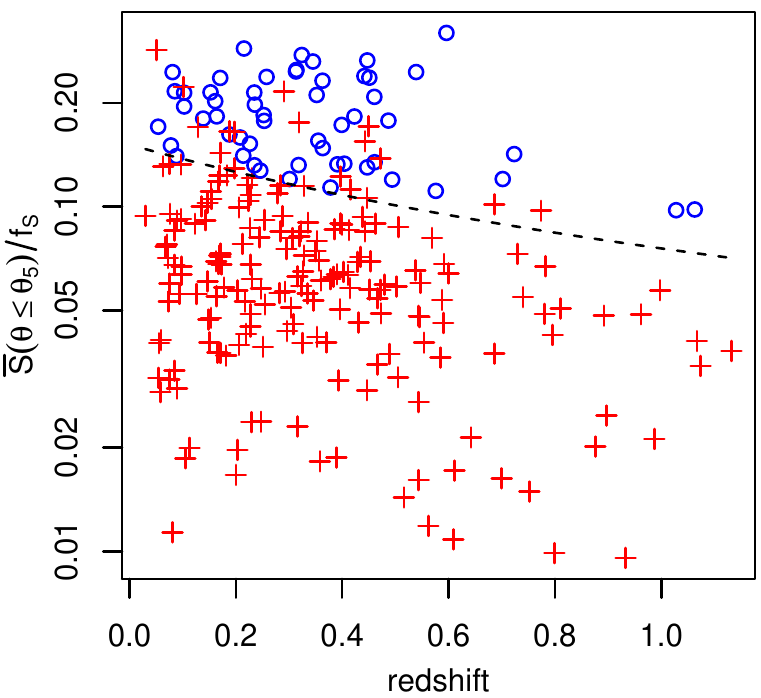}
    }
  \end{minipage}
  \begin{minipage}[c]{0.6\textwidth}
    \caption[]{
      Average central surface brightness in scaled units as a function of redshift from our \Chandra{} analysis. Clusters that we ultimately classify as relaxed (\secref~\ref{sec:morph_criterion}) are shown as blue circles, and others as red crosses. A net decreasing trend can be seen, qualitatively in agreement with observations based on similar surface brightness measurements \citep{Santos0802.1445, Santos1008.0754, McDonald1305.2915}. Our peakiness measure incorporates a $1+z$ weighting to approximately compensate for this evolution in core brightness; the dashed line corresponds to the constant-peakiness threshold used to define the relaxed sample in \secref~\ref{sec:morph_criterion}.
    }
  \end{minipage}
  \label{fig:censbav}
\end{figure}

\begin{figure*}
  \centering
  \hspace{17mm}
  \includegraphics[scale=0.7]{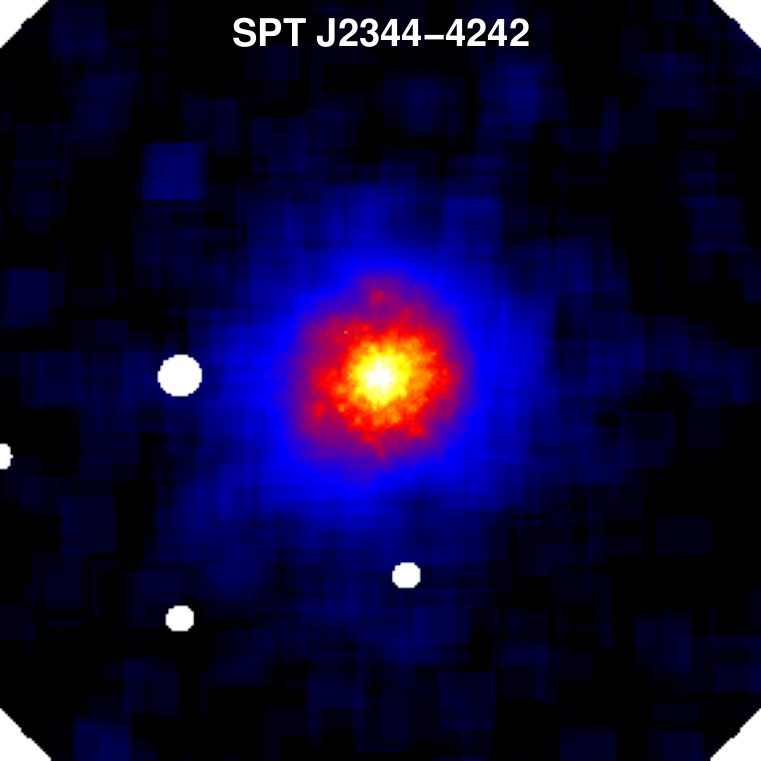}
  \hspace{2mm}
  \includegraphics[scale=0.7]{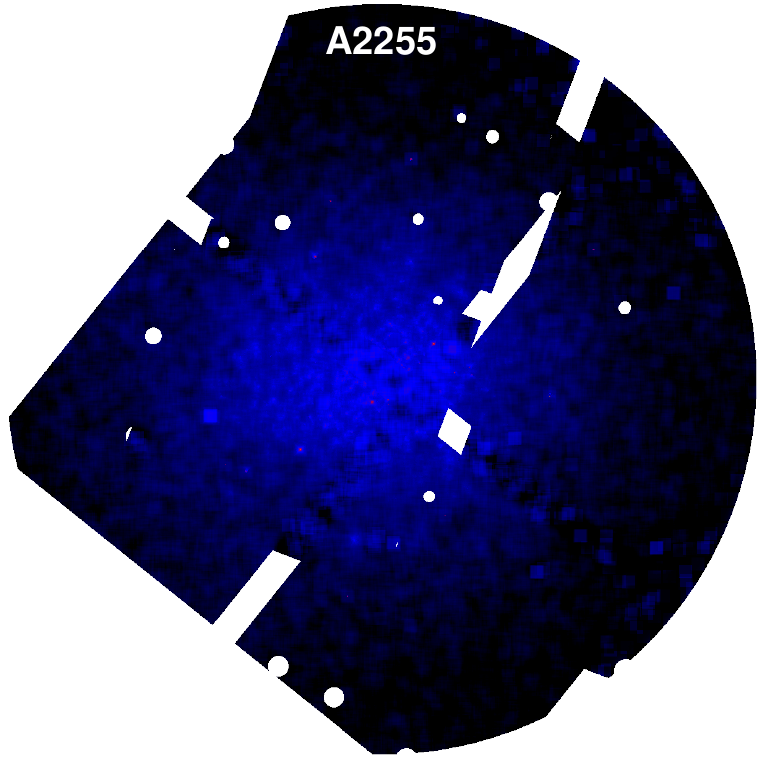}
  \hspace{2mm}
  \includegraphics[scale=0.7]{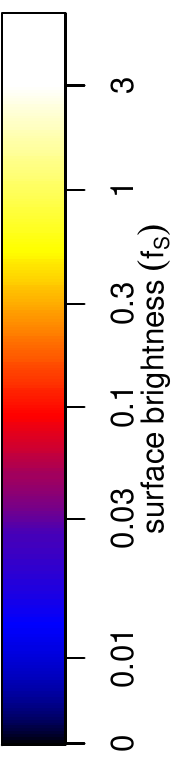}
  \vspace{2mm}\\
  \hspace{17mm}
  \includegraphics[scale=0.7]{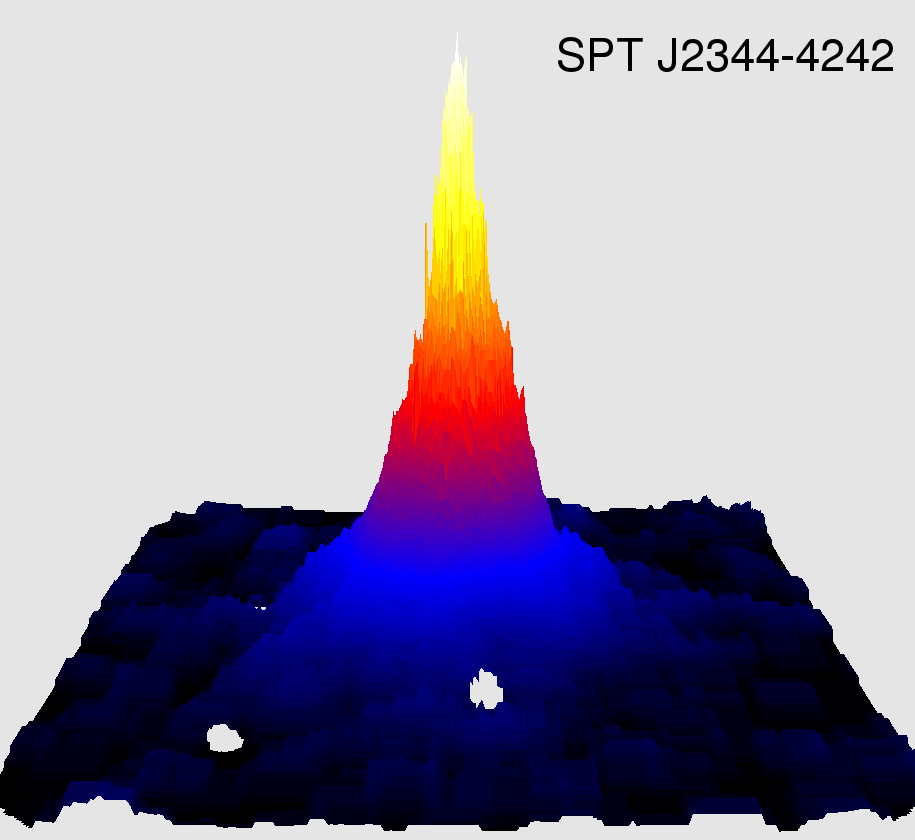}
  \hspace{2mm}
  \includegraphics[scale=0.7]{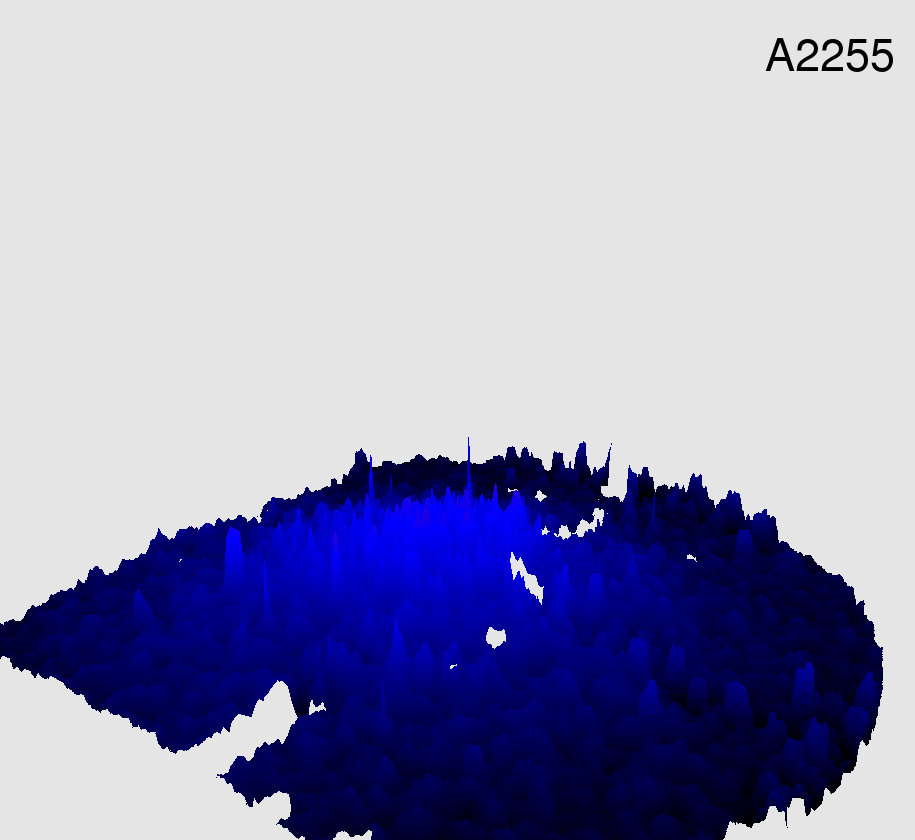}
  \hspace{17mm}
  \caption[]{
    The most and least peaky clusters in our \Chandra{} sample: SPT\,J2344$-$4242 ($p=-0.29$; left column) and Abell\,2255 ($p=-1.91$; right column). The top row shows the adaptively smoothed images, scaled according to \eqnref~\ref{eq:sbscal}, used in our isophote analysis (\secref~\ref{sec:ellipses}), and the bottom row shows the same images as 3-dimensional surfaces to emphasize the contrast in central brightness. The vertical direction in the bottom row, and the colors throughout, use a common logarithmic scaling.
  }
 \label{fig:extremep}
\end{figure*}

This average, scaled central surface brightness, $\bar{S}(\theta\leq\theta_5)/f_S$, shows an overall downward trend with redshift across the data set, as seen in \figref~\ref{fig:censbav}. This is expected; qualitatively similar trends have been reported in measurements of surface brightness ``concentration'' (\citealt{Santos0802.1445}; see also \citealt{ Santos1008.0754, McDonald1305.2915}), which our measurements are closely related to (see \secref~\ref{sec:otherx}). Physically, this increase of brightness with time, particularly at the high central brightness end, presumably corresponds to non-self-similar evolution in the development of cool cores in relaxed clusters. Since our procedure is intended to select morphologically relaxed clusters at any redshift, we include a redshift weighting, which in the absence of precise predictions from hydrodynamical simulations, we assume to be linear.\footnote{We have not attempted to fine-tune the redshift dependence further, since the motivation for doing so is questionable and since it would provide an opportunity to tailor the high-redshift content of our final relaxed sample (potentially biasing the cosmological results of \cosmopaper{}). However, a posteriori, it is interesting to note that the $1+z$ weighting results in an approximately constant fraction of peaky clusters with redshift (\secref~\ref{sec:trends}), seemingly in good agreement with the constant cool-core fraction predicted from simulations \citep{Burns0708.1954}. This is encouraging, as it suggests we are selecting dynamically similar clusters at each redshift.} Taking the logarithm for convenience, the surface brightness peakiness, $p$, is thus defined as

\begin{equation} \label{eq:peakiness}
  p = \log_{10} \left[ (1+z) \frac{\bar{S}(\theta \leq \theta_5)}{f_S} \right].
\end{equation}
To the extent that cluster surface brightness profiles are self-similar at radii greater than $\theta_5$, this quantity contains as much information as the ratio of flux in small and large apertures, while being measured more precisely. The particular value of $S_5$ (\eqnref~\ref{eq:sblevels}) was chosen for exactly this purpose; the divergence of the surface brightness profiles of Abell\,1835 (bright core) and Abell\,2163 (non-bright core) at radii $<\theta_5$ seen in \figref~\ref{fig:sbprof} is typical (see also \figref~\ref{fig:allprofiles}). A more extreme contrast can be seen in \figref~\ref{fig:extremep}, which compares the clusters with the lowest and highest values of $p$ from our analysis.

\subsection{Elliptical Isophote Fitting and Statistics} \label{sec:ellipses}

\begin{figure*}
  \centering
  \includegraphics[scale=0.7]{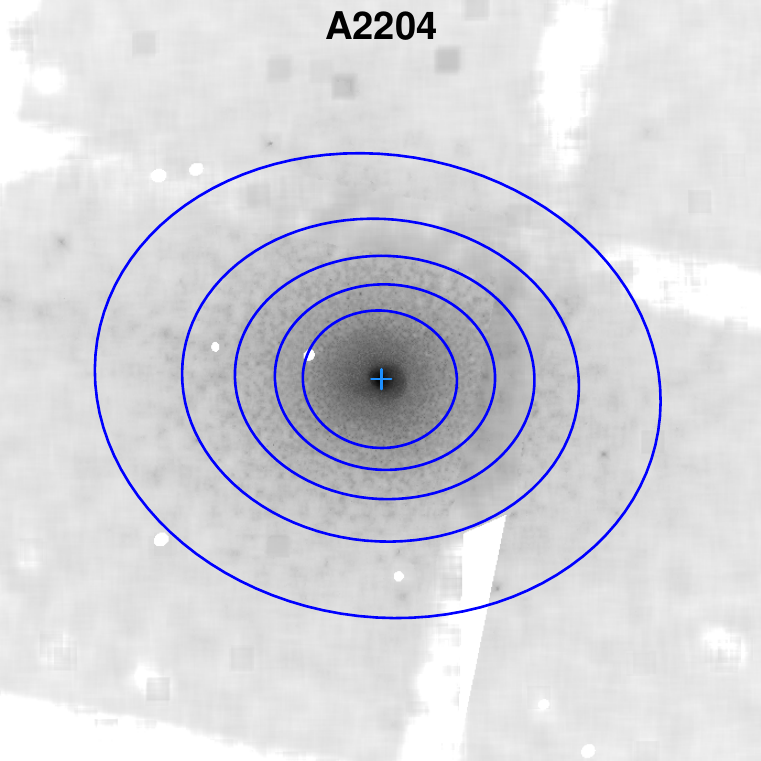}
  \hspace{4mm}
  \includegraphics[scale=0.7]{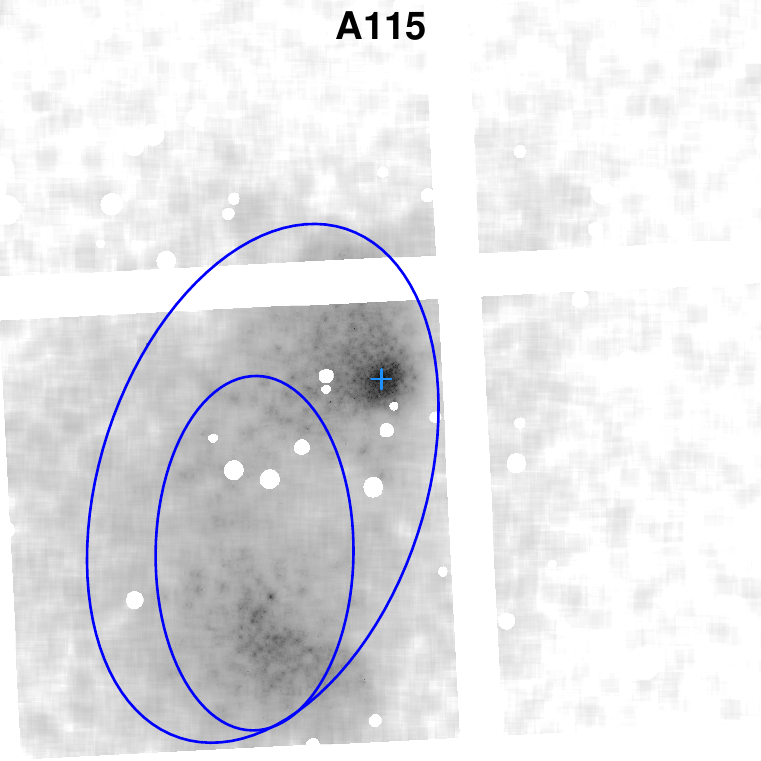}
  \hspace{4mm}
  \includegraphics[scale=0.7]{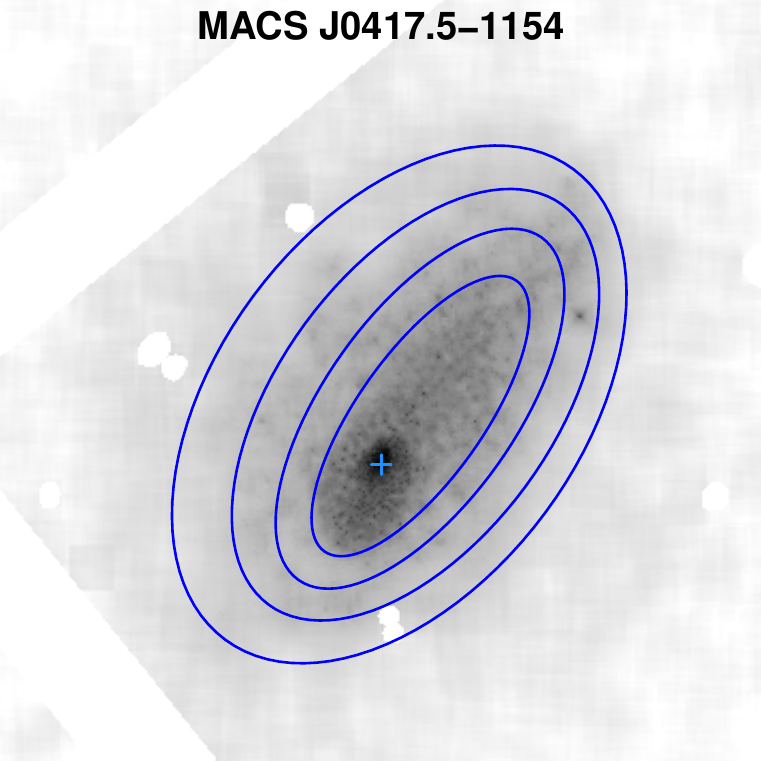}
 \caption[]{
   Left, center: Smoothed images of clusters representing the extremes of symmetry  from our \Chandra{} analysis, with elliptical isophote fits superimposed.  The clusters also represent extremes in alignment, as these quantities are strongly correlated in practice (\figref~\ref{fig:morph_cuts}). The left panel shows Abell\,2204 ($s=1.64$, $a=1.55$), and the center panel shows Abell\,115 ($s=-0.04$, $a=0.45$). In Abell\,115, the low symmetry value is due to the strong disagreement between the ellipse centers and the global center (blue cross), which is located in the cool core of one of the merging sub-clusters. The low alignment value follows from the disagreement of the ellipse centers with one another. Note that the ellipses shown correspond to isophotes that truly are (roughly) elliptical in shape (the $j=0$--1 and 1--2 isophotes in \eqnref~\ref{eq:sblevels}). The three brighter isophotes are disjoint, containing well separated groupings of pixels in the two X-ray bright clumps; our algorithm flags them as being non-elliptical and excludes them from the analysis. The right panel shows MACS\,J0417.5$-$1154, a good example of a merging cluster that has acceptable alignment but poor symmetry.
 }
 \label{fig:extremesa}
\end{figure*}

Our other morphological measurements aim to quantify the two-dimensional structure of clusters. Here again we avoid algorithms which assume complete imaging coverage, such as the centroid variance \citep{Mohr1993ApJ...413..492} and various other measures of substructure and asymmetry (e.g.\ \citealt{Nurgaliev1309.7044, Rasia1211.7040}), as masked point sources or the gaps between adjacent CCDs often impinge on cluster images in practice. (Indeed, \figref s~\ref{fig:ellipses}, \ref{fig:extremep}, \ref{fig:extremesa} and \ref{fig:triangle} all provide examples of this.)

Instead, our approach fits elliptical shapes to the 5 isophotes defined by the brightness levels in \eqnref~\ref{eq:sblevels}. This analysis does not use the ``filled-in'' image introduced in \secref~\ref{sec:peakiness}, since azimuthal symmetry is assumed in the production of those images. Instead, to reduce Poisson noise, we apply an adaptive boxcar smoothing algorithm to the original flat-fielded image, with a maximum kernel radius of 10 pixels and target signal-to-noise of two, enforcing that pixels masked in the original image remain masked in the final product. To prevent very distant pixels with large noise fluctuations from influencing our results, these smoothed images are cropped beyond the radius corresponding to $0.1 S_0$. We then identify pixels in the smoothed image with values in each of the 5 brightness ranges (isophotes) $S_j<S<S_{j+1}$. An elliptical shape is fit to each of these isophotes, where the fit minimizes the sum of absolute distances from the ellipse to each pixel in the isophote along the line passing through the pixel and the ellipse center.

To automatically catch cases where the ellipse fit is suspect, we compute the following two quantities. The first, \fel{}, is straightforwardly the fraction of the ellipse which falls on unmasked pixels; this is useful for identifying cases where the ellipse fit should not be trusted because most of the true azimuthal extent of the isophote was not imaged. The second quantity is $\bel = \left\langle e^{i\phi} \right\rangle$, where $\phi$ is the angle between the major axis of the ellipse and a ray from the ellipse center to a given pixel, and the average is over pixels in the corresponding isophote. This statistic measures how balanced the distribution of isophote pixels is with respect to the fitted ellipse center, and efficiently finds cases where the best-fitting ellipse simply passes as closely as possible to a very non-elliptical distribution of pixels.\footnote{Note that \bel{} is similar to the displacement between the ellipse center and the isophote centroid, under the assumption that all pixels in the isophote have exactly the same brightness.} For a given isophote and bootstrap iteration, if $\fel<0.5$ or either the real or imaginary part of \bel{} has magnitude $>0.4$, the fit is considered to have failed, and the isophote is discarded. In addition, no fit is attempted for isophotes where the lower end of the brightness range lies in the outer portion of the surface brightness profile (where the target signal-to-noise was not achieved), for isophotes where the upper end of the brightness range is greater than the central point in the surface brightness profile, or for isophotes consisting of $<100$ pixels. For an isophote to contribute to the final set of statistics for a cluster, we require it to be successfully fit in $>3/4$ of bootstrap iterations.

\begin{figure}
  \begin{minipage}[c]{0.5\textwidth}
    {\centering
      \includegraphics[scale=1]{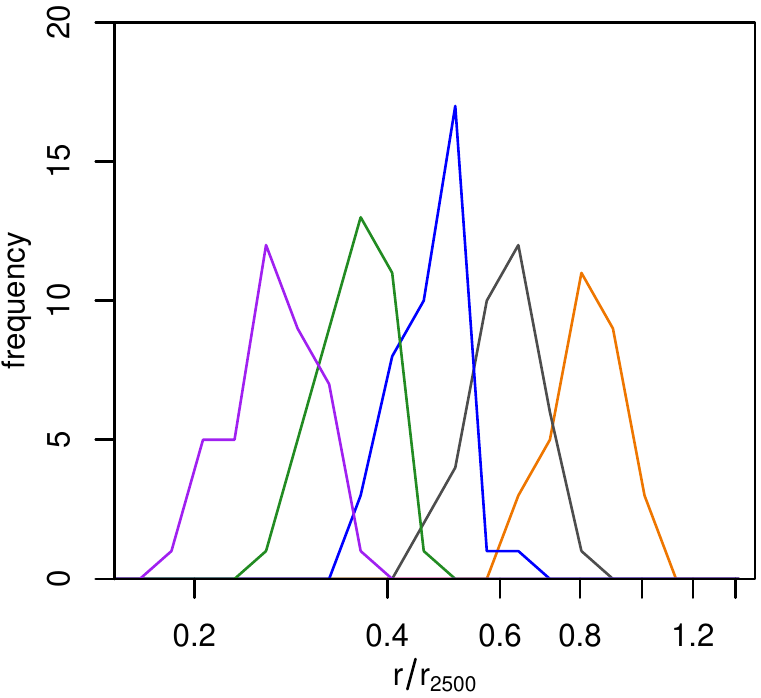}
    }
  \end{minipage}
  \begin{minipage}[c]{0.5\textwidth}
    \caption[]{
      Histograms of the average of the semi-major and semi-minor axes of ellipses corresponding to the five isophotes used in our analysis, as a fraction of $r_{2500}$ (shown as lines, for clarity). Only 40 clusters which are classified as highly relaxed here and for which we can reliably determined mass profiles (hence $r_{2500}$; see \cosmopaper{}) are used here. The cluster region used in our isophote analysis typically spans radii $0.2\ltsim r/r_{2500}\ltsim1$.
    }
  \end{minipage}
  \label{fig:isoph_radii}
\end{figure}

In \cosmopaper{}, mass profiles are derived for a sample of 40 relaxed clusters identified in the present work. Histograms of the mean of the semi-major and semi-minor axes in units of $r_{2500}$ are shown for these clusters in \figref~\ref{fig:isoph_radii}. As expected, the isophotes in units of $f_S$ broadly map onto comparable radii in units of $r_{2500}$.

From this set of ellipses, we calculate two statistics, which we refer to as alignment, $a$, and symmetry, $s$. These are defined to have the same sense as the peakiness, i.e.\ more positive (negative) values being typical of more (less) relaxed clusters.

The alignment is defined as
\begin{equation}
  a  = -\log_{10}\left[\frac{1}{\Nel-1} \sum_{j=1}^{\Nel-1} \frac{\delta_{j,j+1}}{\langle b\rangle_{j,j+1}}\right],
\end{equation}
where \Nel{} is the number of ellipses and the sum is over pairs of ``adjacent'' ellipses, i.e.\ those corresponding to progressively higher surface brightness. Here $\delta_{j,j+1}$ is the distance between the centers of two ellipses, and $\langle b\rangle_{j,j+1}$ is the average of the four ellipse axis lengths (major and minor axes of both ellipses).

The symmetry statistic is
\begin{equation}
  s = -\log_{10} \left[ \frac{1}{\Nel} \sum_{j=1}^{\Nel} \frac{\delta_{j,\mathrm{c}}}{\langle b\rangle_j} \right],
\end{equation}
where $\delta_{j,\mathrm{c}}$ is the distance between the center of the $j$th ellipse and the global center identified in \secref~\ref{sec:sbprofile}, and $\langle b\rangle_j$ is the average of the major and minor axes of the ellipse.

These quantities provide complementary measurements of cluster substructure. The alignment is sensitive to shifts in the center of emission at the relatively large scales probed by our set of isophotes, whereas the symmetry parameter measures the overall agreement of those isophotes with the global center. Note that, by design, the brightness range covered by this analysis does not extend to the brightest (spatially central) regions of cool core clusters (left panel of \figref~\ref{fig:sbprof}), where complex, non-elliptical features such as cavities and small-scale sloshing are ubiquitous, even in more globally relaxed clusters. \figref~\ref{fig:ellipses} shows smoothed images and isophote ellipse fits to the unmodified (i.e.\ not bootstrapped) data for the example clusters A1835 and A2163, which respectively have relatively high and low values of both alignment and symmetry. Clusters representing even more extreme values of $a$ and $s$ are on display in \figref~\ref{fig:extremesa}.

\section{Results} \label{sec:results}

\begin{figure*}
 \centering
 \includegraphics[scale=0.75]{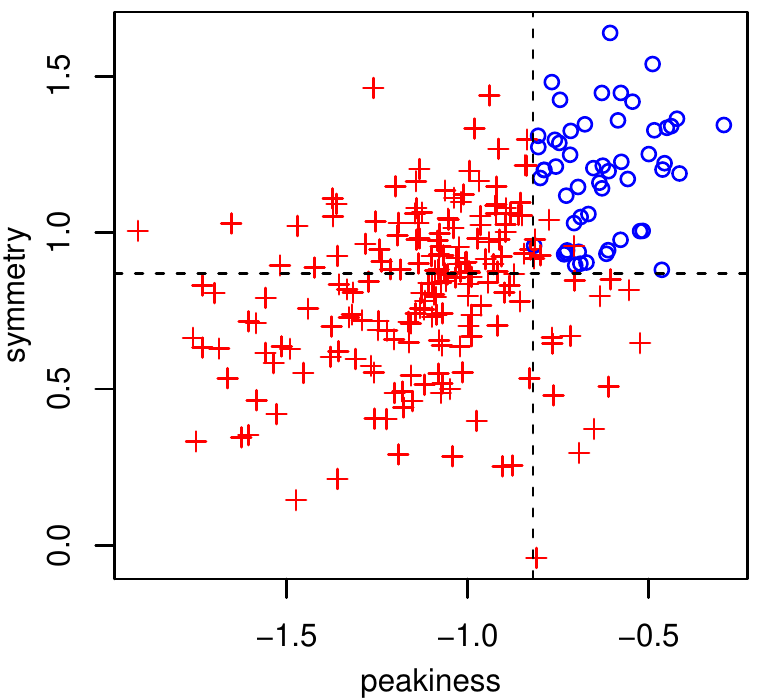}
 \includegraphics[scale=0.75]{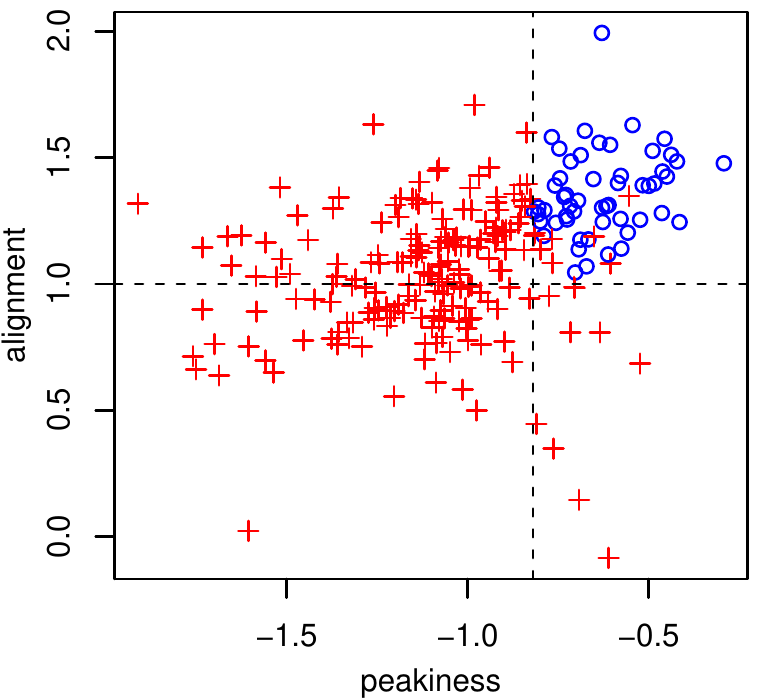}
 \includegraphics[scale=0.75]{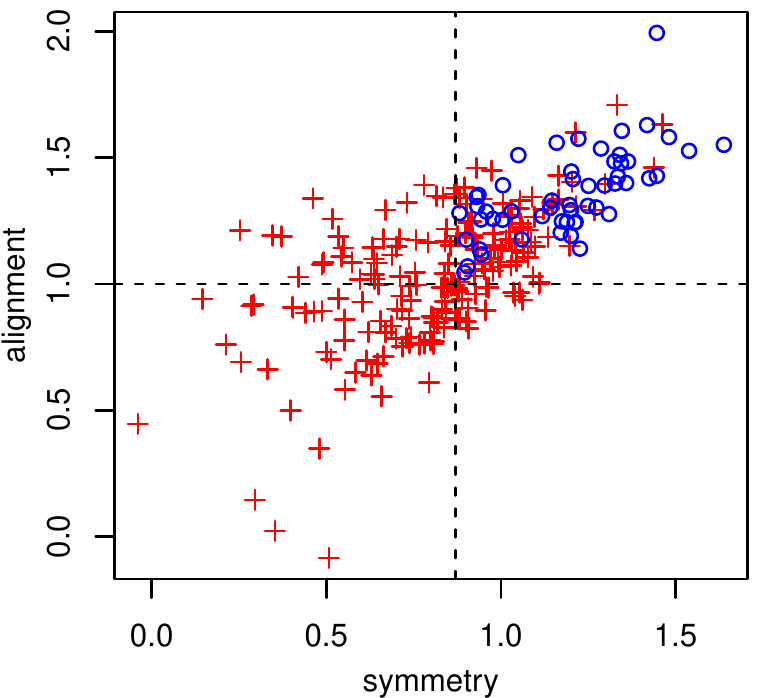}
 \caption[]{
   Distributions of measured morphological values from our \Chandra{} analysis, with the $s$-$p$-$a$ cuts defining the relaxed sub-sample shown as dashed lines. Clusters in this relaxed sub-sample (identified with a $\surd$ symbol in \tabref{}s~\ref{tab:morph} and \ref{tab:allresults}) are shown as blue circles, and others as red crosses. To be considered relaxed a cluster must simultaneously pass all three cuts at $>50$ per cent confidence (see \secref~\ref{sec:morph_criterion}).
 }
 \label{fig:morph_cuts}
\end{figure*}

The procedure of \secref~\ref{sec:procedure} was applied to obtain morphological statistics from 1000 bootstrap simulations of the clusters identified in \secref~\ref{sec:data}. Results are tabulated in \tabref{}s~\ref{tab:morph} and \ref{tab:allresults}. For the \Chandra{} sample, these are also shown in \figref~\ref{fig:morph_cuts}.

\begin{table*}
  \begin{center}
    \caption{
      Abridged results of our morphological analysis. The remaining clusters can be found in \tabref~\ref{tab:allresults}.
      [1] Catalog that each cluster was drawn from, abbreviating the BCS, REFLEX, CIZA, MACS, 400d and SPT catalogs, or none of the above ($\star$). A $\star$ appended to a catalog identifier means that the cluster can be found in the indicated sample, but does not satisfy the X-ray luminosity threshold normally applied in \secref~\ref{sec:data}.
      [2] Cluster name. Prefixes that are implied by column [1] have been suppressed.
      [3] Observatory used to produce the listed results, either \Chandra{} (C) or ROSAT (R).
      [4] Flags indicating whether a cluster is part of the \arsemf{} (a), CLASH relaxed (c), LoCuSS low centroid variance (l), or CCCP low central entropy (p) samples (see \secref{}s~\ref{sec:fgascompare}--\ref{sec:othersel}). An $f$ indicates that the cluster is used in \cosmopaper{}.
      [5--7] Symmetry, peakiness and alignment measurements. Note that some or all may be missing, dependent on data quality (see \secref~\ref{sec:results}). In general, when the data were inadequate to measure $s$ and $a$, we did not carry through the bootstrapping procedure to obtain uncertainties on $p$.
      [8] Indicates whether the cluster is relaxed according to the SPA criterion introduced in \secref~\ref{sec:morph_criterion}.
      [9] Mean ellipticity of the isophotes employed in our analysis.
      [10--11] J2000 coordinates of the global center measured from the X-ray analysis (without bootstrapping).
      [12--13] J2000 coordinates of the BCG identified in \secref~\ref{sec:bcgs}.
    }
    \label{tab:morph}
    \vspace{1ex}
    \footnotesize
    \begin{tabular}{@{}c@{\hspace{0ex}}c@{}c@{\hspace{1ex}}c@{\hspace{1ex}}r@{}c@{}l@{\hspace{2ex}}r@{}c@{}l@{\hspace{2ex}}r@{}c@{}l@{\hspace{1ex}}c@{\hspace{1ex}}r@{}c@{}l@{\hspace{2ex}}r@{\hspace{1ex}}r@{\hspace{1ex}}r@{\hspace{1ex}}r@{}}
      \hline
      Cat. & Name & Obs. & flags & \multicolumn{3}{c}{$s$} & \multicolumn{3}{c}{$p$} & \multicolumn{3}{c}{$a$} & Rel. & \multicolumn{3}{c}{ellip.} & \multicolumn{1}{c}{RA$_\mathrm{X}$} & \multicolumn{1}{c}{Dec$_\mathrm{X}$} & \multicolumn{1}{c}{RA$_\mathrm{BCG}$} & \multicolumn{1}{c}{Dec$_\mathrm{BCG}$} \\
      \hline
      B  &  Abell~1068  &  C  &  &  $  1.050$    &       $\pm$  &      $0.033$   &        $-0.688$  &         $\pm$  &      $0.019$  &        $  1.51$  &      $\pm$  &      $0.07$  &       $\surd$  &  0.256     &      $\pm$  &         0.013  &         160.1859  &         $         39.9531$  &         160.1854  &         $  39.9531$  \\
B  &  Abell~1132  &  C  &  &  $  1.034$    &       $\pm$  &      $0.217$   &        $-1.369$  &         $\pm$  &      $0.064$  &        $  1.25$  &      $\pm$  &      $0.15$  &       ~        &  0.192     &      $\pm$  &         0.027  &         164.6091  &         $         56.7950$  &         164.5986  &         $  56.7949$  \\
B  &  Abell~115   &  C  &  p  &  $-0.040$  &       $\pm$  &      $0.029$   &        $-0.810$  &         $\pm$  &      $0.023$  &        $  0.45$  &      $\pm$  &      $0.08$  &       ~        &  0.406     &      $\pm$  &         0.019  &         13.9598   &         $         26.4098$  &         13.9609   &         $  26.4104$  \\
B  &  Abell~1201  &  C  &  &  $  0.488$    &       $\pm$  &      $0.032$   &        $-1.074$  &         $\pm$  &      $0.019$  &        $  1.08$  &      $\pm$  &      $0.10$  &       ~        &  0.505     &      $\pm$  &         0.019  &         168.2264  &         $         13.4351$  &         168.2271  &         $  13.4358$  \\
B  &  Abell~1204  &  C  &  &  $  1.171$    &       $\pm$  &      $0.073$   &        $-0.558$  &         $\pm$  &      $0.024$  &        $  1.20$  &      $\pm$  &      $0.09$  &       $\surd$  &  0.194     &      $\pm$  &         0.020  &         168.3354  &         $         17.5945$  &         168.3354  &         $  17.5947$  \\
B  &  Abell~1246  &  C  &  &  ~  &         &       ~      &      $-1.390$  &        &         ~         &      ~      &        &        ~  &      ~      &      ~      &       &       ~        &  170.9906  &      $      21.4810$  &      170.9947  &         $         21.4794$  \\
B  &  Abell~1413  &  C  &  a  &  $         1.333$  &      $\pm$  &         $0.050$  &         $-0.981$  &      $\pm$  &        $0.007$  &  $      1.71$  &      $\pm$  &       $0.08$  &        ~  &         0.307  &      $\pm$     &      0.005     &         178.8247  &         $         23.4050$  &         178.8250  &  $         23.4049$  \\
B  &  Abell~1423  &  C  &  c  &  $         0.900$  &      $\pm$  &         $0.116$  &         $-1.024$  &      $\pm$  &        $0.026$  &  $      0.85$  &      $\pm$  &       $0.16$  &        ~  &         0.231  &      $\pm$     &      0.032     &         179.3217  &         $         33.6112$  &         179.3222  &  $         33.6109$  \\
B  &  Abell~1553  &  C  &  &  $  0.621$    &       $\pm$  &      $0.097$   &        $-1.357$  &         $\pm$  &      $0.314$  &        $  0.81$  &      $\pm$  &      $0.17$  &       ~        &  0.146     &      $\pm$  &         0.027  &         187.6972  &         $         10.5530$  &         187.7036  &         $  10.5464$  \\
B  &  Abell~1682  &  C  &  &  ~  &         &       ~      &      $-1.421$  &        &         ~         &      ~      &        &        ~  &      ~      &      ~      &       &       ~        &  196.7088  &      $      46.5579$  &      196.6904  &         $         46.5585$  \\
B  &  Abell~1758  &  C  &  &  ~  &         &       ~      &      $-1.733$  &        &         ~         &      ~      &        &        ~  &      ~      &      ~      &       &       ~        &  203.1737  &      $      50.5472$  &      203.1601  &         $         50.5599$  \\

      \hline
    \end{tabular}
  \end{center}
\end{table*}

We note that there are cases where our morphology code fails outright. For example, for flat-core (low $p$) clusters in very shallow images, we are sometimes unable to constrain even two isophote ellipses, which is necessary for the calculation of alignment; however, in these cases, it is generally still possible to measure peakiness. The great majority of these can be classified as unrelaxed according to the criterion introduced in \secref~\ref{sec:morph_criterion} based solely on peakiness. In yet lower signal-to-noise data, it is sometimes impossible to obtain meaningful constraints on the surface brightness profile, and thus even peakiness cannot be measured. Subjectively speaking, this small minority of clusters appears unambiguously unrelaxed, and we classify them as such.

Note that there is a strong correlation between symmetry and alignment (\figref~\ref{fig:morph_cuts}), by virtue of their similar definitions in terms of isophote properties. Somewhat weaker correlations exists between symmetry or alignment on one hand and peakiness on the other; these presumably reflect the role of mergers in either destroying or preventing the formation of cool cores.

\subsection{Comparison with Other X-ray Morphology Statistics} \label{sec:otherx}

To provide some context, we now compare our morphological statistics to typical estimators used in the literature. Specifically, we have chosen the surface brightness concentration parameter of \citet{Santos0802.1445} and the centroid variance \citep{Mohr1993ApJ...413..492}, defined by
\begin{eqnarray}
  \csb &=& \frac{F(r<40\kpc)}{F(r<400\kpc)}, \\
  w^2 &=& \frac{1}{r_{500}^2} \mathrm{Var}(\Delta), \nonumber
\end{eqnarray}
where we estimate $r_{500}$ from the temperature--mass relation of \citet{Mantz0909.3099}. The distances $\Delta$ are calculated between our global centers and the centroids of emission in our ``filled-in'' images within apertures of radius $(0.1j)r_{500}$ ($j=1,2,\ldots,10$) about the global centers. We additionally compute the power ratio $P_3/P_0$ \citep{Buote9502002}, again using the filled-in images.

\begin{figure*}
  \centering
  \includegraphics[scale=0.75]{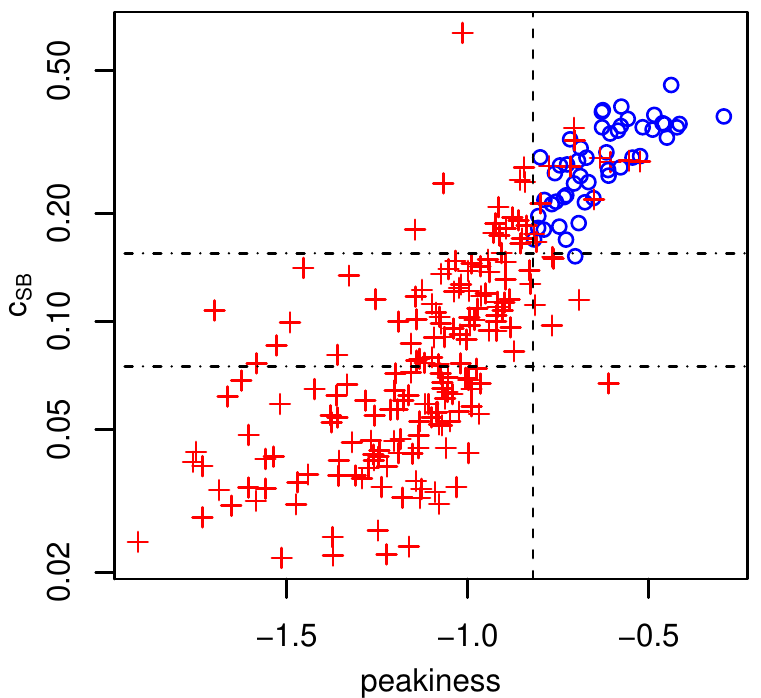}
  \includegraphics[scale=0.75]{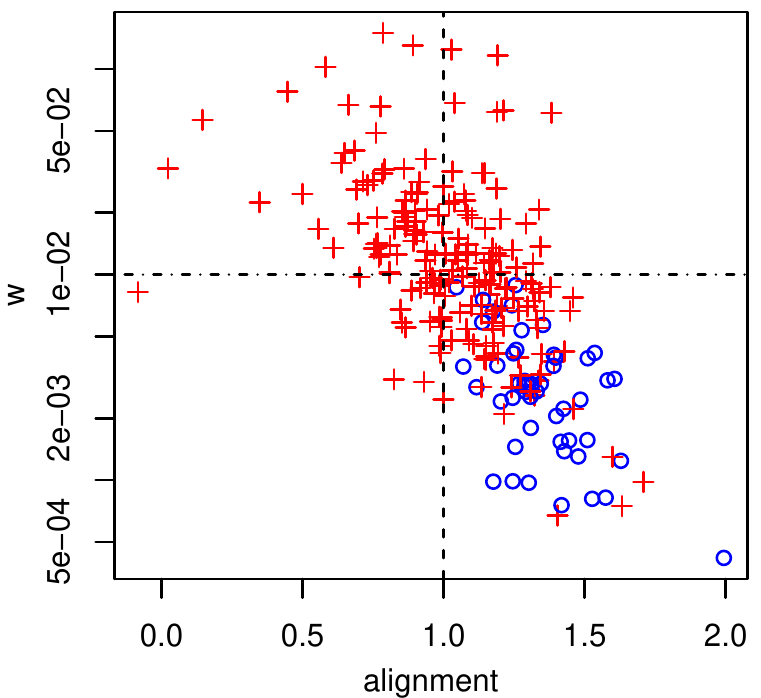}
  \includegraphics[scale=0.75]{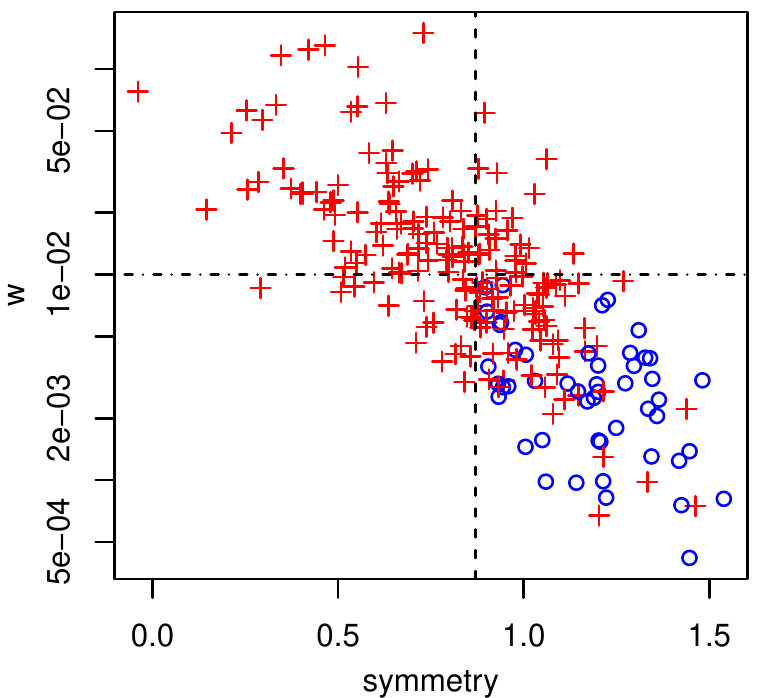}
  \caption[]{
    Comparison of our morphology statistics (peakiness, alignment, symmetry) from \Chandra{} data with surface brightness concentration (\csb) and centroid shift ($w$). Clusters that we classify as relaxed in \secref~\ref{sec:morph_criterion} are shown as blue circles, and others as red crosses. As expected, our peakiness metric correlates with \csb{} and both alignment and symmetry anti-correlate with the centroid shift. Dot-dashed lines show the \csb{} thresholds defining ``moderate'' and ``strong'' cool cores in the work of \citet{Santos0802.1445} and the $w$ cut used by \citet{Bohringer0912.4667} to distinguish relaxed systems, while dashed lines show the thresholds we adopt in \secref~\ref{sec:morph_criterion}. (The latter were determined without reference to either \csb{} or $w$.)
  }
  \label{fig:morph_lit}
\end{figure*}

We compare our morphological statistics to these alternatives in \figref~\ref{fig:morph_lit}. Not surprisingly, peakiness correlates most strongly with \csb{},\footnote{Consequently, we can also conclude that peakiness correlates with other cool-core indicators, such as central cooling time, which have been observed to correlate with \csb{} \citep{Santos0802.1445}.} while both alignment and symmetry anti-correlate strongly with centroid variance. The power ratio correlates less well with our statistics. While there are important differences, it is clear that our statistics measure similar image features to these other quantities. In fact, the cuts in $s$, $p$ and $a$ that we use to define a relaxed sample in \secref~\ref{sec:morph_criterion}, which were determined before we had even calculated \csb{} and $w$, correspond surprisingly well to the cuts used by \citet{Santos0802.1445} and \citet{Bohringer0912.4667} to define strong cool cores and low centroid variance, respectively. Note, however, that our final selection appears to be somewhat more conservative than these cuts on \csb{} and $w$ would be, as one might generically expect given the use of a third, non-degenerate measurement in our selection.

\subsection{Comparison with BCG/X-ray Offsets} \label{sec:bcgs}

A simple metric that has been used to try to distinguish between relaxed and unrelaxed clusters is the distance in projection between the center of the X-ray emission and the location of the brightest cluster galaxy (BCG). This approach is potentially appealing because in principle the X-ray data need not be deep enough to provide peakiness measurements, let alone the more challenging alignment and symmetry measurements. This may be the case for, e.g., X-ray snapshots of distant SZ- or IR-selected clusters, whose X-ray brightness is not well known prior to the observations. At the same time, optical or IR imaging is still commonly used to confirm the presence of a galaxy overdensity at the location of a candidate cluster, and to study the properties of cluster galaxies, and so a BCG identification may be readily available.

Where available, we use the BCGs identified in the Weighing the Giants project (54 clusters; \citealt{von-der-Linden1208.0597}) or for the SPT survey (18 clusters in common with our sample; \citealt{Song1207.4369}). For the remaining clusters, we query the DR7 and DR10 catalog and imaging databases of the Sloan Digital Sky Survey\footnote{Querying both databases is advantageous since bright galaxies are masked in the DR10 catalog processing.} (SDSS; \citealt{Abazajian0812.0649, Ahn1307.7735}), which provides BCGs for an additional 123 clusters. The clusters considered here span a wide redshift range, and several are known to have central galaxies bluer than the red sequence (e.g.\ \citealt{Crawford9903057}), making simple algorithmic identification schemes difficult to implement. We therefore verify each BCG candidate by eye, considering galaxies up to 1\,Mpc from the X-ray center. For each cluster, the initial BCG candidates are taken as the brightest objects likely to be elliptical galaxies (in the SDSS {\sc Galaxies} catalog, with concentration $R_{90}/R_{50}>2.3$, and where a de~Vaucouleur profile is a better fit than an exponential) within two apertures (50\,kpc and 500\,kpc) from the X-ray center. For 73 clusters, the two apertures select the same galaxy; in 69 clusters, it also passes visual verification (in the remaining 4 clusters the initial candidate is a foreground galaxy). For 38 clusters, the two apertures select different BCG candidates; in 21 (17) clusters, we select the candidate within 50\,kpc (500\,kpc). For 12 clusters, the BCG is not one of these two candidates for a variety of reasons (e.g.\ nearby BCGs are de-blended into several detections). In total, this yields 195 BCG positions.

\begin{figure*}
  \centering
  \includegraphics[scale=0.75]{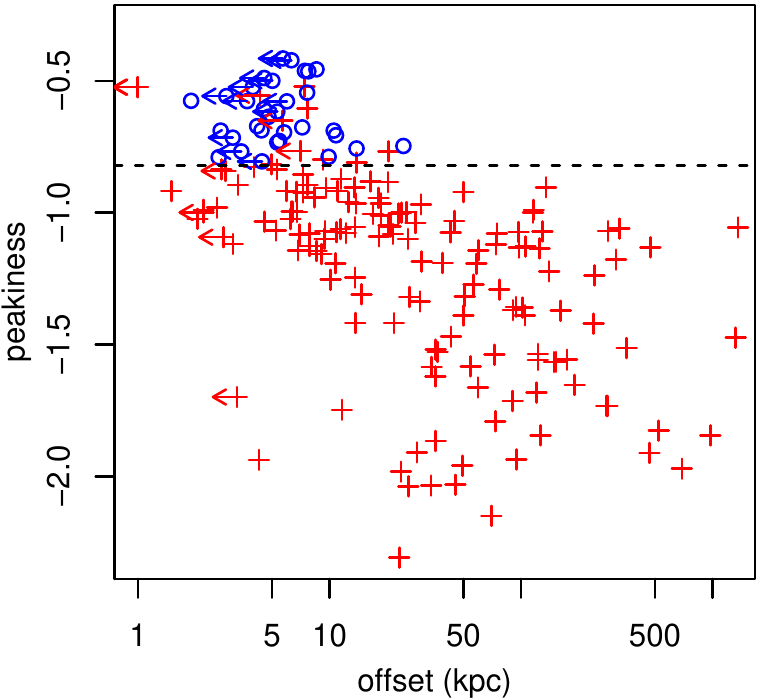}
  \includegraphics[scale=0.75]{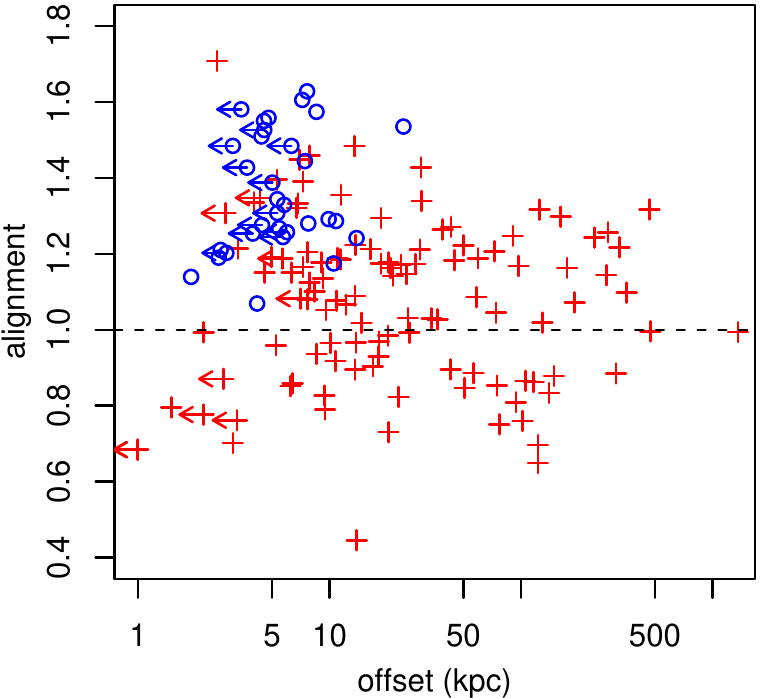}
  \includegraphics[scale=0.75]{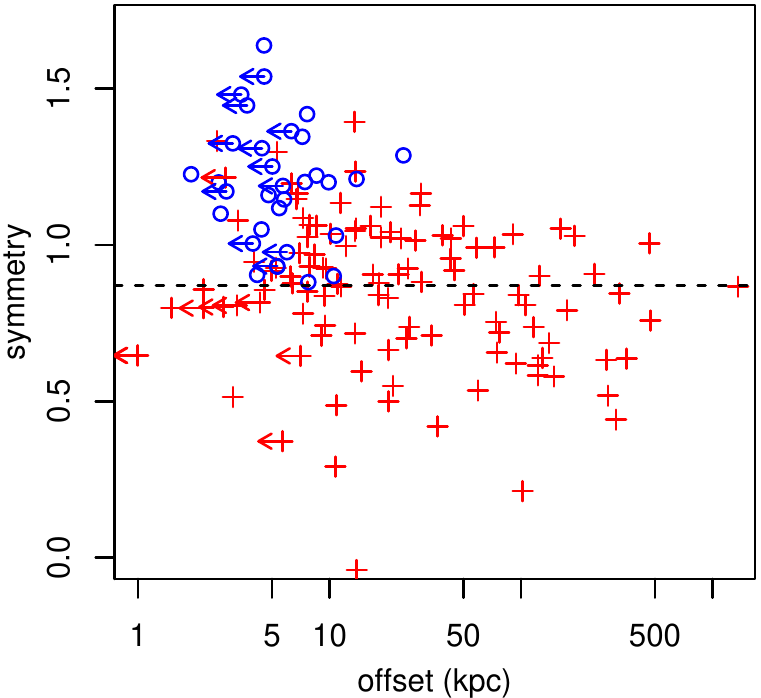}
  \caption[]{
    Comparison of our morphology statistics (peakiness, alignment, symmetry) from \Chandra{} data with the projected offset in kpc between our measure of the global X-ray center and the location of the brightest cluster galaxy. Clusters that we classify as relaxed in \secref~\ref{sec:morph_criterion} are shown as blue circles, and others as red crosses (dashed lines show the cuts associated with this classification). Arrows label offsets which are $<1''$ in projection, i.e.\ below the resolution of our X-ray images (distances in kpc for these are calculated assuming $1''$ offsets). The BCG offsets are visibly correlated with peakiness, which is intuitive, since both measurements are most sensitive to activity in the cluster center.
  }
  \label{fig:morph_bcg}
\end{figure*}

\figref~\ref{fig:morph_bcg} shows the projected distance between these BCG locations and the global X-ray centers defined in \secref~\ref{sec:sbprofile} versus the corresponding measurements of X-ray symmetry, peakiness and alignment. Note that a large fraction of the $<10\kpc$ offsets translate to $<1''$ in angular distance (i.e.\ less than the resolution of our X-ray images), and so are uncertain in detail. (Conversely, offsets $>10\kpc$ are resolved, i.e.\ $>1''$, for the entire data set.) Nevertheless, there is a clear correlation between the BCG/X-ray offset and peakiness, while in contrast there is not such a pronounced trend between the offset and either alignment or symmetry. This makes physical sense, since merger activity generically should produce BCG/X-ray offsets as well as a reduction in peakiness at some level. At the same time, while the offsets for clusters that we ultimately classify as morphologically relaxed (\secref~\ref{sec:morph_criterion}) are generally small, there is a range in offsets, reaching 24\kpc{} in the most extreme case.\footnote{The relaxed cluster with the largest BCG/X-ray offset (24\,kpc) is MACS\,J1311.0$-$0311. This cluster fails the additional cuts required for inclusion in our cosmology sample, although for reasons of data quality rather than morphology (\cosmopaper{}). The other relaxed clusters all have BCG/X-ray offsets $<14$\,kpc.} This scatter has a natural explanation in sloshing of the ICM due to merger events; the small-scale displacement of the ICM from the precise center of the gravitational potential may persist for Gyr, even as the effect on X-ray emission on the larger scales probed by the symmetry and alignment measurements is muted \citep{ZuHone1108.4427}.

Based on the distributions in \figref~\ref{fig:morph_bcg}, it is not clear that measurements of the BCG offset contribute much in addition to the full set of X-ray morphological measurements, particularly peakiness. On the other hand, given BCG locations and  relatively poor X-ray data -- sufficient to find an X-ray center, but not to measure even peakiness, e.g.\ from a shallow survey -- a suitable cut on the BCG offset clearly would eliminate a large fraction of unrelaxed clusters.

\subsection{Comparison with Radio Halo/Relic Samples} \label{sec:radiohalos}

\begin{figure*}
  \centering
  \includegraphics[scale=0.75]{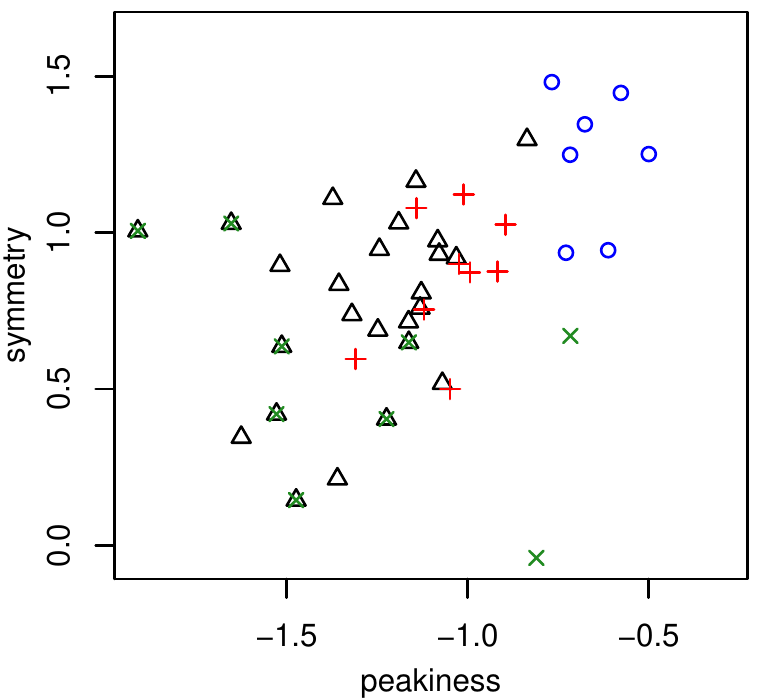}
  \includegraphics[scale=0.75]{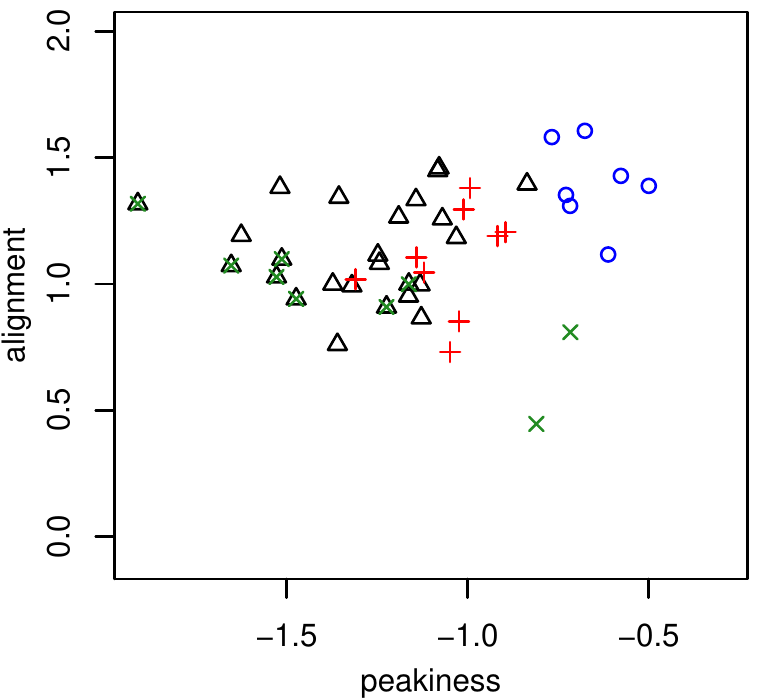}
  \includegraphics[scale=0.75]{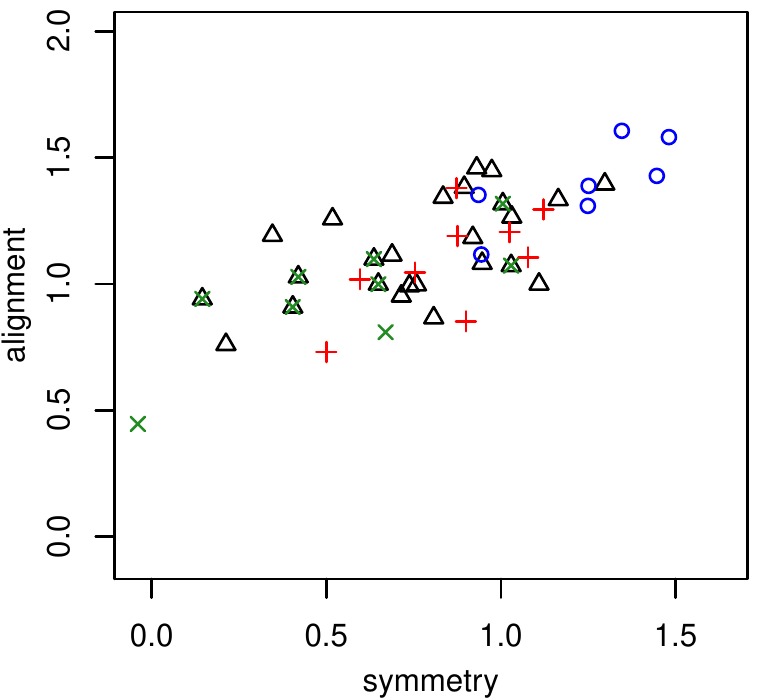}
  \caption[]{
    As \figref~\ref{fig:morph_cuts}, but showing only a subset of clusters. Those with radio halos listed in the compilations of \citet{Feretti1205.1919} or \citet{Cassano1306.4379}, all of which we identify as unrelaxed based on their X-ray morphology, are shown as black triangles. Clusters for which \citet{Cassano1306.4379}  list strong upper limits on the radio halo power are shown as red crosses (unrelaxed) or blue circles (relaxed). Green $\times$ symbols indicate clusters with radio relics compiled by \citet{Feretti1205.1919}. Our findings are consistent with radio halos and relics occurring exclusively in morphologically unrelaxed clusters.
  }
  \label{fig:morph_halos}
\end{figure*}

Radio halos, low surface brightness synchrotron emission located in the central regions of clusters, have been associated with merging activity, although not all merging clusters display radio halos (see \citealt{Feretti1205.1919} and references therein). \figref~\ref{fig:morph_halos} shows our morphological measurements for clusters with detected radio halos \citep{Feretti1205.1919, Cassano1306.4379}. Also shown are clusters for which strong upper limits have been placed on the radio power without detecting a halo \citep{Cassano1306.4379}. The radio halo clusters are uniformly unrelaxed according to our X-ray morphological analysis (\secref~\ref{sec:morph_criterion}), while the clusters with only upper limits split between being relaxed and unrelaxed. These trends are consistent with previous work comparing the incidence of radio halos with other morphological estimators, namely power ratios, surface brightness concentration and/or centroid variance \citep{Buote0104211, Cassano1008.3624}. Similarly, all the clusters in our analysis which host radio relics (emission localized to cluster outskirts) according to the compilation of \citet{Feretti1205.1919} are found to be unrelaxed.

\subsection{The SPA Criterion for Relaxation} \label{sec:morph_criterion}

An interesting extension of this work would be to test our morphological statistics against the actual dynamical state of simulated clusters using mock X-ray images, as in \citet{Bohringer0912.4667} and \citet{Meneghetti1404.1384}, although we note that overcooling in simulations has historically limited the applicability of this approach. For the moment, we are concerned only with selecting the most morphologically relaxed group of clusters, rather than clusters that meet a specific criterion in terms of non-thermal support. We therefore use the subjective determinations of \arsemf{} as a broad guide for identifying the ranges of $p$, $a$ and $s$ corresponding to the most relaxed clusters. Note that the \arsemf{} selection, though subjective, has previously survived ``double-blind'' tests; i.e., the same clusters were independently selected as the most relaxed by multiple viewers, with cluster identities hidden. The advantage of this work is that it provides a practical and evenhanded way to compare a large number of clusters, putting the \arsemf{} selection in a wider context.

\begin{figure*}
  \centering
  \includegraphics[scale=0.75]{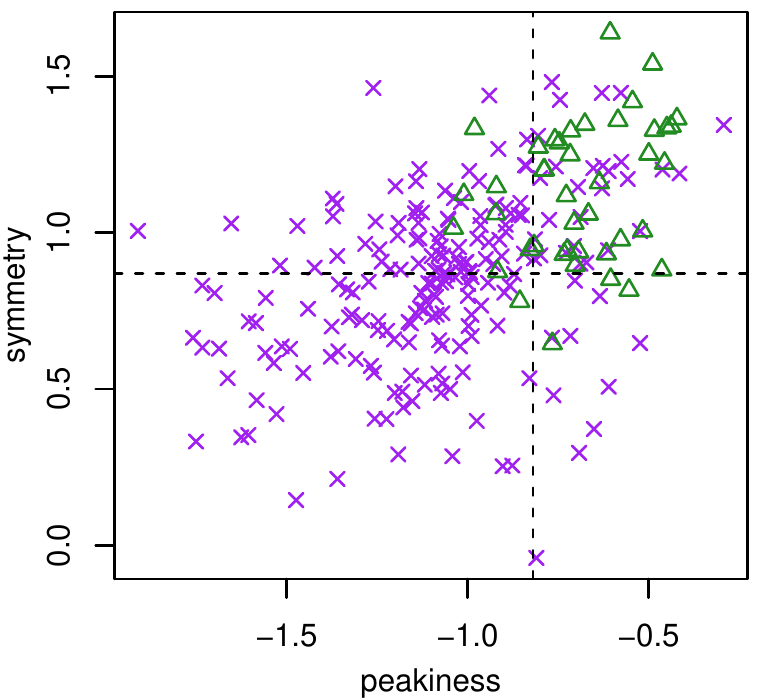}
  \hspace{10mm}
  \includegraphics[scale=0.75]{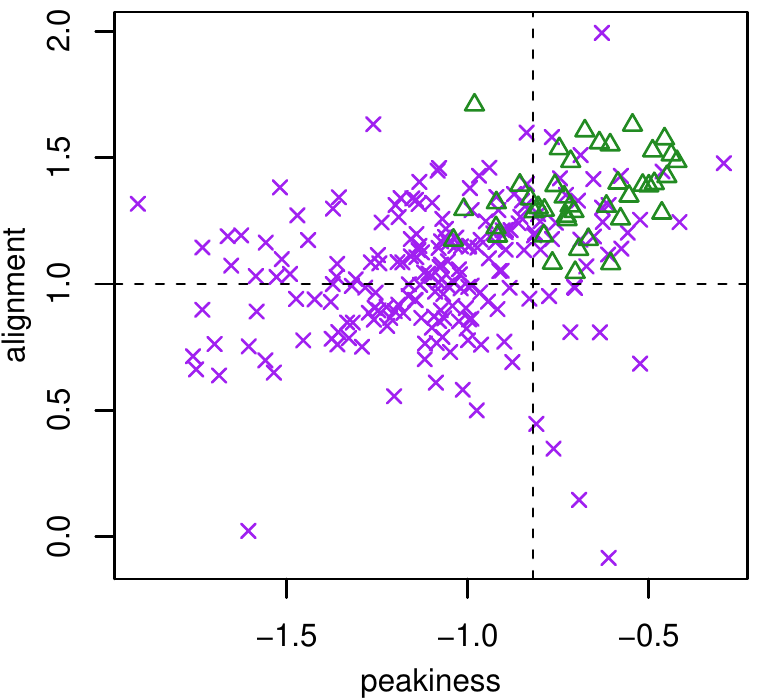}
  \caption[]{
    Peakiness--symmetry and peakiness--alignment distributions from our \Chandra{} analysis. Clusters used in \arsemf{} are shown as green triangles, and others as purple $\times$s. Our criterion for relaxation is motivated by but more strict than (in terms of these quantities) the subjective determinations of \arsemf{}. To be classified relaxed, a cluster must simultaneously exceed thresholds in all three quantities (dashed lines) at $>50$ per cent confidence (see \secref~\ref{sec:morph_criterion}).
  }
  \label{fig:morph_A08}
\end{figure*}

\figref{}~\ref{fig:morph_A08} shows the distribution of peakiness, alignment and symmetry for the large sample of analyzed clusters as purple `$\times$' symbols, with clusters from \arsemf{} shown as green triangles. Clearly, the morphological statistics introduced above are related to the subjective determinations used by \arsemf{}. At the same time, within the context of the large, homogeneously analyzed sample, it is clear that not all of the \arsemf{} clusters belong to a well defined locus in the most relaxed corner of parameter space. Introducing cuts based on our morphology measurements may thus produce a more rigorously defined relaxed sample.

Motivated by the distributions in \figref{}~\ref{fig:morph_A08}, we introduce the Symmetry-Peakiness-Alignment (SPA) criterion for cluster relaxation. Namely, we define simple cuts in these three parameters, as depicted in \figref{}~\ref{fig:morph_cuts}: $s>0.87$, $p>-0.82$, and $a>1.00$ (\figref~\ref{fig:morph_cuts}).\footnote{A posteriori, these cuts appear well matched to thresholds in surface brightness concentration and centroid variance, respectively used by \citet{Santos0802.1445} and \citet{Bohringer0912.4667}, as noted in \secref~\ref{sec:otherx}.} We categorize a cluster as relaxed if $>50$ per cent of the $s$-$p$-$a$ triplets from the cluster's bootstrap analysis simultaneously satisfy all three of these cuts.\footnote{There is a straightforward degeneracy between the location of the cuts themselves and the fraction of passing bootstrap samples required for to be classified as relaxed. While essentially the same selection could be obtained with an ostensibly stricter threshold (given slightly shifted cuts), the 50 per cent threshold is convenient because it makes plots of the bootstrap mean for each cluster simpler to interpret (e.g.\ \figref{}s~\ref{fig:morph_cuts}--\ref{fig:morph_A08}). Note, however, that this 50 per cent criterion is not identical to only requiring the bootstrap mean to satisfy all three cuts, even assuming a symmetric bootstrap distribution.} \tabref~\ref{tab:morph} lists whether each cluster was classified as relaxed. Our intent is to generate a \emph{conservative} (i.e.\ as pure as possible) sample of relaxed clusters, even at the expense of excluding some legitimately relaxed systems; however, for convenience, we will use the term ``unrelaxed'' to refer to clusters that do not meet the SPA criterion. We compare the resulting selection to similarly motivated samples in the literature in \secref{}s~\ref{sec:fgascompare} and \ref{sec:othersel}, below.

\begin{figure*}
  \centering
  \includegraphics[width=\textwidth]{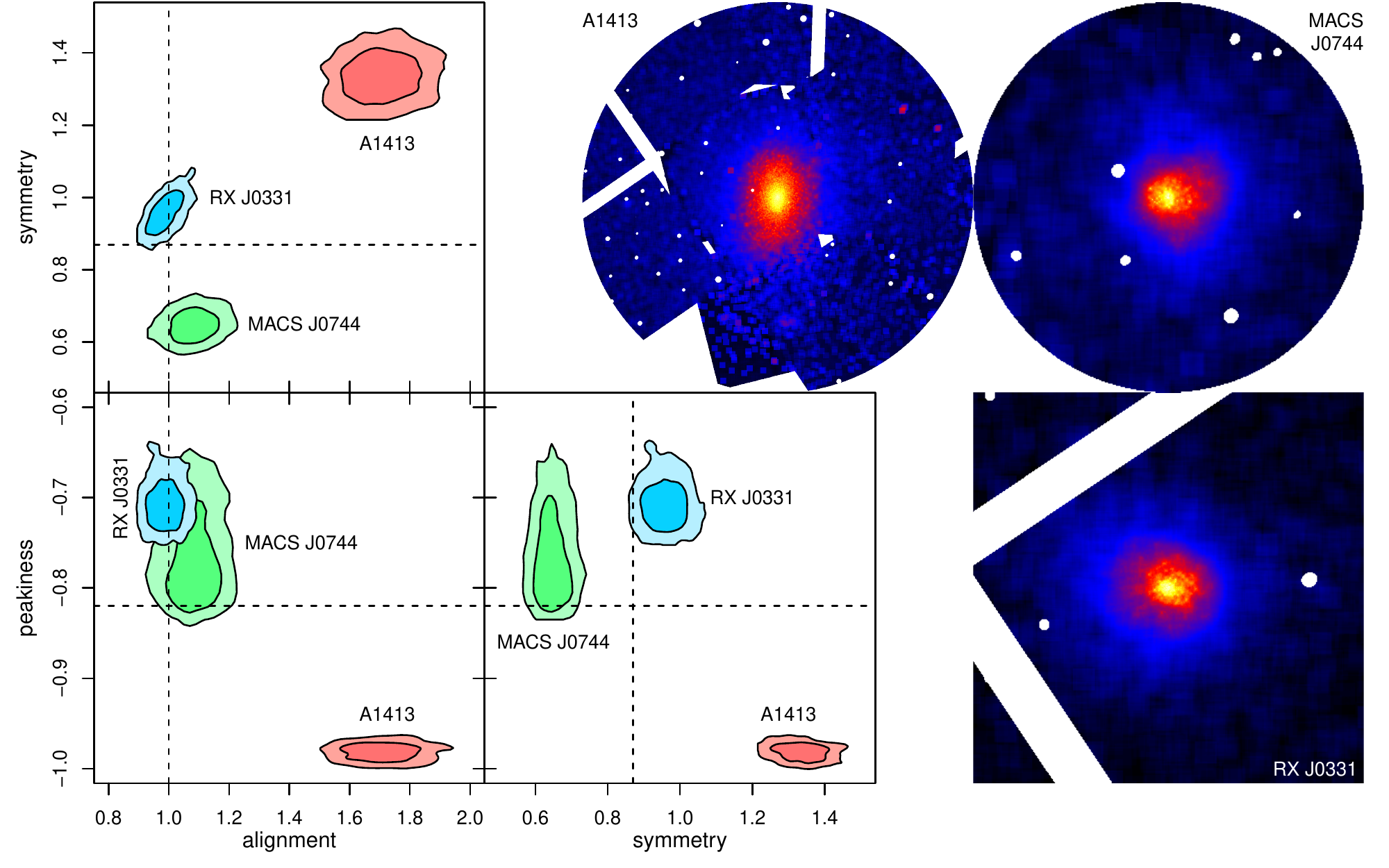}
  \caption[]{
    Left triangle: 68.3 and 95.4 per cent confidence regions of our morphological statistics for Abell\,1413, MACS\,J0744.8+3927 and RX\,J0331.1$-$2100 from the bootstrap analysis. Dashed lines show the cut applied to each statistic to identify relaxed clusters: clusters are classified as relaxed only if $>50$ per cent of their bootstrap samples exceed the cuts on all three quantities (i.e.\ relaxed clusters must reside in the upper-right quadrant of all panels). Right triangle: Smoothed images of the three clusters. Each of these clusters is classified as unrelaxed due to (only) one of the morphological indicators: Abell\,1413 due to peakiness, MACS\,J0744.8+3927 due to symmetry and RX\,J0331.1$-$2100 due to alignment. Note that the color scaling is chosen independently for each image to maximize the dynamic range shown, unlike \figref{}s~\ref{fig:extremep} and \ref{fig:gallery}.
  }
 \label{fig:triangle}
\end{figure*}

\figref~\ref{fig:triangle} shows the SPA cuts in relation to the bootstrap confidence regions associated with three example clusters, Abell\,1413, MACS\,J0744.8+3927 and RX\,J0331.1$-$2100, along with smoothed images. Each of these clusters is classified as unrelaxed due to only one of the SPA criteria (i.e., each would be classified as relaxed if only two of the cuts were applied to the bootstrap distributions). Specifically, the emission from Abell\,1413 is very regular, but not strongly peaked; MACS\,J0744.8+3927 has a strong peak and acceptable alignment, but fails the symmetry requirement; and RX\,J0331.1$-$2100 has acceptable peakiness and symmetry, but low alignment.

\subsection{Differences from the \arsemf{} Sample} \label{sec:fgascompare}

\begin{sidewaysfigure}
  \centering
  \includegraphics[scale=1]{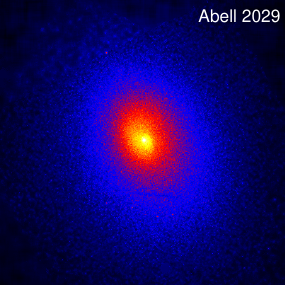}\hspace{0.3mm}%
  \includegraphics[scale=1]{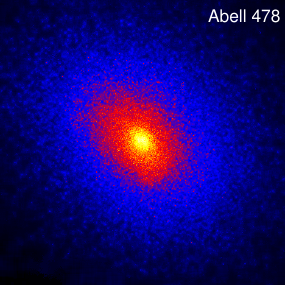}\hspace{0.3mm}%
  \includegraphics[scale=1]{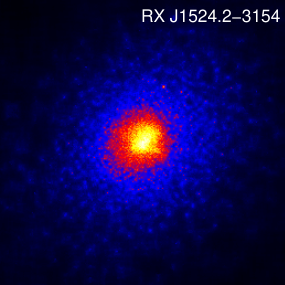}\hspace{0.3mm}%
  \includegraphics[scale=1]{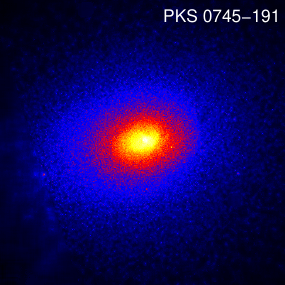}\hspace{0.3mm}%
  \includegraphics[scale=1]{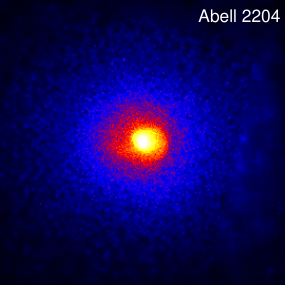}\hspace{0.3mm}%
  \includegraphics[scale=1]{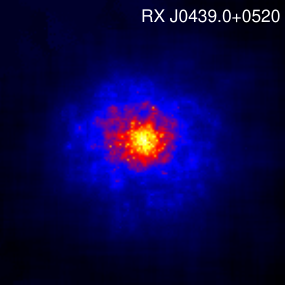}\hspace{0.3mm}%
  \includegraphics[scale=1]{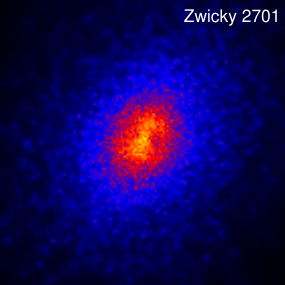}\hspace{0.3mm}%
  \includegraphics[scale=1]{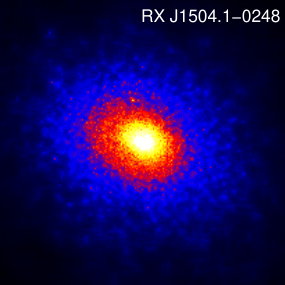}
  \includegraphics[scale=1]{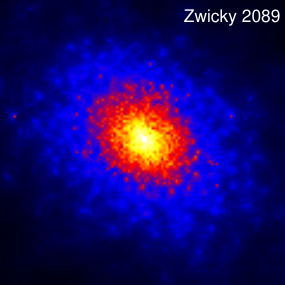}\hspace{0.3mm}%
  \includegraphics[scale=1]{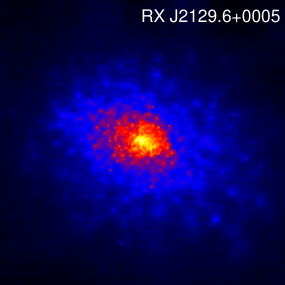}\hspace{0.3mm}%
  \includegraphics[scale=1]{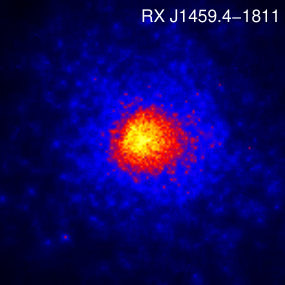}\hspace{0.3mm}%
  \includegraphics[scale=1]{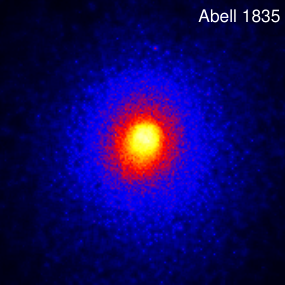}\hspace{0.3mm}%
  \includegraphics[scale=1]{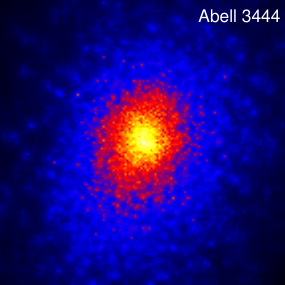}\hspace{0.3mm}%
  \includegraphics[scale=1]{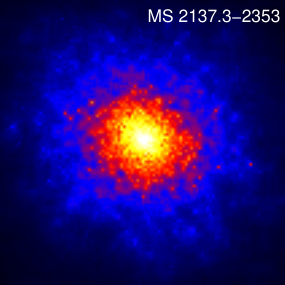}\hspace{0.3mm}%
  \includegraphics[scale=1]{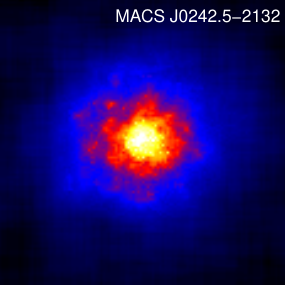}\hspace{0.3mm}%
  \includegraphics[scale=1]{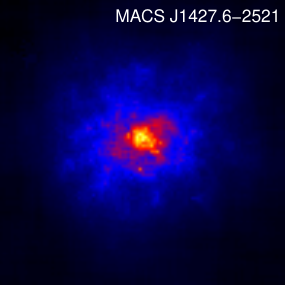}
  \includegraphics[scale=1]{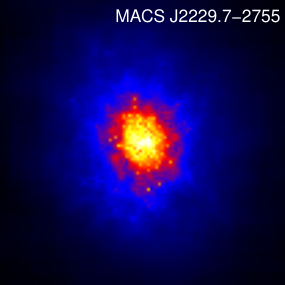}\hspace{0.3mm}%
  \includegraphics[scale=1]{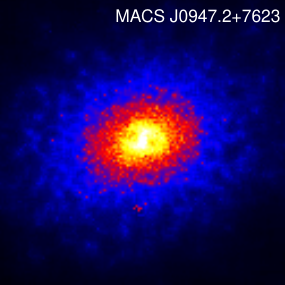}\hspace{0.3mm}%
  \includegraphics[scale=1]{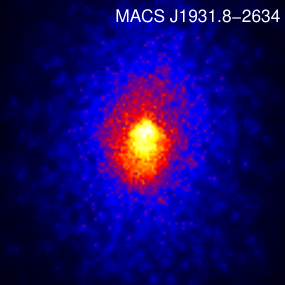}\hspace{0.3mm}%
  \includegraphics[scale=1]{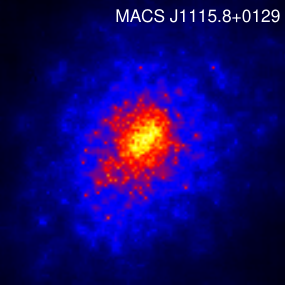}\hspace{0.3mm}%
  \includegraphics[scale=1]{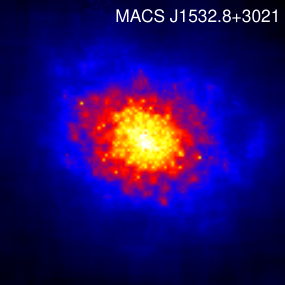}\hspace{0.3mm}%
  \includegraphics[scale=1]{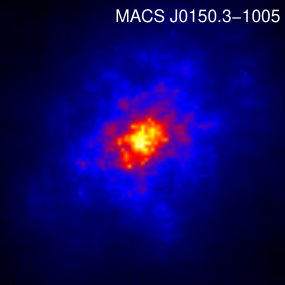}\hspace{0.3mm}%
  \includegraphics[scale=1]{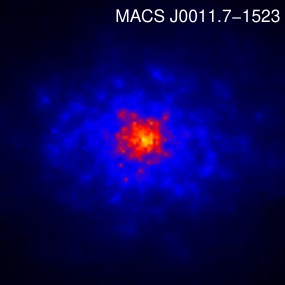}\hspace{0.3mm}%
  \includegraphics[scale=1]{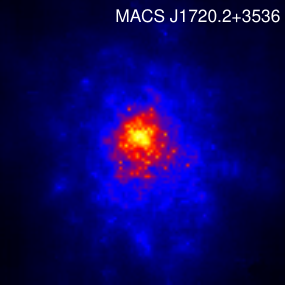}
  \includegraphics[scale=1]{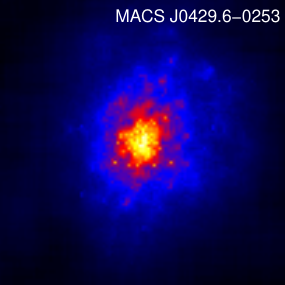}\hspace{0.3mm}%
  \includegraphics[scale=1]{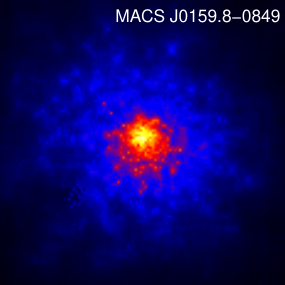}\hspace{0.3mm}%
  \includegraphics[scale=1]{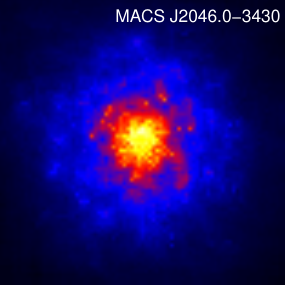}\hspace{0.3mm}%
  \includegraphics[scale=1]{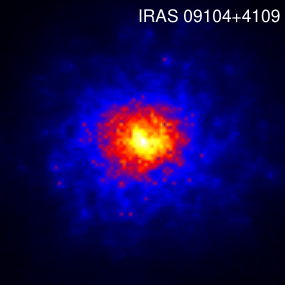}\hspace{0.3mm}%
  \includegraphics[scale=1]{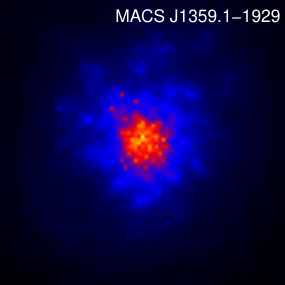}\hspace{0.3mm}%
  \includegraphics[scale=1]{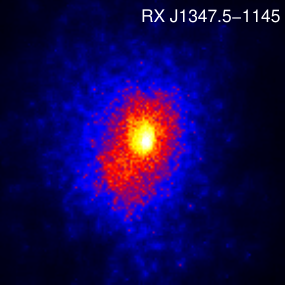}\hspace{0.3mm}%
  \includegraphics[scale=1]{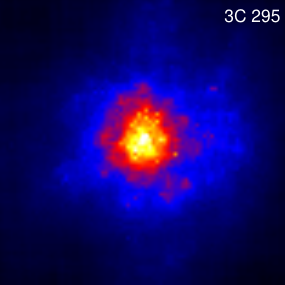}\hspace{0.3mm}%
  \includegraphics[scale=1]{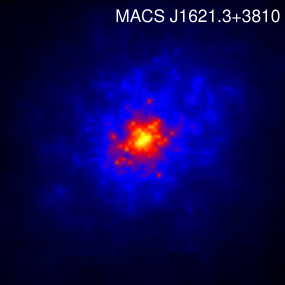}
  \includegraphics[scale=1]{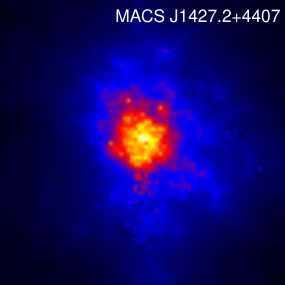}\hspace{0.3mm}%
  \includegraphics[scale=1]{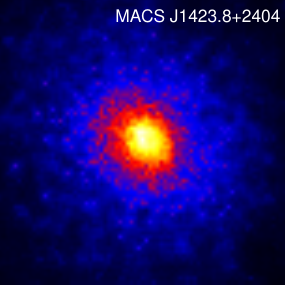}\hspace{0.3mm}%
  \includegraphics[scale=1]{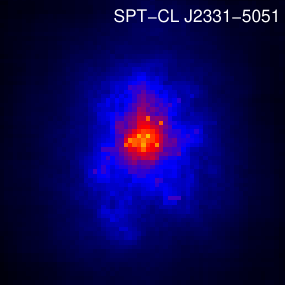}\hspace{0.3mm}%
  \includegraphics[scale=1]{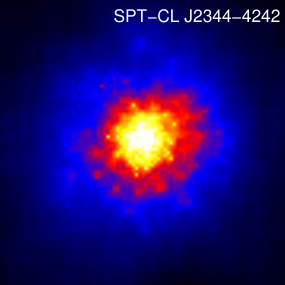}\hspace{0.3mm}%
  \includegraphics[scale=1]{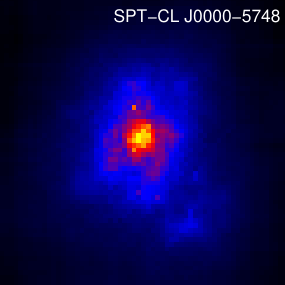}\hspace{0.3mm}%
  \includegraphics[scale=1]{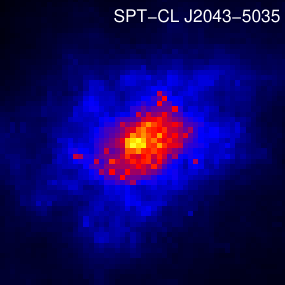}\hspace{0.3mm}%
  \includegraphics[scale=1]{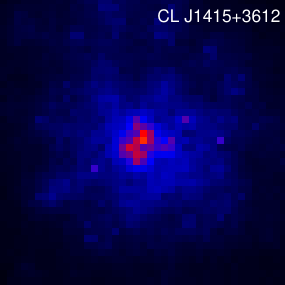}\hspace{0.3mm}%
  \includegraphics[scale=1]{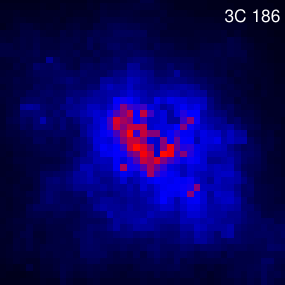}
  \caption[]{
    Smoothed, flat-fielded images of clusters satisfying the SPA criterion for relaxation, which are used in \cosmopaper{} to constrain cosmological models. Clusters ordered by increasing redshift, left-to-right and top-to-bottom. All images are scaled by the factor $f_S$ given in \eqnref~\ref{eq:sbscal} and use the same logarithmic color scale. Each square image is $r_{2500}$ on a side, according to the analysis of \cosmopaper{}. Masked regions have been filled in by randomly sampling values from nearby, unmasked pixels.
  }
  \label{fig:gallery}
\end{sidewaysfigure}

One motivation for this work is to identify a relaxed cluster sample to be used for cosmological studies of the gas mass fraction, as in \arsemf{} and \cosmopaper{}. The cosmological sample must meet additional criteria to those discussed here, regarding the cluster temperature and data quality (see \cosmopaper{} for details). Nevertheless, we note here the differences between the two cosmology samples which are due to morphological considerations. Specifically, Abell\,1795, Abell\,1413, Abell\,963, Abell\,2390, Abell\,611, Zw\,3146, Abell\,2537, MACS\,J0329.7$-$0212, MACS\,J0744.9+3927, \\MS\,1137.5+6625, and CL\,J1226.9+3332 were used in \arsemf{} but are excluded from the sample used in \cosmopaper{} (henceforth \SPAc{}) by the present analysis. (This analysis adds an equal number of clusters to the \SPAc{} sample, on the basis of data taken since 2008.) A gallery of clusters in the \SPAc{} sample appears in \figref~\ref{fig:gallery}.

The intrinsic scatter in cluster gas mass fractions, \fgas{}, is a useful metric for determining the effect of our more stringent morphological criteria compared to \arsemf{}. To the extent that dynamical state is the main difference between the \SPAc{} and \arsemf{} samples, the intrinsic scatter in \fgas{} can be interpreted as a surrogate for scatter in non-thermal support, since other systematics affecting the \fgas{} measurements should be roughly equivalent across the two samples. We use the gas mass fraction measured in a spherical shell at radii $0.8<r/r_{2500}<1.2$, as discussed in detail in \cosmopaper{}, and compare the intrinsic scatter of \fgas{} for the \SPAc{} sample to that of \SPAc{} plus the clusters which were included in \arsemf{} but are classified as unrelaxed on morphological grounds in this work. Marginalizing over a complete model, including cosmological terms appropriate for non-flat \LCDM{} models and various astrophysical and calibration nuisance parameters (see \cosmopaper{}), yields intrinsic scatters of $7.4\pm2.3$ and $13.5\pm2.4$ per cent for these two samples. We conclude that adopting the more stringent selection criteria motivated by our morphological analysis results in a quantitatively more relaxed cluster sample. The smaller intrinsic scatter of the \SPAc{} sample translates directly into tighter cosmological constraints on dark energy parameters (\cosmopaper{}). Note that this check was performed a posteriori, and did not influence the construction of the \SPAc{} sample itself.

\subsection{Caveats Regarding ROSAT Observations} \label{sec:rosatres}

Image resolution potentially affects many stages of our morphology analysis. Low resolution generically results in flatter surface brightness peaks, rounder isophotes, and a diminished sensitivity to structure that would otherwise influence the global center and isophote centers. These limitations should be kept in mind when interpreting our results based on ROSAT PSPC data, although their effect should be negligible for the largest, most nearby clusters such as Perseus and Coma.

For 17 clusters spanning redshifts $0.04<z<0.1$, we directly compared the SPA values obtained from ROSAT and \Chandra{}. As expected, the peakiness values from ROSAT are lower, although only by $\sim0.04\pm0.03$ (mean and intrinsic scatter). Alignment and symmetry values are higher by $0.08\pm0.23$ and $0.11\pm0.18$, respectively. Somewhat surprisingly, there is no clear trend with redshift over the range probed (i.e.\ as a function of how well resolved the clusters are), although in the cases of alignment and symmetry a trend could easily be lost in the scatter.

Among the 24 clusters for which we only use ROSAT data, only three are classified as relaxed: Abell\,133, Abell\,780 and Perseus. Each of these meets the SPA criteria with sufficient margin that the above scatter should not affect this determination.

\subsection{Comparison with Other X-ray Image-Based Samples} \label{sec:othersel}

\begin{figure*}
  \centering
  \includegraphics[scale=0.75]{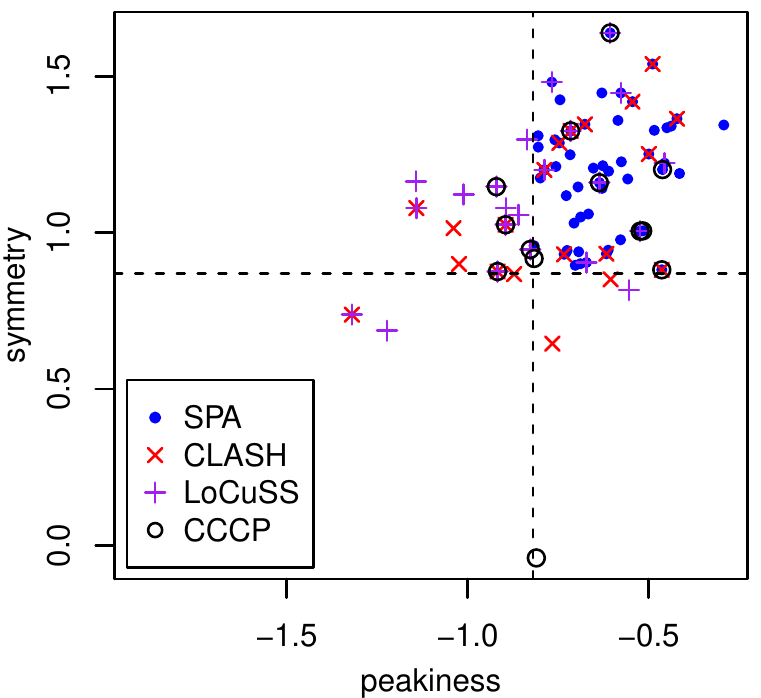}
  \hspace{10mm}
  \includegraphics[scale=0.75]{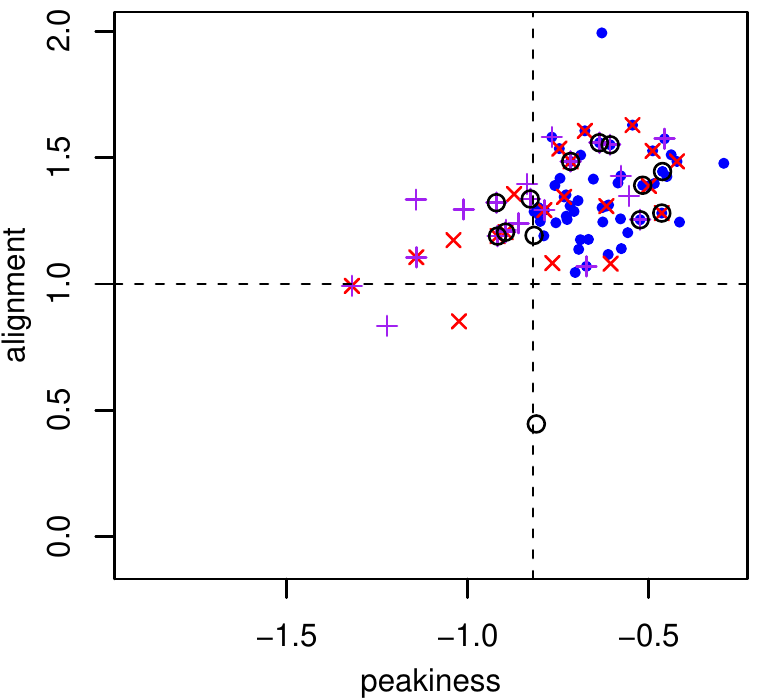}
  \caption[]{
    Peakiness--symmetry and peakiness--alignment distributions from our \Chandra{} analysis. Shown are clusters selected according to our SPA criterion, ``relaxed'' clusters from CLASH \citep{Postman1106.3328}, low centroid variance clusters from LoCuSS \citep{Martino1406.6831}, and low central entropy clusters from the CCCP \citep{Mahdavi1210.3689}.
  }
  \label{fig:morph_other}
\end{figure*}

For reference, we show in \figref~\ref{fig:morph_other} the morphological quantities from our analysis for clusters which have been selected by broadly similar criteria to ours, specifically subsets of the Cluster Lensing And Supernova survey with Hubble (CLASH; \citealt{Postman1106.3328}), the Local Cluster Substructure Survey (LoCuSS; \citealt{Martino1406.6831}), and the Canadian Cluster Comparison Project (CCCP; \citealt{Mahdavi1210.3689}). Significantly, only in the case of CLASH is the cluster selection explicitly described as targeting relaxed systems (in this case, a majority are selected from \arsemf{}). The LoCuSS and CCCP clusters considered here are instead selected based on a single measurement, respectively the centroid variance and central entropy. While only $\sim50$ per cent of the clusters selected in these independent samples typically meet our SPA criterion, they clearly are close to relaxation (by our metric) compared with the cluster population as a whole, as one would expect. (The most obvious outlier in \figref~\ref{fig:morph_other} is Abell~115, selected in CCCP due to the cool core in the northern sub-cluster.)

\subsection{Additional X-ray Morphological Statistics}

\begin{figure*}
  \centering
  \includegraphics[scale=0.75]{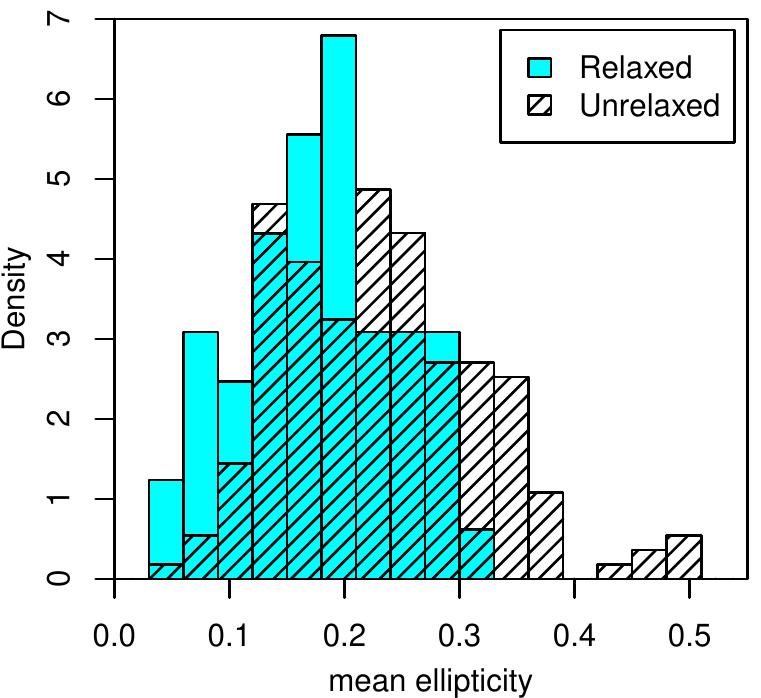}
  \includegraphics[scale=0.75]{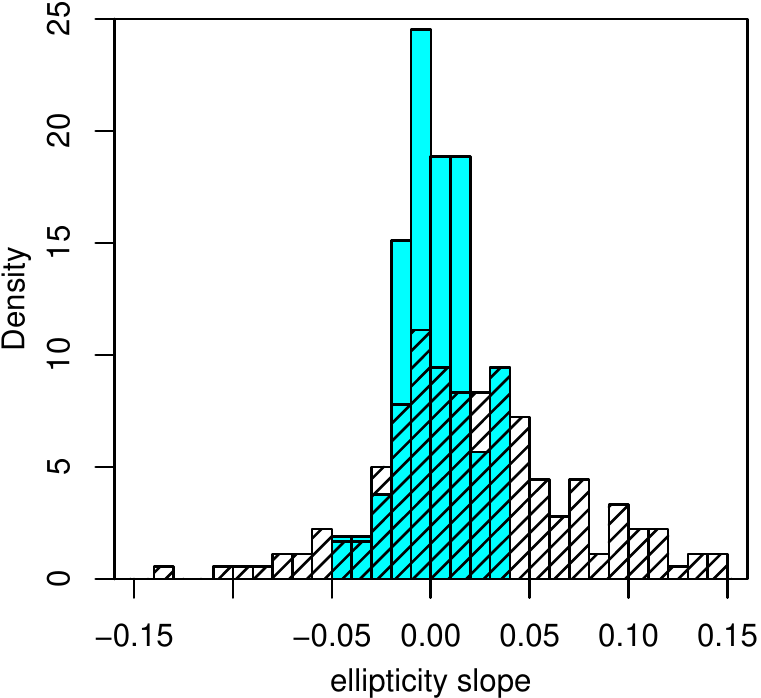}
  \includegraphics[scale=0.75]{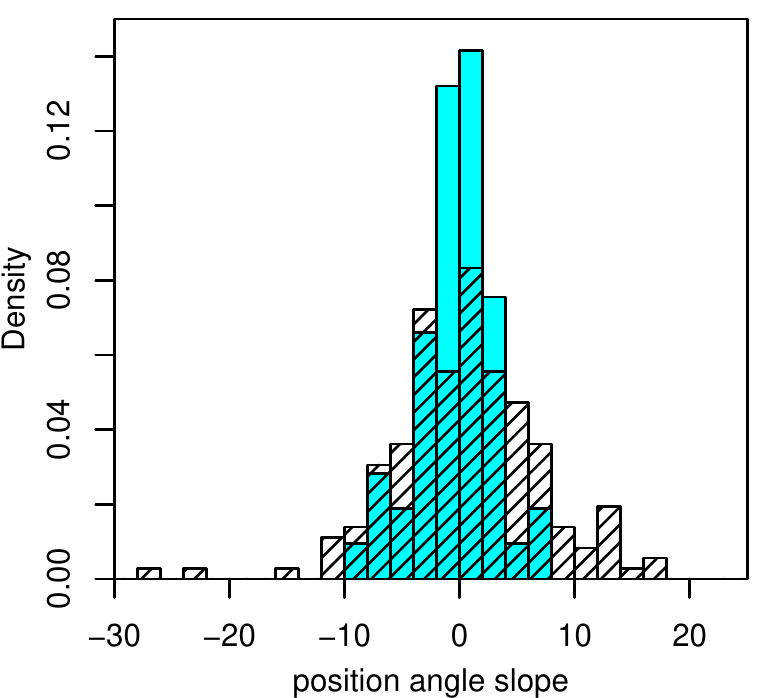}\smallskip\\
  \includegraphics[scale=0.88]{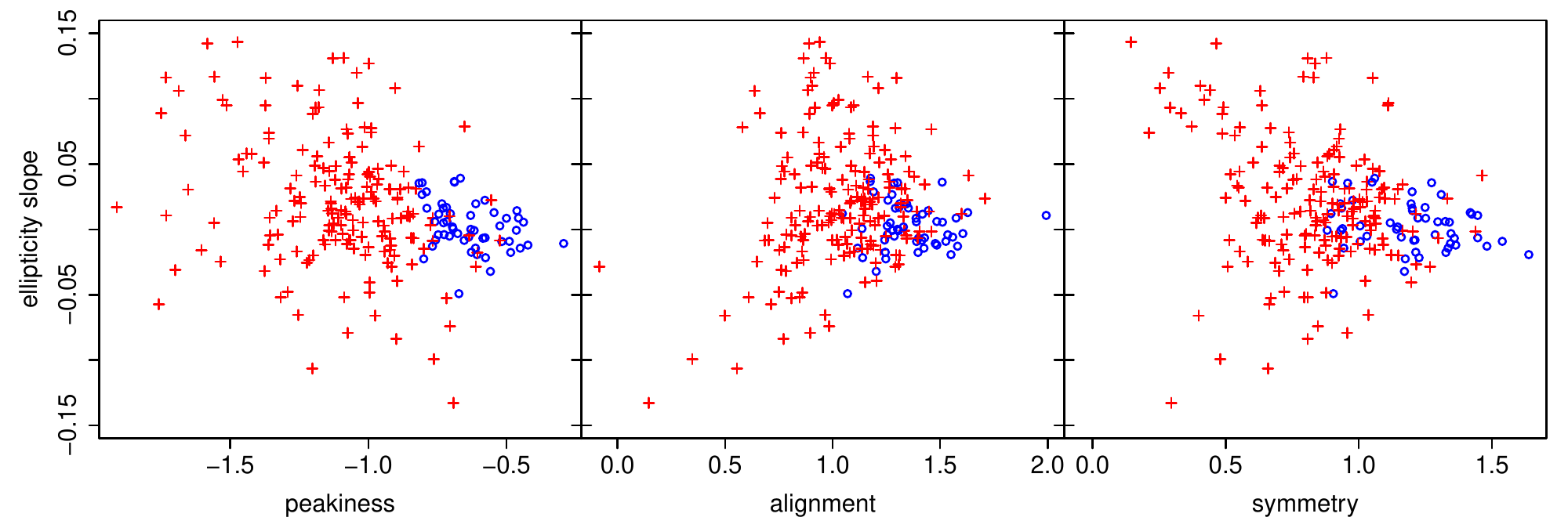}
  \caption[]{
    Top row: Histograms (normalized by sample size) of the mean ellipticity, ellipticity slope (i.e.\ the trend with isophote/radius) and position angle slope for clusters classified as relaxed or unrelaxed, based on the set of elliptical isophote fits generated by our analysis. The relaxed sample has slightly lower (but consistent) ellipticity compared with the unrelaxed sample, and has more consistent ellipticities and position angles as a function of radius.
    Bottom: Ellipticity slope is plotted against our three morphological statistics, with relaxed clusters shown as blue circles, and unrelaxed clusters as red crosses. The least relaxed clusters in terms of alignment and symmetry tend to also be outliers in ellipticity slope.
  }
  \label{fig:ellippa}
  \label{fig:elslope}
\end{figure*}

In this section, we consider three additional morphological quantities which are potentially of interest, but which do not inform our criterion for relaxation. Each of these is a function of the elliptical isophote model fits described in \secref~\ref{sec:ellipses}, namely (1) their mean ellipticity, (2) the change of ellipticity with brightness, and (3) the change of position angle with brightness. The latter two cases we quantify with a ``slope'' obtained by regressing ellipticity or position angle against the index of the isophotes, which is effectively the logarithm of the surface brightness (\eqnref~\ref{eq:sblevels}). \figref~\ref{fig:ellippa} compares histograms of relaxed and unrelaxed clusters for these three quantities.

While the lowest mean ellipticity clusters are relaxed, and the highest unrelaxed, the two distributions overlap considerably. In particular, the excess density of the relaxed distribution at the lowest ellipticities corresponds to only 3 clusters. At large ellipticities, the heavy tail seen in the unrelaxed cluster distribution consists of messy mergers rather than simple, prolate ellipsoids seen in the plane of the sky, and is thus not replicated in the relaxed sample. Discounting this tail, we thus see no evidence that the SPA selection of relaxed clusters is particularly biased towards lower than typical projected ellipticities, i.e.\ clusters likely to be elongated along the line of sight as opposed to in the plane of the sky. This is by construction, since our morphological estimators do not penalize clusters for having ellipsoidal rather than circular shapes in projection. For all clusters, the mean ellipticity is 0.22, with an intrinsic (Gaussian)  scatter of 0.08.

The distributions of ellipticity slope and position angle slope peak near zero for both relaxed and unrelaxed clusters, but are more sharply peaked for relaxed clusters. The difference is particularly evident for the ellipticity slope, which for unrelaxed clusters is asymmetric and has a heavy tail towards positive values (larger ellipticity at smaller radius/greater brightness). The ellipticity slope is plotted against each of the SPA measurements in \figref~\ref{fig:elslope}, which shows that the clusters with the lowest alignment and symmetry also tend to have large absolute values of the ellipticity slope. This is intuitive, as all three indicators should be sensitive to the effects of ongoing merger activity on cluster emission.

\subsection{Trends with Redshift, Temperature and Parent Sample} \label{sec:trends}

\begin{figure}
  \begin{minipage}[c]{0.45\textwidth}
    {\centering
      \includegraphics[scale=0.9]{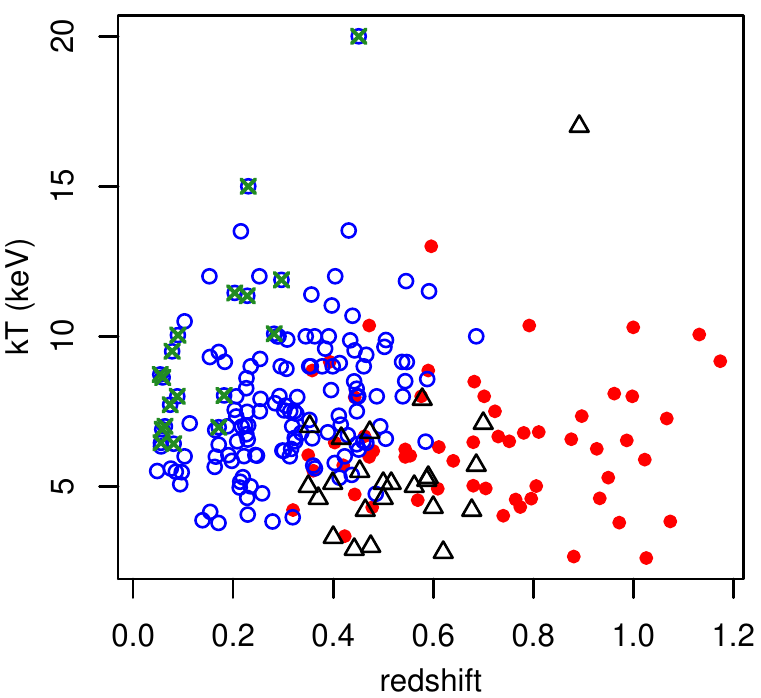}
    }
  \end{minipage}
  \begin{minipage}[c]{0.55\textwidth}
    \caption[]{
      The redshift--temperature distribution of four differently selected cluster populations within our data set: those detected in the X-ray flux-limited ROSAT All-Sky Survey (blue, open circles), the smaller 400 square degree ROSAT survey (black triangles), the SZ-selected SPT cluster survey (red, filled circles), and an SZ-selected subset of the \Planck{} Early SZ catalog (green crosses).
    }
  \end{minipage}
  \label{fig:morph_samples}
\end{figure}

The fraction of clusters that are relaxed as a function of mass and redshift has important implications for cluster cosmology, in addition to astrophysical significance. In this section, we consider four subsets of the data set, defined according to how they were originally selected: from the X-ray flux-limited ROSAT All-Sky Survey (RASS),\footnote{Strictly speaking, the BCS, REFLEX, CIZA and MACS samples, which we collectively call RASS here, were also constructed using different methods to detect cluster emission. However, particularly given the exhaustive optical follow-up and confirmation employed for the RASS samples, these differences are relatively minor.} the 400d ROSAT survey, the SPT-SZ cluster survey, and the \Planck{} Early SZ sample \citep{Planck1303.5089}. Here we remove from consideration the 400d detections at $z<0.35$, for which \Chandra{} follow-up is neither extensive nor systematic. For the \Planck{} sample, we consider only the 30 most significant SZ detections in terms of signal-to-noise, all of which were previously known in our source X-ray catalogs. The resulting sample is thus well represented in our data set, while nevertheless being SZ rather than X-ray selected. To good approximation, this \Planck{} sample, and the \Chandra{} follow-up of SPT clusters, can be considered fair selections of SZ signal-to-noise limited surveys, with the effective mass limit of the \Planck{} sample being somewhat higher. The distribution of each of these samples in redshift and temperature is shown in \figref~\ref{fig:morph_samples}. Note that in this section we use only clusters where our temperatures are based on spectral measurements, as opposed to being estimated using an X-ray luminosity--temperature relation.

In principle, X-ray selected samples should be biased in favor of detecting strongly peaked clusters, due to the enhanced X-ray surface brightness that this implies, and we therefore expect the yield of relaxed clusters to be higher than in other samples. In contrast, SZ selection is not directly dependent on any of the X-ray surface brightness features we have measured. Merging could plausibly affect the SZ detectability of a cluster: in most cases we expect a decrease in the SZ signal for a given mass, since the ICM takes some time to reach its post-merger virial temperature, but the generation of a strong shock could significantly if briefly boost the SZ signal from a merging cluster. A variety of hydrodynamical simulations indicate that the net bias of SZ samples due to mergers should be relatively small \citep{Yang1010.0249,  Rasia1012.4027, Battaglia1109.3709, Krause1107.5740}, although the dependence of these predictions on complex gas physics is such that they must be treated with caution. The uncertain effect of X-ray and SZ selection biases, as well as the relatively large statistical uncertainties, should be kept in mind throughout the following discussion.

\begin{figure*}
  \centering
  \includegraphics[scale=0.8]{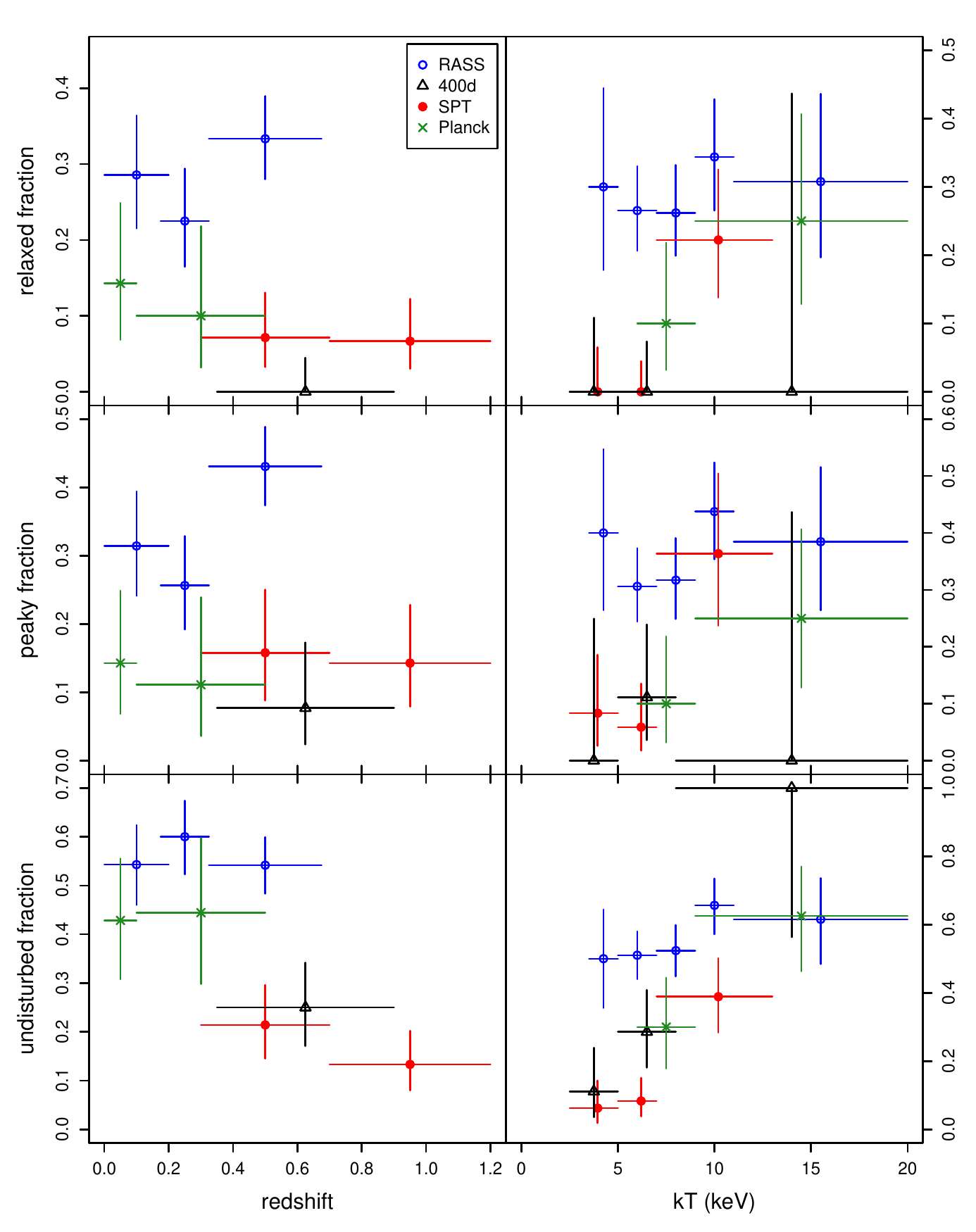}
  \caption[]{
    The fraction of relaxed, peaky and undisturbed clusters as a function of redshift and temperature, as determined for four differently selected cluster populations: those detected in the X-ray flux-limited ROSAT All-Sky Survey (high X-ray luminosity at redshifts $<0.7$), the smaller 400 square degree ROSAT survey (lower luminosities at redshifts $0.35<z<0.9$), the SZ-selected SPT cluster survey, and an SZ-selected subset of the \Planck{} Early SZ catalog. Horizontal bars indicate bins in redshift or temperature, points the fraction in each bin, and vertical bars the corresponding 68.3 per cent confidence intervals (for equivalently selected clusters). In the right panels, SPT points have been offset slightly in $kT$ for clarity. ``Peaky'' refers to clusters which satisfy our peakiness cut, irrespective of symmetry and alignment, and conversely ``undisturbed'' refers to clusters satisfying cuts in symmetry and alignment, irrespective of peakiness.
  }
  \label{fig:relaxed_z_kT}
\end{figure*}

With that caveat in mind, \figref~\ref{fig:relaxed_z_kT} shows, for each cluster sample, the redshift and temperature dependence of three quantities: the fraction of relaxed clusters, the fraction of peaky clusters (satisfying our cut in peakiness, irrespective of symmetry or alignment), and the fraction of ``undisturbed'' clusters (satisfying cuts in symmetry and alignment, irrespective of peakiness). Horizontal bars in the figure show the bins in $z$ or $kT$, points the relaxed, peaky or undisturbed fraction in each bin, and vertical bars the corresponding 68.3 per cent confidence intervals.\footnote{We adopt a uniform prior between 0 and 1 on the fraction of relaxed (or peaky or undisturbed) clusters in a given redshift or temperature bin. With this choice, for a bin where $x$ clusters are found to be relaxed and $y$ unrelaxed, the posterior for the relaxed fraction is the Beta distribution with shape parameters $x+1$ and $y+1$.} In choosing the bins, we have endeavored to make the results for different samples as straightforward to compare as possible, while still having a statistically useful number of clusters in each bin.\footnote{In practice, we aimed to have $\geq10$ clusters in each bin. Matching the approximate redshift and temperature binning across samples sometimes resulted in there being significantly more, $\sim70$ in the case of the most populated bin. The exception is the highest-$kT$ bin for the 400d sample, which contains only 1 cluster.}

Due to selection effects, we expect the X-ray samples to contain a larger fraction of peaky clusters than SZ samples at any redshift or temperature. In fact, since there is also a correlation between peakiness and both symmetry and alignment, this preference should also hold for the undisturbed and relaxed fractions. For the RASS sample this is indeed the case; the relaxed, peaky and undisturbed fractions uniformly exceed those of SZ samples. They are, in addition, approximately constant as a function of both redshift and mass (with the possible exception of the peaky fraction as a function of $z$). Overall, the relaxed cluster fraction of RASS is 29 per cent.

However, the situation is markedly different for the 400d sample, which in all respects appears more similar to the SZ samples (below) than to the RASS sample. In particular, the fraction of peaky clusters in the 400d sample is significantly smaller than in RASS, as has been remarked on previously \citep{Vikhlinin0611438, Mantz2009PhDT........18M, Santos1008.0754}. We find no relaxed clusters in the 400d sample. Note that, while the RASS and 400d samples are essentially disjoint in the X-ray luminosity--redshift plane (e.g.\ \citealt{Allen1103.4829}), they do overlap in both redshift and temperature (a more reliable tracer of mass than luminosity; see the right panel of \figref~\ref{fig:relaxed_z_kT}). The level of disagreement between the two X-ray samples suggests two possible explanations: either the relaxed cluster fraction drops precipitously at relatively high redshifts and low masses, or the selection properties of the two samples are significantly different. For example, wavelet-based detection algorithms designed to automatically reject point-like sources, which the 400d sample employs, could plausibly be biased against finding peaky clusters near the flux limit \citep{Santos1008.0754}.

Taking the SPT and \Planck{} samples together, the relaxed cluster fraction in SZ samples is consistent with being constant with redshift; this behavior is similar to the RASS sample, but the SZ relaxed fraction is lower (8.5 per cent overall). The SZ relaxed fraction is consistent with RASS at high temperatures, $kT \gtsim 10\keV$, but appears to decrease down to zero for cooler clusters, $kT \ltsim 6\keV$. As a function of temperature, the peaky and undisturbed fractions behave similarly, increasing from $\ltsim0.1$ at low temperatures to values comparable to the RASS sample at $\gtsim 10\keV$. In contrast, their trends with redshift differ; the peaky fraction is consistent with a constant, while the undisturbed fraction decreases with $z$. The latter is, however, largely an artifact of the observed $kT$ dependence combined with the differing redshift--temperature distributions of the \Planck{} and SPT samples. Restricting the SPT sample to $kT>6\keV$ (i.e.\ to the range spanned by the \Planck{} clusters) increases its undisturbed fraction to 26 per cent, reducing the evidence of a trend with redshift, while not significantly changing the picture for the peaky fraction.

Both the absolute value of the SZ peaky fraction (14 per cent overall) and its constant behavior with redshift are consistent with the predictions of hydrodynamical simulations \citep{Burns0708.1954, Planelles0906.4024}. However, the same simulations predict a decreasing cool-core fraction with cluster mass, which contradicts the increasing fraction of peaky clusters with temperature observed for the SZ sample. The increase in the undisturbed fraction with temperature, and its decrease with redshift (if real), are also seemingly in contradiction with simulations, which predict a mildly decreasing relaxed fraction (increasing fraction of merging clusters) as a function of mass and a constant merging fraction with redshift \citep{Planelles0906.4024, Fakhouri1001.2304}. Note, however, that these simulations contain relatively few clusters in the mass range of our data set, and generally combine these into a single bin of masses $\gtsim 10^{14}\Msun$. Hence, the simulation results reflect trends with mass between cluster and group scales, not necessarily within the mass range probed by our data.

A strong SZ selection bias favoring mergers, though contrary to expectations, could account for the lack of relaxed clusters at low temperatures in our SZ sample. However, the close agreement of the SZ and 400d results poses a problem for this explanation, since it would need the 400d X-ray selection to be similarly biased in favor of mergers. A simpler scenario is simply that the 400d selection is not biased towards finding strongly peaked clusters, as speculated above, and thus finds clusters morphologically similar to SZ searches. Note that, according to this picture, the lack of cool cores in the 400d sample compared to RASS is not due to its higher redshift coverage (as suggested by \citealt{Vikhlinin0611438}), but rather its lower mass range in combination with different selection effects.

Assuming that the temperature trends seen in the SZ sample are indeed real, they have potentially interesting implications for cool core formation and survival. Specifically, the increasing peaky fraction implies that cool core disruption is more efficient in less massive halos. There are several known examples of cool cores being destroyed by ram pressure stripping as they oscillate (slosh) about the bottom of the cluster potential following a merger (\citealt{Markevitch0001269, Mazzotta0102291, Million0910.0025, Ehlert1010.0253, Ichinohe1410.1955}, Canning et~al., in prep.), a process also observed in hydrodynamic simulations (e.g.\ \citealt{Burns0708.1954, ZuHone1108.4427}). Hence a possible explanation is that mergers with the necessary mass ratio and impact parameter to destroy a hosted cool core via sloshing are relatively less common for the most massive clusters, despite these clusters having a larger merger rate overall; this would be qualitatively consistent with the larger undisturbed fraction we observe for the most massive clusters. Since cool core development is manifestly a non-self-similar phenomenon, it may also be the case that cool cores formed in more massive clusters are intrinsically more resilient to ram pressure stripping by the ambient ICM.

Regardless of the reasons underlying the observed trends, we can make some broad statements about the best strategy for finding new relaxed clusters. Overall, the greatest yield of relaxed clusters can be obtained from an all-sky X-ray survey with greater sensitivity than RASS (such as eROSITA; \citealt{Predehl1001.2502}), provided that the cluster detection algorithm does not reject peaky cool-core clusters. Assuming optical/IR follow-up observations exist, a first cut for selecting relaxed clusters can be made using the X-ray/BCG position offset in all cases. For a fraction of the discovered clusters, it should be possible to make additional, preliminary cuts from the X-ray survey data based on peakiness alone or, for the brightest systems, using the full suite of SPA measurements (adjusting appropriately for image resolution). However, the similarity of the RASS and SZ relaxed fractions at high temperatures strongly suggests that targeted X-ray snapshots of the most significant detections in SZ surveys would be an efficient complement for finding relaxed clusters, particularly at high redshifts where X-ray survey data suffer more from cosmological dimming.

\section{Summary} \label{sec:conclusion}

We have presented a new suite of image measurements used to assess the X-ray morphology of galaxy clusters. These estimators are designed to provide a fair basis for comparison over a wide range in redshift, to avoid strong assumptions regarding the background cosmology and cluster scaling relations, and to be as robust as possible against incomplete images (due to CCD gaps, point-source masks, etc.). The three statistics we use respectively probe the {\it peakiness} of the cluster surface brightness profile, the degree of {\it alignment} between isophotes at intermediate radii, and the {\it symmetry} of those isophotes with respect to a globally determined center. Uncertainties are propagated faithfully by bootstrap sampling the original images and varying the background normalization.

These measurements were performed for a sample of 361 galaxy clusters, selected from several X-ray and SZ cluster surveys, using a combination of archival \Chandra{} and ROSAT observations. There are clear correlations between the new measurements and more traditional X-ray estimators, indicating that they are sensitive to similar features, as expected. Intuitively, our peakiness measure also correlates clearly with the metric distance separating the X-ray center and the BCG. Motivated by trends in the data and comparison with the earlier relaxed cluster sample of \arsemf{}, we define a requirement for a cluster to be considered morphologically relaxed in terms of the symmetry, peakiness and alignment measurements. The fraction of relaxed clusters identified this way is strongly dependent on the selection of the parent sample. We find a higher relaxed fraction in clusters selected from the RASS compared with SZ samples (respectively 0.29 and 0.085), as expected due to the strong dependence of X-ray detectability on surface brightness peakiness. Furthermore, the relaxed fraction in RASS is consistent with being constant with both redshift and ICM temperature, whereas an increasing trend with temperature is observed in the SZ-selected sample.

The relaxed sample identified here, with some refinements based on cluster temperature and data quality, is used to derive cosmological constraints from cluster gas mass fractions in \cosmopaper{}. As described in that work, significant improvements in dark energy constraints using this method will require the efficient identification and follow-up of relaxed clusters discovered in new cluster surveys. The algorithms introduced here provide a widely useful tool for identifying relaxed systems in new data, and for quantifying the morphological states of cluster samples in general.

\section*{Acknowledgments}

ABM was supported by National Science Foundation grants AST-0838187 and AST-1140019. We acknowledge support from the U.S. Department of Energy under contract number DE-AC02-76SF00515, and from the National Aeronautics and Space Administration (NASA) through Chandra Award Numbers GO8-9118X and TM1-12010X, issued by the Chandra X-ray Observatory Center, which is operated by the Smithsonian Astrophysical Observatory for and on behalf of NASA under contract NAS8-03060.

\def \aap {A\&A} 
\def \aapr {A\&AR} 
\def \aaps {A\&AS} 
\def \statisci {Statis. Sci.} 
\def \physrep {Phys. Rep.} 
\def \pre {Phys.\ Rev.\ E} 
\def \sjos {Scand. J. Statis.} 
\def \jrssb {J. Roy. Statist. Soc. B} 
\def \pan {Phys. Atom. Nucl.} 
\def \epja {Eur. Phys. J. A} 
\def \epjc {Eur. Phys. J. C} 
\def \jcap {J. Cosmology Astropart. Phys.} 
\def \ijmpd {Int.\ J.\ Mod.\ Phys.\ D} 
\def \nar {New Astron. Rev.} 
\def \araa {ARA\&A}
\def \aj {AJ}
\def \aar {A\&AR}
\def \apj {ApJ}
\def \apjl {ApJL}
\def \apjs {ApJS}
\def \asl {Adv. Sci. Lett.} 
\def \mnras {MNRAS}
\def \nat {Nat}
\def \pasj {PASJ}
\def \pasp {PASP}
\def \science {Sci}
\def \gca {Geochim.\ Cosmochim.\ Acta}
\def \npa {Nucl.\ Phys.\ A}
\def \plb {Phys.\ Lett.\ B}
\def \prc {Phys.\ Rev.\ C}
\def \prd {Phys.\ Rev.\ D}
\def \prl {Phys.\ Rev.\ Lett.}
\begin{multicols}{2}

\end{multicols}

\appendix

\section{X-ray Data}

\tabref{}s~\ref{tab:data} and \ref{tab:rosatdata} provides details of the \Chandra{} and ROSAT observations employed here.

\begin{table*}
  \begin{center}
    \caption{
      \Chandra{} data used in this work:
      [1] cluster name (ordered as in \tabref~\ref{tab:morph});
      [2] observation ID;
      [3] date of observation;
      [4] blank-sky background epoch;
      [5] detector (ACIS-I or ACIS-S);
      [6] data mode (VFAINT or FAINT; V$\star$ indicates VFAINT reduced in FAINT mode);
      [7] nominal exposure length, ks;
      [8] good time interval remaining after filtering, ks (E indicates an excluded exposure).
    }
    \label{tab:data}
    \vspace{1ex}
    \footnotesize
\begin{tabular}{lrccccrr}
      \hline
      Cluster & ObsID & Date & B & D & M & exp. & gti \\
      \hline
      Abell~1068           &  1652        &  2001-02-04  &  D  &  S  &  F         &  26.8   &  24.3   \\
Abell~1068           &  3068        &  2002-03-11  &  D  &  S  &  F         &  6.0    &  E      \\
Abell~1132           &  13376       &  2011-08-25  &  E  &  I  &  V         &  8.9    &  8.7    \\
Abell~115            &  3233        &  2002-10-07  &  D  &  I  &  V         &  49.7   &  39.7   \\
Abell~1201           &  4216        &  2003-11-01  &  D  &  S  &  V         &  39.7   &  16.4   \\
Abell~1201           &  7697        &  2007-03-18  &  E  &  I  &  V         &  4.9    &  4.7    \\
Abell~1201           &  9616        &  2008-04-06  &  E  &  I  &  V         &  47.4   &  45.6   \\
\hline
    \end{tabular}
  \end{center}
\end{table*}

\begin{table*}
  \begin{center}
    \addtocounter{table}{-1}\caption{continued}
    \vspace{1ex}
    \footnotesize
    \begin{tabular}{lrccccrr}
      \hline
      Cluster & ObsID & Date & B & D & M & exp. & gti \\
      \hline
      Abell~1795           &  14272       &  2012-03-25  &  E  &  I  &  V         &  14.6   &  13.3   \\
Abell~1795           &  14268       &  2012-03-26  &  E  &  S  &  V         &  9.9    &  9.9    \\
Abell~1795           &  14273       &  2012-03-26  &  E  &  I  &  V         &  14.6   &  12.8   \\
Abell~1795           &  14274       &  2012-04-02  &  E  &  I  &  V         &  14.9   &  14.1   \\
Abell~1795           &  14275       &  2012-04-07  &  E  &  I  &  V         &  14.9   &  12.8   \\
Abell~1795           &  14269       &  2012-04-08  &  E  &  S  &  V         &  9.9    &  9.9    \\
Abell~1835           &  495         &  1999-12-11  &  B  &  S  &  F         &  19.5   &  E      \\
\hline
    \end{tabular}
  \end{center}
\end{table*}

\begin{table*}
  \begin{center}
    \addtocounter{table}{-1}\caption{continued}
    \vspace{1ex}
    \footnotesize
    \begin{tabular}{lrccccrr}
      \hline
      Cluster & ObsID & Date & B & D & M & exp. & gti \\
      \hline
      Abell~267            &  1448        &  1999-10-16  &  B  &  I  &  F         &  7.9    &  7.1    \\
Abell~267            &  1517        &  2000-01-29  &  C  &  I  &  V$\star$  &  1.2    &  E      \\
Abell~267            &  3580        &  2003-12-07  &  D  &  I  &  V         &  19.9   &  15.5   \\
Abell~401            &  518         &  1999-09-17  &  B  &  I  &  F         &  18.0   &  17.5   \\
Abell~401            &  2309        &  2000-11-03  &  C  &  I  &  F         &  11.6   &  11.6   \\
Abell~401            &  14024       &  2011-12-14  &  E  &  I  &  V         &  135.1  &  133.5  \\
Abell~478            &  1669        &  2001-01-27  &  D  &  S  &  F         &  42.4   &  36.2   \\
\hline
    \end{tabular}
  \end{center}
\end{table*}

\begin{table*}
  \begin{center}
    \addtocounter{table}{-1}\caption{continued}
    \vspace{1ex}
    \footnotesize
    \begin{tabular}{lrccccrr}
      \hline
      Cluster & ObsID & Date & B & D & M & exp. & gti \\
      \hline
      Zwicky~3146          &  9371        &  2008-01-18  &  E  &  I  &  V         &  40.2   &  34.5   \\
Zwicky~5029          &  4184        &  2003-07-28  &  D  &  I  &  V         &  12.1   &  9.7    \\
Zwicky~5247          &  539         &  2000-03-23  &  C  &  I  &  V$\star$  &  9.3    &  8.2    \\
Zwicky~5247          &  11727       &  2009-11-20  &  E  &  I  &  V         &  20.9   &  19.6   \\
Zwicky~7160          &  543         &  2000-05-19  &  C  &  I  &  F         &  9.9    &  7.9    \\
Zwicky~7160          &  4192        &  2003-09-05  &  D  &  I  &  V         &  91.9   &  71.6   \\
Zwicky~7160          &  7709        &  2007-03-23  &  E  &  I  &  V         &  7.1    &  6.8    \\
\hline
    \end{tabular}
  \end{center}
\end{table*}

\begin{table*}
  \begin{center}
    \addtocounter{table}{-1}\caption{continued}
    \vspace{1ex}
    \footnotesize
    \begin{tabular}{lrccccrr}
      \hline
      Cluster & ObsID & Date & B & D & M & exp. & gti \\
      \hline
      Abell~545            &  13998       &  2011-12-16  &  E  &  S  &  F         &  29.6   &  29.6   \\
Abell~754            &  577         &  1999-10-30  &  B  &  I  &  F         &  44.2   &  25.6   \\
Abell~754            &  6794        &  2006-01-27  &  E  &  I  &  V         &  9.9    &  9.1    \\
Abell~754            &  10743       &  2009-02-17  &  E  &  I  &  V         &  94.1   &  88.0   \\
Abell~907            &  535         &  2000-06-29  &  C  &  I  &  V$\star$  &  11.0   &  10.2   \\
Abell~907            &  3185        &  2002-06-14  &  D  &  I  &  V         &  48.0   &  41.9   \\
Abell~907            &  3205        &  2002-10-30  &  D  &  I  &  V         &  47.1   &  31.5   \\
\hline
    \end{tabular}
  \end{center}
\end{table*}

\begin{table*}
  \begin{center}
    \addtocounter{table}{-1}\caption{continued}
    \vspace{1ex}
    \footnotesize
    \begin{tabular}{lrccccrr}
      \hline
      Cluster & ObsID & Date & B & D & M & exp. & gti \\
      \hline
      MACS~J0140.0$-$0555  &  5013        &  2004-06-04  &  D  &  I  &  V         &  10.2   &  8.9    \\
MACS~J0140.0$-$0555  &  12243       &  2010-09-15  &  E  &  I  &  V         &  19.3   &  18.4   \\
MACS~J0150.3$-$1005  &  11711       &  2009-09-14  &  E  &  I  &  V         &  26.8   &  26.1   \\
MACS~J0152.5$-$2852  &  3264        &  2002-09-17  &  D  &  I  &  V         &  17.5   &  14.7   \\
MACS~J0159.0$-$3412  &  5818        &  2006-02-19  &  E  &  I  &  V         &  9.4    &  8.7    \\
MACS~J0159.8$-$0849  &  3265        &  2002-10-02  &  D  &  I  &  V         &  17.9   &  14.6   \\
MACS~J0159.8$-$0849  &  6106        &  2004-12-04  &  D  &  I  &  V         &  35.3   &  31.0   \\
\hline
    \end{tabular}
  \end{center}
\end{table*}

\begin{table*}
  \begin{center}
    \addtocounter{table}{-1}\caption{continued}
    \vspace{1ex}
    \footnotesize
    \begin{tabular}{lrccccrr}
      \hline
      Cluster & ObsID & Date & B & D & M & exp. & gti \\
      \hline
      MACS~J1359.1$-$1929  &  5811        &  2005-03-17  &  D  &  I  &  V         &  9.9    &  8.9    \\
MACS~J1359.1$-$1929  &  9378        &  2008-03-21  &  E  &  I  &  V         &  49.4   &  45.5   \\
MACS~J1423.8+2404    &  1657        &  2001-06-01  &  D  &  I  &  V         &  18.5   &  16.2   \\
MACS~J1423.8+2404    &  4195        &  2003-08-18  &  D  &  S  &  V         &  115.6  &  106.5  \\
MACS~J1427.2+4407    &  6112        &  2005-02-12  &  D  &  I  &  V         &  9.4    &  8.4    \\
MACS~J1427.2+4407    &  9380        &  2008-01-14  &  E  &  I  &  V         &  25.8   &  23.3   \\
MACS~J1427.2+4407    &  9808        &  2008-01-15  &  E  &  I  &  V         &  14.9   &  13.1   \\
\hline
    \end{tabular}
  \end{center}
\end{table*}

\begin{table*}
  \begin{center}
    \addtocounter{table}{-1}\caption{continued}
    \vspace{1ex}
    \footnotesize
    \begin{tabular}{lrccccrr}
      \hline
      Cluster & ObsID & Date & B & D & M & exp. & gti \\
      \hline
      CL~J0106+3209      &  3555        &  2003-08-06  &  D  &  S  &  F         &  5.1    &  5.1    \\
CL~J0124+0932      &  6778        &  2006-08-29  &  E  &  S  &  V         &  13.7   &  8.1    \\
CL~J0141$-$3034      &  5778        &  2005-06-04  &  D  &  I  &  V         &  29.7   &  25.3   \\
CL~J0152$-$1357      &  913         &  2000-09-08  &  C  &  I  &  F         &  36.5   &  31.9   \\
CL~J0159+0030        &  5777        &  2005-09-02  &  E  &  I  &  V         &  19.9   &  18.1   \\
CL~J0216$-$1747      &  5760        &  2005-09-07  &  E  &  I  &  V         &  40.0   &  8.7    \\
CL~J0216$-$1747      &  6393        &  2005-10-04  &  E  &  I  &  V         &  26.6   &  21.8   \\
\hline
    \end{tabular}
  \end{center}
\end{table*}

\begin{table*}
  \begin{center}
    \addtocounter{table}{-1}\caption{continued}
    \vspace{1ex}
    \footnotesize
    \begin{tabular}{lrccccrr}
      \hline
      Cluster & ObsID & Date & B & D & M & exp. & gti \\
      \hline
      CL~J1231+4137      &  4726        &  2004-11-20  &  D  &  S  &  V         &  39.6   &  36.6   \\
CL~J1236+1224      &  8098        &  2008-02-24  &  E  &  S  &  F         &  5.1    &  5.1    \\
CL~J1248$-$0548    &  784         &  2000-01-15  &  B  &  S  &  F         &  39.3   &  38.2   \\
CL~J1248$-$0548    &  4727        &  2003-12-26  &  D  &  S  &  V         &  39.9   &  35.8   \\
CL~J1248$-$0548    &  4728        &  2004-01-06  &  D  &  S  &  V         &  35.7   &  17.3   \\
CL~J1248$-$0548    &  4729        &  2004-02-12  &  D  &  S  &  V         &  38.1   &  E      \\
CL~J1248$-$0548    &  4730        &  2004-08-18  &  D  &  S  &  V         &  40.0   &  33.9   \\
\hline
    \end{tabular}
  \end{center}
\end{table*}

\begin{table*}
  \begin{center}
    \addtocounter{table}{-1}\caption{continued}
    \vspace{1ex}
    \footnotesize
    \begin{tabular}{lrccccrr}
      \hline
      Cluster & ObsID & Date & B & D & M & exp. & gti \\
      \hline
      SPT-CL~J0441$-$4854  &  13475       &  2011-12-10  &  E  &  I  &  V         &  24.7   &  21.5   \\
SPT-CL~J0446$-$5849  &  13482       &  2012-10-11  &  E  &  I  &  V         &  32.9   &  29.4   \\
SPT-CL~J0446$-$5849  &  15560       &  2012-10-14  &  E  &  I  &  V         &  20.7   &  17.1   \\
SPT-CL~J0449$-$4901  &  13473       &  2012-06-29  &  E  &  I  &  V         &  51.0   &  44.8   \\
SPT-CL~J0456$-$5116  &  13474       &  2012-05-08  &  E  &  I  &  V         &  50.2   &  44.6   \\
SPT-CL~J0509$-$5342  &  9432        &  2009-06-03  &  E  &  I  &  V         &  28.8   &  26.7   \\
SPT-CL~J0528$-$5300  &  9341        &  2009-01-26  &  E  &  I  &  V         &  14.8   &  14.3   \\
\hline
    \end{tabular}
  \end{center}
\end{table*}

\begin{table*}
  \begin{center}
    \caption{
      ROSAT data used in this work:
      [1] cluster name
      [2] observation ID;
      [3] date of observation;
      [4] nominal exposure length, ks.
    }
    \label{tab:rosatdata}
    \vspace{1ex}
    \footnotesize
    \begin{tabular}{lccr}
      \hline
      Cluster & ObsID & Date & exp. \\
      \hline
      Abell~119        &  rp800251n00  &  1992-07-11  &  14.7  \\
Abell~133        &  rp800319n00  &  1992-07-05  &  18.7  \\
Abell~1651       &  rp800353n00  &  1992-07-18  &  7.2   \\
Abell~1795       &  rp700145a00  &  1991-07-01  &  17.7  \\
Abell~1795       &  rp800055n00  &  1991-07-01  &  25.0  \\
Abell~1795       &  rp800105n00  &  1992-01-04  &  35.1  \\
Abell~1795       &  rp700145a01  &  1992-01-06  &  1.9   \\
\hline
    \end{tabular}
  \end{center}
\end{table*}

\begin{table*}
  \begin{center}
    \addtocounter{table}{-1}\caption{continued}
    \vspace{1ex}
    \footnotesize
    \begin{tabular}{lccr}
      \hline
      Cluster & ObsID & Date & exp. \\
      \hline
      Abell~85         &  rp800174a00  &  1991-12-20  &  2.1   \\
Abell~85         &  rp800174a01  &  1992-06-11  &  3.3   \\
Abell~85         &  rp800250n00  &  1992-07-01  &  9.9   \\
AWM~7        &  rp800168n00  &  1992-01-28  &  12.7  \\
Coma        &  rp800006n00  &  1991-06-16  &  20.8  \\
Coma        &  rp800009n00  &  1991-06-16  &  19.7  \\
Coma        &  rp800005n00  &  1991-06-17  &  20.4  \\
Coma        &  rp800013n00  &  1991-06-18  &  20.7  \\
Cygnus~A     &  rp701152n00  &  1992-11-06  &  3.9   \\
Cygnus~A     &  rp800622n00  &  1993-10-10  &  9.1   \\
Hydra~A      &  rp800318n00  &  1992-11-08  &  17.8  \\
Ophiuchus   &  rp300214n00  &  1992-09-20  &  13.5  \\
Ophiuchus   &  rp800279n00  &  1992-09-20  &  3.8   \\
Perseus     &  rp800033n00  &  1991-02-09  &  14.6  \\
Perseus     &  rp800034a00  &  1991-02-09  &  9.9   \\
Perseus     &  rp800035a00  &  1991-02-09  &  5.8   \\
Perseus     &  rp800035a01  &  1991-08-18  &  23.7  \\
Perseus     &  rp800034a01  &  1991-08-19  &  23.3  \\
Perseus     &  rp800032a01  &  1991-09-02  &  29.6  \\
Perseus     &  rp800186n00  &  1992-02-02  &  4.5   \\
PKS~0745$-$191  &  rp800623n00  &  1993-10-15  &  10.1  \\
RX~J2344.2$-$0422    &  rp800333n00  &  1992-06-17  &  4.8   \\
RX~J2344.2$-$0422     &  rp800333a01  &  1992-12-09  &  9.8   \\
Triangulum  &  rp800280n00  &  1992-09-26  &  7.1   \\
Zwicky~5029    &  rp600166n00  &  1991-12-14  &  12.3  \\

      \hline
    \end{tabular}
  \end{center}
\end{table*}

\section{Complete Results}

\tabref~\ref{tab:allresults} extends the listing of results in \tabref~\ref{tab:morph} to the entire data set.

\begin{table*}
  \begin{center}
    \caption{
      Continuation of \tabref~\ref{tab:morph}.
    }
    \label{tab:allresults}
    \vspace{1ex}
    \footnotesize
    \begin{tabular}{@{}c@{\hspace{0ex}}c@{}c@{\hspace{1ex}}c@{\hspace{1ex}}r@{}c@{}l@{\hspace{2ex}}r@{}c@{}l@{\hspace{2ex}}r@{}c@{}l@{\hspace{1ex}}c@{\hspace{1ex}}r@{}c@{}l@{\hspace{2ex}}r@{\hspace{1ex}}r@{\hspace{1ex}}r@{\hspace{1ex}}r@{}}
      \hline
      Cat. & Name & Obs. & flags & \multicolumn{3}{c}{$s$} & \multicolumn{3}{c}{$p$} & \multicolumn{3}{c}{$a$} & Rel. & \multicolumn{3}{c}{ellip.} & \multicolumn{1}{c}{RA$_\mathrm{X}$} & \multicolumn{1}{c}{Dec$_\mathrm{X}$} & \multicolumn{1}{c}{RA$_\mathrm{BCG}$} & \multicolumn{1}{c}{Dec$_\mathrm{BCG}$} \\
      \hline
      B         &  Abell~1763         &  C  &  l       &  $       0.686$  &      $\pm$  &         $0.097$   &         $-1.223$  &        $\pm$    &        $0.076$  &        $      0.83$  &      $\pm$   &       $0.05$  &        ~        &         0.361     &           $\pm$       &         0.015     &           203.8230    &           $           41.0011$  &         203.8337    &           $         41.0011$  \\
B         &  Abell~1795         &  C  &  a       &  $       0.780$  &      $\pm$  &         $0.008$   &         $-0.856$  &        $\pm$    &        $0.004$  &        $      1.39$  &      $\pm$   &       $0.02$  &        ~        &         0.247     &           $\pm$       &         0.003     &           207.2192    &           $           26.5913$  &         207.2188    &           $         26.5929$  \\
B         &  Abell~1835         &  C  &  alp$f$  &  $       1.160$  &      $\pm$  &         $0.020$   &         $-0.636$  &        $\pm$    &        $0.010$  &        $      1.56$  &      $\pm$   &       $0.04$  &        $\surd$  &         0.122     &           $\pm$       &         0.007     &           210.2583    &           $           2.8783$   &         210.2586    &           $         2.8785$   \\
\hline
    \end{tabular}
  \end{center}
\end{table*}

\begin{table*}
  \begin{center}
    \addtocounter{table}{-1}\caption{continued}
    \vspace{1ex}
    \footnotesize
    \begin{tabular}{@{}c@{\hspace{0ex}}c@{}c@{\hspace{1ex}}c@{\hspace{1ex}}r@{}c@{}l@{\hspace{2ex}}r@{}c@{}l@{\hspace{2ex}}r@{}c@{}l@{\hspace{1ex}}c@{\hspace{1ex}}r@{}c@{}l@{\hspace{2ex}}r@{\hspace{1ex}}r@{\hspace{1ex}}r@{\hspace{1ex}}r@{}}
      \hline
      Cat. & Name & Obs. & flags & \multicolumn{3}{c}{$s$} & \multicolumn{3}{c}{$p$} & \multicolumn{3}{c}{$a$} & Rel. & \multicolumn{3}{c}{ellip.} & \multicolumn{1}{c}{RA$_\mathrm{X}$} & \multicolumn{1}{c}{Dec$_\mathrm{X}$} & \multicolumn{1}{c}{RA$_\mathrm{BCG}$} & \multicolumn{1}{c}{Dec$_\mathrm{BCG}$} \\
      \hline
      R         &  Abell~3112         &  C  &  &       $  0.933$  &       $\pm$  &      $0.023$   &         $-0.844$  &         $\pm$    &        $0.014$  &        $        1.13$  &      $\pm$  &       $0.04$  &       ~        &        0.283     &         $\pm$       &           0.006     &         49.4899     &           $-44.2384$  &           ~         &         ~           \\
R         &  Abell~3158         &  C  &  &       $  1.092$  &       $\pm$  &      $0.143$   &         $-1.362$  &         $\pm$    &        $0.185$  &        $        1.03$  &      $\pm$  &       $0.06$  &       ~        &        0.189     &         $\pm$       &           0.010     &         55.7225     &           $-53.6296$  &           ~         &         ~           \\
R         &  Abell~3266         &  R  &  &       $  0.843$  &       $\pm$  &      $0.055$   &         $-1.546$  &         $\pm$    &        $0.234$  &        $        1.07$  &      $\pm$  &       $0.05$  &       ~        &        0.226     &         $\pm$       &           0.015     &         67.8061     &           $-61.4545$  &           ~         &         ~           \\
\hline
    \end{tabular}
  \end{center}
\end{table*}

\begin{table*}
  \begin{center}
    \addtocounter{table}{-1}\caption{continued}
    \vspace{1ex}
    \footnotesize
    \begin{tabular}{@{}c@{\hspace{0ex}}c@{}c@{\hspace{1ex}}c@{\hspace{1ex}}r@{}c@{}l@{\hspace{2ex}}r@{}c@{}l@{\hspace{2ex}}r@{}c@{}l@{\hspace{1ex}}c@{\hspace{1ex}}r@{}c@{}l@{\hspace{2ex}}r@{\hspace{1ex}}r@{\hspace{1ex}}r@{\hspace{1ex}}r@{}}
      \hline
      Cat. & Name & Obs. & flags & \multicolumn{3}{c}{$s$} & \multicolumn{3}{c}{$p$} & \multicolumn{3}{c}{$a$} & Rel. & \multicolumn{3}{c}{ellip.} & \multicolumn{1}{c}{RA$_\mathrm{X}$} & \multicolumn{1}{c}{Dec$_\mathrm{X}$} & \multicolumn{1}{c}{RA$_\mathrm{BCG}$} & \multicolumn{1}{c}{Dec$_\mathrm{BCG}$} \\
      \hline
      M         &  J0140.0$-$0555     &  C  &  &       $  0.656$  &       $\pm$  &      $0.072$   &         $-1.079$  &         $\pm$    &        $0.076$  &        $        0.85$  &      $\pm$  &       $0.08$  &       ~        &        0.301     &         $\pm$       &           0.033     &         25.0067     &           $           -5.9188$    &         25.0033   &           $           -5.9174$  \\
M         &  J0150.3$-$1005     &  C  &  $f$     &  $       1.146$  &      $\pm$  &         $0.103$   &         $-0.695$  &        $\pm$    &        $0.035$  &        $      1.33$  &      $\pm$   &       $0.13$  &        $\surd$  &         0.198     &           $\pm$       &         0.036     &           27.5887     &           $-10.0915$  &         27.5887   &           $-10.0918$  \\
M         &  J0152.5$-$2852     &  C  &  &       $  0.703$  &       $\pm$  &      $0.064$   &         $-0.918$  &         $\pm$    &        $0.044$  &        $        0.90$  &      $\pm$  &       $0.08$  &       ~        &        0.274     &         $\pm$       &           0.031     &         28.1436     &           $-28.8937$  &           ~         &         ~           \\
\hline
    \end{tabular}
  \end{center}
\end{table*}

\begin{table*}
  \begin{center}
    \addtocounter{table}{-1}\caption{continued}
    \vspace{1ex}
    \footnotesize
    \begin{tabular}{@{}c@{\hspace{0ex}}c@{}c@{\hspace{1ex}}c@{\hspace{1ex}}r@{}c@{}l@{\hspace{2ex}}r@{}c@{}l@{\hspace{2ex}}r@{}c@{}l@{\hspace{1ex}}c@{\hspace{1ex}}r@{}c@{}l@{\hspace{2ex}}r@{\hspace{1ex}}r@{\hspace{1ex}}r@{\hspace{1ex}}r@{}}
      \hline
      Cat. & Name & Obs. & flags & \multicolumn{3}{c}{$s$} & \multicolumn{3}{c}{$p$} & \multicolumn{3}{c}{$a$} & Rel. & \multicolumn{3}{c}{ellip.} & \multicolumn{1}{c}{RA$_\mathrm{X}$} & \multicolumn{1}{c}{Dec$_\mathrm{X}$} & \multicolumn{1}{c}{RA$_\mathrm{BCG}$} & \multicolumn{1}{c}{Dec$_\mathrm{BCG}$} \\
      \hline
      4         &  3C~288             &  C  &  &       $  1.175$  &       $\pm$  &      $0.093$   &         $-0.800$  &         $\pm$    &        $0.048$  &        $        1.25$  &      $\pm$  &       $0.09$  &       $\surd$  &        0.126     &         $\pm$       &           0.028     &         204.7078    &           $           38.8531$    &         ~         &           ~           \\
4         &  3C~88              &  C  &  &       $  0.554$  &       $\pm$  &      $0.199$   &         $-1.014$  &         $\pm$    &        $0.060$  &        $        0.58$  &      $\pm$  &       $0.28$  &       ~        &        0.331     &         $\pm$       &           0.155     &         51.9757     &           $           2.5617$     &         ~         &           ~           \\
4         &  Abell~1775         &  C  &  &       $  0.291$  &       $\pm$  &      $0.012$   &         $-1.192$  &         $\pm$    &        $0.024$  &        $        0.92$  &      $\pm$  &       $0.03$  &       ~        &        0.116     &         $\pm$       &           0.010     &         205.4527    &           $           26.3732$    &         205.4548  &           $           26.3735$  \\
\hline
    \end{tabular}
  \end{center}
\end{table*}

\begin{table*}
  \begin{center}
    \addtocounter{table}{-1}\caption{continued}
    \vspace{1ex}
    \footnotesize
    \begin{tabular}{@{}c@{\hspace{0ex}}c@{}c@{\hspace{1ex}}c@{\hspace{1ex}}r@{}c@{}l@{\hspace{2ex}}r@{}c@{}l@{\hspace{2ex}}r@{}c@{}l@{\hspace{1ex}}c@{\hspace{1ex}}r@{}c@{}l@{\hspace{2ex}}r@{\hspace{1ex}}r@{\hspace{1ex}}r@{\hspace{1ex}}r@{}}
      \hline
      Cat. & Name & Obs. & flags & \multicolumn{3}{c}{$s$} & \multicolumn{3}{c}{$p$} & \multicolumn{3}{c}{$a$} & Rel. & \multicolumn{3}{c}{ellip.} & \multicolumn{1}{c}{RA$_\mathrm{X}$} & \multicolumn{1}{c}{Dec$_\mathrm{X}$} & \multicolumn{1}{c}{RA$_\mathrm{BCG}$} & \multicolumn{1}{c}{Dec$_\mathrm{BCG}$} \\
      \hline
      4         &  J1701+6414         &  C  &  &       $  0.877$  &       $\pm$  &      $0.150$   &         $-0.996$  &         $\pm$    &        $0.113$  &        $        0.86$  &      $\pm$  &       $0.24$  &       ~        &        0.118     &         $\pm$       &           0.045     &         255.1738    &           $           64.2166$    &         255.1740  &           $           64.2163$  \\
4         &  J1807+6946         &  C  &  &       ~  &       &       ~      &      ~         &         &         ~         &        ~        &        &        ~        &      ~      &      ~       &       &       ~        &        271.7308  &         $           69.7967$    &         ~         &           ~           \\
4         &  J2137+0027         &  C  &  &       $  0.647$  &       $\pm$  &      $0.188$   &         $-0.524$  &         $\pm$    &        $0.137$  &        $        0.68$  &      $\pm$  &       $0.30$  &       ~        &        0.210     &         $\pm$       &           0.060     &         324.2849    &           $           0.4464$     &         324.2849  &           $           0.4464$   \\
\hline
    \end{tabular}
  \end{center}
\end{table*}

\end{document}